\documentclass{iopjournal}
\usepackage[utf8]{inputenc}
\usepackage[version=4]{mhchem}
\usepackage{rotating}
\usepackage{pdflscape}

\usepackage{booktabs}
\usepackage{tabularx}

\usepackage[most]{tcolorbox}

\newtcolorbox{terminologybox}{
  colback=gray!7,
  colframe=gray!55,
  boxrule=0.6pt,
  arc=2pt,
  left=6pt,
  right=6pt,
  top=3pt,
  bottom=3pt,
  boxsep=0.5mm,
  before skip=6pt,
  after skip=8pt,
  fontupper=\small,
  fonttitle=\bfseries\small,
  title=Terminology in this review
}

\usepackage{aas_macros, amssymb, graphicx}
\usepackage{amsmath}
\usepackage[export]{adjustbox}

\providecommand{\ion}[2]{#1\,\textsc{\romannumeral#2}}

\def\mr{\mathrm}
\def\d{\mr{d}}

\def\mc{\mathcal}

\def\me{m_{\rm e}}
\def\mproton{m_{\rm p}}

\def\kB{k_{\rm B}}

\def\angstrom{\text{\r{A}}}

\newcommand{\good}{\ensuremath{\checkmark}}
\newcommand{\partialok}{\ensuremath{\triangle}}
\newcommand{\bad}{\ensuremath{\times}}
\newcommand{\unknown}{\ensuremath{?}}

\newcommand{\lrb}[1]{\left({#1}\right)}
\newcommand{\lrsb}[1]{\left[{#1}\right]}

\newcommand{\lara}[1]{\left\langle{#1}\right\rangle}

\def\msunyr{M_\odot\mr{\,yr^{-1}}}

\newcommand{\farcs}{%
  \mbox{%
    \kern0.1em
    \rlap{.}%
    \kern0.05em
    $^{\prime\prime}$%
  }%
}

\begin{document}

\articletype{Review} 

\title{Cows, Tasmanian Devils, and Other Mysteries: \\
Diversity and Origins of Luminous Fast Blue Optical Transients}

\author{Anna Y. Q. Ho$^{1*}$\orcid{0000-0000-0000-0000} and Wenbin Lu$^{2*}$\orcid{0000-0000-0000-0000}}

\affil{$^1$Department of Astronomy, Cornell University, Ithaca, NY 14853, USA}

\affil{$^2$Department of Astronomy, University of California, Berkeley, CA 94720-3411, USA}

\affil{$^*$Author to whom any correspondence should be addressed.}

\email{annayqho@cornell.edu, wenbinlu@berkeley.edu}


\begin{abstract}

By repeatedly mapping the sky, optical telescopes have unveiled a diverse landscape of transients, ephemeral celestial events arising from the destruction of stars and stellar remnants. 
In this review, we focus on a recently identified phenomenon referred to as luminous fast blue optical transients (LFBOTs) and typified by the transient AT2018cow (``The Cow''). AT2018cow was distinguished by luminous emission at X-ray and radio wavelengths in addition to luminous, rapidly evolving, and blue optical emission. 
Certain LFBOTs have properties that have not been seen in any other class of transients, such as minutes-duration optical flaring in AT2022tsd (``The Tasmanian Devil'').
Determining the physical origin of LFBOTs is likely important for understanding stellar evolution and the formation and growth of compact objects such as black holes. 
We describe how LFBOTs were discovered, their basic observational properties, and the models that have been proposed to explain them.
We conclude by highlighting promising avenues for resolving the mystery of their origins.  

\end{abstract}

\renewcommand{\contentsname}{Outline}
\setcounter{tocdepth}{3} 
\tableofcontents

\section{Introduction to Fast Blue Optical Transients and The Cow}
\label{sec:introduction}

For a century, optical telescopes have been used to 
search the night sky for supernovae \cite{1958HDP....51..766Z}, with over 10,000 discovered and classified to date (e.g., \cite{2020ApJ...904...35P}). Supernovae are classified based on their optical spectra into two predominant types: 
thermonuclear-powered supernovae (explosions of white dwarfs; Type~Ia), and supernovae triggered by the gravitational collapse of massive stars (``core-collapse supernovae'' of Type~II if hydrogen is present in the spectrum or Type~Ibc if hydrogen is absent) \cite{1997ARA&A..35..309F}. Supernovae are rare (about one per century per galaxy) and transpire on short timescales. For example, in a core collapse supernova the star's core collapses in about one second\footnote{The collapse duration is set by the gravitational free-fall time $t_\mathrm{ff}\sim (G\rho)^{-1/2}$, where $G$ is the gravitational constant and $\rho$ is the density of the core.}. Triggered by the collapse, a shockwave propagates through the outer layers of the star in about one hour, unbinding the star in an explosion \cite{2005NatPh...1..147W}. 

\begin{figure}[tb]
    \centering
    \includegraphics[width=0.35\linewidth, valign=c]{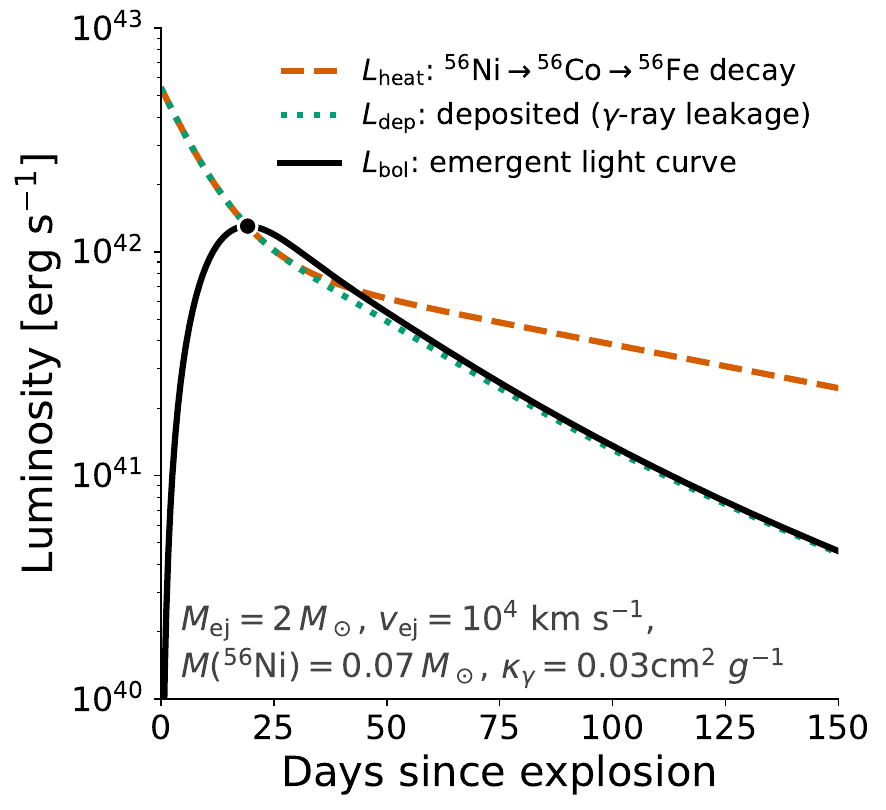}
    \hfill
    \includegraphics[width=0.64\linewidth, valign=c]{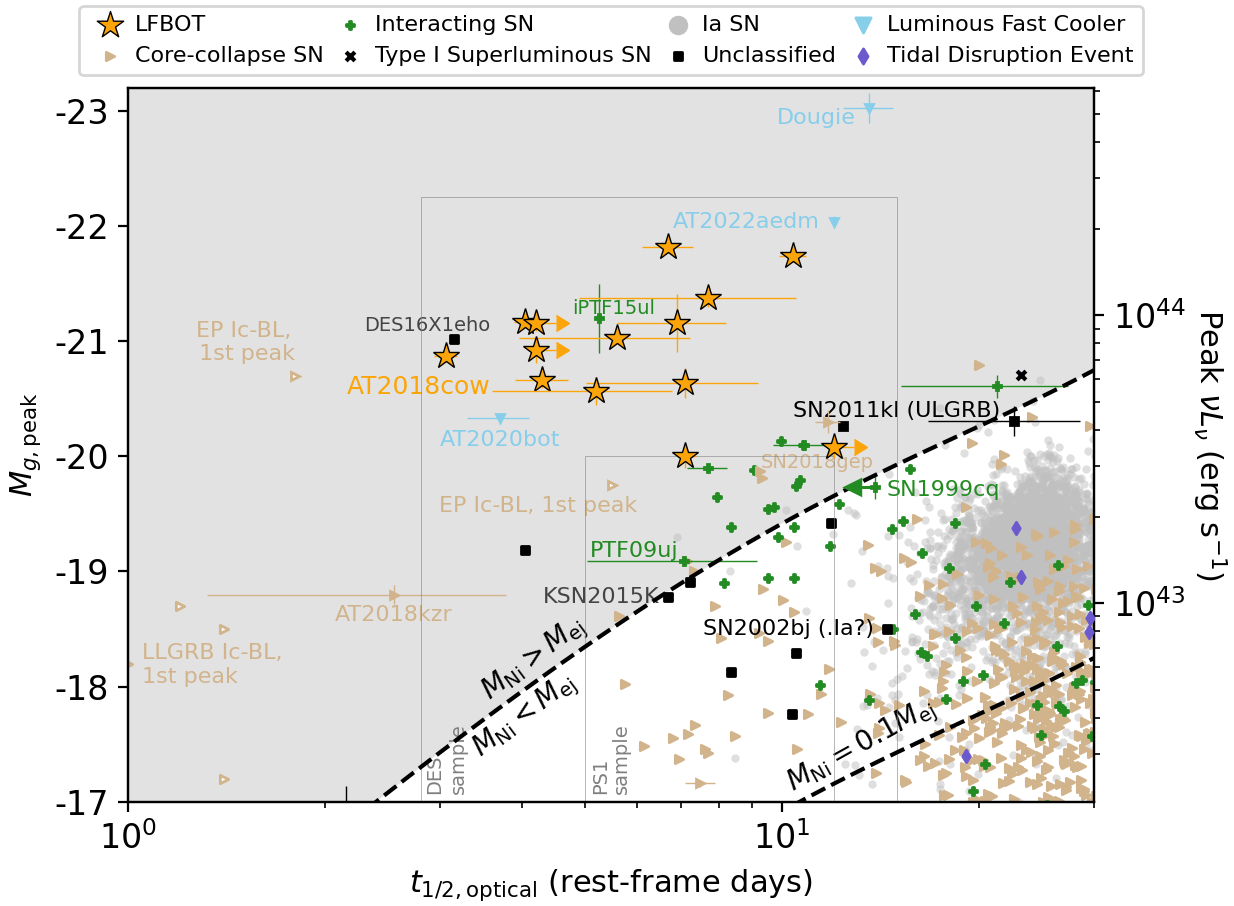}
    \vspace{-0.5em}
        \caption{\emph{Left:} Model light curve for the supernova of a compact star. Duration is set by photon diffusion through optically thick ejecta, and luminosity is set by the mass of $^{56}$Ni synthesized along with a gamma-ray trapping efficiency factor. Figure modified from Daniel Kasen (UC Berkeley).  \emph{Right:} Phase space of optical transients that are thermal in origin. A limit is set by requiring that the mass of ejected material ($M_\mathrm{ej}$) exceed the mass of synthesized $^{56}$Ni ($M_\mathrm{Ni}$). The majority of known supernovae (SNe) lie in the ``permitted'' part of the phase space. The boxes show selection criteria for ``fast blue optical transients'' (FBOTs) used by historical searches (DES \cite{2018MNRAS.481..894P}, PS1 \cite{2014ApJ...794...23D}). 
        FBOTs with luminous X-ray and/or radio emission are marked with stars, and are often referred to as luminous FBOTs (LFBOTs). 
        Some well-observed LFBOTs show particularly strong resemblance to AT2018cow. 
        Measurements obtained primarily from Ho et al. (2023) \cite{2023ApJ...949..120H}, Yao et al. (2023) \cite{2023ApJ...955L...6Y}, and the Zwicky Transient Facility Bright Transient Survey \cite{2020ApJ...904...35P}. For reference we show the first peak of X-ray flashes with double peaked optical light curves \cite{2025ApJ...988L..14E,2018A&A...619A..66D,2006Natur.442.1008C}. References for other individual-object measurements are given in the text. 
        }
        \vspace{-0.1in}
    \label{fig:phase-space}
\end{figure}

The optical light observed from a supernova is thermal radiation from the expanding debris (the ``ejecta''). The ejecta has some initial energy from the shockwave that disrupted the star, and is subsequently heated by the radioactive decay of isotopes (particularly $^{56}$Ni)\footnote{The reason $^{56}$Ni is most abundant is that it is the most bound nucleus with equal numbers of protons and neutrons.} that were synthesized in the explosion. As shown in the left panel of Figure~\ref{fig:phase-space}, the duration of the optical emission is typically several weeks, set by the time it takes photons to diffuse out of the ejecta. The luminosity is set by the energy deposited into the ejecta. For extended progenitors the deposited energy is primarily from the initial shock. For compact stars energy is primarily deposited after the initial shock, by radioactive decay---with some ``leakage'' of gamma rays decreasing the resulting luminosity \cite{2017hsn..book..939K}.  

In the past two decades, the increased size of charge-coupled device (CCD) detectors---along with faster readout times and computer processing speeds---enabled optical telescopes to search larger areas of sky more quickly. As a result, the supernova discovery rate increased, and new and rare types of supernovae were discovered \cite{2012PASA...29..482K, 2019NatAs...3..697I}. For example, events with luminosities exceeding traditional supernovae were dubbed ``superluminous supernovae'' \cite{2019ARA&A..57..305G} while those with a much faster rate of brightening or fading were referred to as  ``rapidly evolving transients.'' From the beginning, it was clear that rapidly evolving transients have a variety of origins. 
One explanation proposed for the fast evolution was a low ejecta mass. In variations of Type Ia supernovae, a low ejecta mass could arise from detonations of the helium shell only (``.Ia'', e.g., \cite{2010ApJ...715..767S, 2010Natur.465..322P, 2010Sci...327...58P, 2010ApJ...723L..98K}), the explosion of a low-mass white dwarf (``SN1991bg-like'' \cite{2008MNRAS.385...75T}), or a partial explosion (``Type Iax'' \cite{2013ApJ...767...57F,2017hsn..book..375J}). In core-collapse supernovae, a low ejecta mass could arise from stars whose envelopes were heavily ``stripped'' by interactions with a binary-star companion \cite{2013ApJ...774...58D, 2013ApJ...778L..23T}. Alternatively, there could be a very small amount of radioactive material, and an inflated radius at the time of explosion \cite{2014MNRAS.438..318K}.

At the fastest timescales and highest luminosities, rapidly evolving transients require a different powering mechanism from conventional supernovae. 
For the purposes of this review, we use the term ``fast blue optical transient'' (FBOT) broadly for blue transients evolving substantially faster than ordinary supernovae, particularly those occupying the short-duration and high-luminosity region of Figure~\ref{fig:phase-space}. 
In practice, searches often adopt duration cuts of 10--15\,d in order to reduce the number of ordinary supernovae (e.g., \cite{2019NatAs...3..697I}). Physically, FBOTs cannot have radioactive decay as the powering source for the peak of their optical emission, because their luminosity implies a higher mass of $^{56}$Ni than there is ejecta (as inferred from their duration). FBOTs are also too luminous to be explained by the thermal energy deposited by the initial explosion shockwave \cite{2017hsn..book..939K}. 

\begin{figure}[tb]
    \centering
    \includegraphics[width=\linewidth]{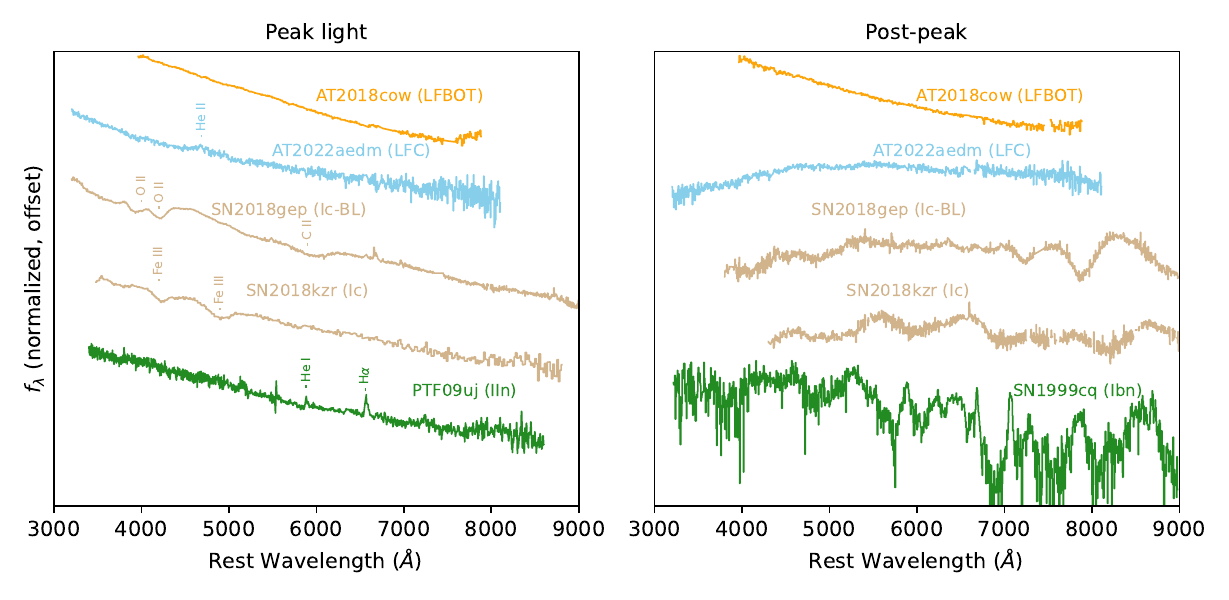}
    \vspace{-2.5em}
    \caption{Sample optical spectra of fast blue optical transients (FBOTs). \emph{Left:} Spectra obtained within a few days of the peak of the optical light curve \cite{2019MNRAS.484.1031P,2023ApJ...954L..28N,2019ApJ...887..169H,2020MNRAS.497..246G,2010ApJ...724.1396O} are typically a hot ($\approx10^{4}\,$K) blackbody, sometimes with features (some are marked) reflecting the composition of the ejecta or the ambient medium. \emph{Right:} At later times, the spectra sometimes develop the broad lines of established supernova classes (SN2018gep as Type~Ic-BL, SN1999cq as Type~Ibn \cite{2000AJ....119.2303M}), and in other cases remain featureless. SN2018kzr is formally classified as Type~Ic based on its spectrum but has been argued to be the merger of a white dwarf with a compact object \cite{2019ApJ...885L..23M,2020MNRAS.497..246G}. 
    The phases shown are $\approx2t_{1/2}$ after peak in the case of AT2018cow, AT2022aedm, and SN1999cq where $t_{1/2}$ is the duration above half-maximum light of the $g$-band light curve; and slightly earlier ($\approx$1--1.5$t_{1/2}$ after peak) in the case of SN2018gep. PTF09uj did not have a post-peak spectrum while SN1999cq did not have a peak-light spectrum, but both are thought to be powered by circumstellar interaction. 
    }
    \vspace{-0.1in}
    \label{fig:spectra}
\end{figure}

FBOTs themselves have diverse origins, reflected in their optical spectra; a selection is shown in Figure~\ref{fig:spectra}. Early discoveries included 
SN\,1999cq and PTF\,09uj\footnote{Transients are named according to the year of discovery, and a string of letters corresponding to the order of discovery. For example, the first astronomical transient (supernova) of this year would be AT(SN)2026a. Before this standardized AT/SN system, transients were named according to the survey that discovered them, e.g., `PTF09uj' for a supernova discovered by the Palomar Transient Factory (PTF).}, 
both likely core-collapse supernovae whose optical spectra exhibited emission lines from helium and hydrogen, respectively \cite{2000AJ....119.2303M,2010ApJ...724.1396O}. 
The emission lines arise from the collision of the supernova ejecta with dense ambient gas (the circumstellar medium, CSM) expelled by the star prior to core collapse \cite{2010ApJ...724.1396O}.
In transients powered by circumstellar interaction, the luminosity is set by the shock energy deposited in the ambient medium rather than the outer layers of the star; the energy can be thermalized and radiated very efficiently. The emission lines reflect the composition of the ambient material \cite{2017hsn..book..403S}. 



It has also been suggested that some FBOTs are not related to supernovae at all. For example, ``Dougie'' was a transient discovered in 2009 as part of an effort to find superluminous supernovae \cite{2015ApJ...798...12V}. Unlike typical superluminous supernovae, its spectra were featureless long after peak light, over a month after discovery (spectra of a similar transient, AT2022aedm, are shown in Figure~\ref{fig:spectra}). The distance and observed $M_R=17.2$ peak implied a very high luminosity of $M = -22.6$\,mag. The discovery team proposed that Dougie was a tidal disruption event (TDE) \cite{2015ApJ...798...12V}. TDEs---the tidal disruption and subsequent accretion of a star by a black hole---had been discovered in galaxy nuclei, presumably from accretion onto the galaxy's central supermassive black hole \cite{2021ARA&A..59...21G}. It was argued that Dougie's fast timescale and off-nuclear location implied\footnote{We argue in Section~\ref{sec:IMBH_TDE} that this model is challenged by the expected efficiency of the debris stream circularization and the peak luminosity.} a lower-mass ($\approx10^{5}\,M_\odot$) black hole \cite{2015ApJ...798...12V}. The term ``luminous fast coolers'' (LFCs) has recently been applied to objects including Dougie and AT2022aedm \cite{2023ApJ...954L..28N}: LFCs have high optical luminosities, cool significantly over time, and are located in galaxies with negligible star formation \cite{2023ApJ...949..120H,2023ApJ...954L..28N}. 

FBOTs garnered significant interest both because of their unusual properties and because their intrinsic rate was inferred to be high, 1--3\% of the supernova rate \cite{2011ApJ...730...89P,2013ApJ...774...58D}. 
Searches of archival data were undertaken to increase the sample size, with a variety of selection criteria. From the Panoramic Survey Telescope and Rapid Response System (Pan-STARRS \cite{2016arXiv161205560C}), 14 objects were identified with durations under 12 days spanning a wide range of luminosities, and an implied rate of 4--7\% of the supernova rate \cite{2014ApJ...794...23D}. A Subaru Hyper Suprime-Cam Transient Survey search yielded 5 additional events on the basis of a rapid rise rate of $>1$\,mag/day in the rest-frame near-ultraviolet (UV), and found an occurrence rate of 1\% of the supernova rate \cite{2016ApJ...819....5T}. Four luminous ($M_\mathrm{r,peak}\approx-20\,$mag) fast rising events were discovered in Supernova Legacy Survey data   \cite{2016ApJ...819...35A}. Finally, 72 short-duration events were discovered in Dark Energy Survey data \cite{2018MNRAS.481..894P}. Indeed, almost all the transients identified in these archival searches had blue colors at peak light, leading to a popular model: the shock breakout and subsequent cooling of a dense shell of material at a large radius from the progenitor, which would have been expelled shortly prior to explosion \cite{2014ApJ...794...23D,2018MNRAS.481..894P,2016ApJ...819....5T}. 






Although archival searches increased the number of FBOTs by orders of magnitude, most were discovered after the transient faded and at relatively large distances (e.g., the median redshift of the Pan-STARRS sample was $z=0.275$). So, detailed single-object studies continued to play an important role. The model of shock breakout and/or cooling was supported by the discovery of certain stripped-envelope supernovae  \cite{2016MNRAS.461.3057S,2017ApJ...836..158H,2019ApJ...875...76N,2018MNRAS.475.2344V} and the transient KSN2015K. The high-cadence \emph{Kepler} light curve of KSN2015K was well explained by shock breakout from a dense shell, but no spectra were obtained \cite{2018NatAs...2..307R}. Finally, there were hints of FBOTs powered by central engines---the rotational spindown of a neutron star \cite{2010ApJ...717..245K} or accretion onto a black hole \cite{2013ApJ...772...30D,2015MNRAS.451.2656K}---from the discovery of a rapidly evolving broad-lined Type~Ic supernova\footnote{Broad-lined Type Ic-BL supernovae are the supernova type associated with long-duration gamma-ray bursts.} \cite{2017ApJ...851..107W} as well as of SN2011kl, a featureless transient with an ultra-long duration gamma-ray burst (GRB) \cite{2013ApJ...766...30G,2014ApJ...781...13L,2015Natur.523..189G,2016ApJ...819...35A,2019A&A...624A.143K}.

\begin{figure}[tb]
    \centering
    \includegraphics[width=\linewidth]{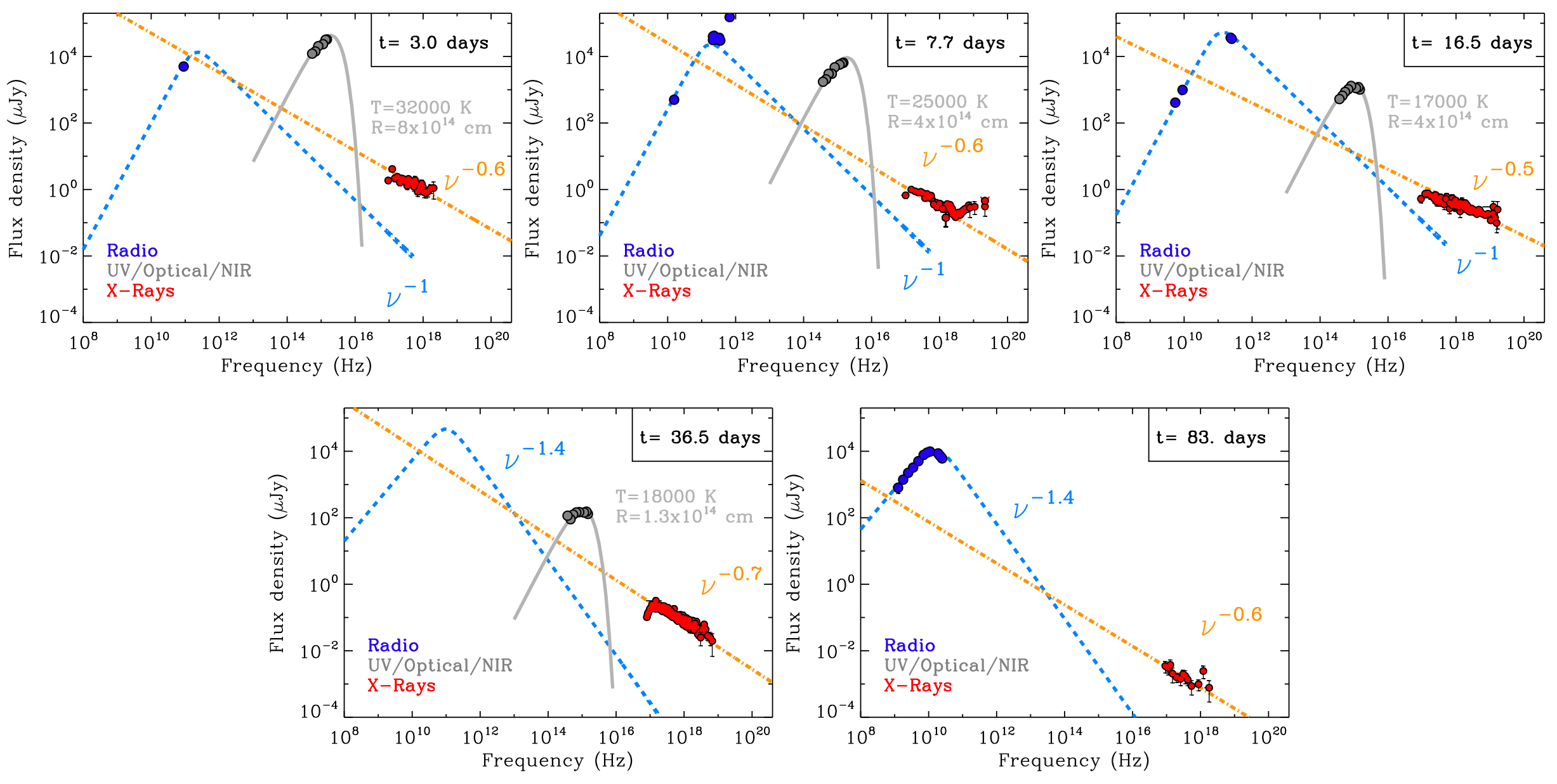}
    \vspace{-2.5em}
    \caption{AT2018cow had luminous emission detected across the electromagnetic spectrum, from radio to X-ray wavelengths. Below the peak of the radio SED, the power law index shown is $F_\nu \propto \nu^2$ before 30\,d, and $F_\nu \propto \nu^{1.2}$ after that time. Figure reproduced from Margutti et al. (2019) \cite{2019ApJ...872...18M}, with permission.} 
    \vspace{-1em}
    \label{fig:cow-intro}
\end{figure}

A significant milestone was the 2018 discovery \cite{2018ApJ...865L...3P} of AT2018cow (nicknamed ``The Cow'') by the Asteroid Terrestrial-impact Last Alert System (ATLAS \cite{2018PASP..130f4505T}). Unlike almost all previously discovered FBOTs, The Cow had luminous emission detected from X-ray to radio wavelengths\footnote{Most earlier FBOTs did not have X-ray or radio observations, so similar behavior cannot be ruled out.} (Figure~\ref{fig:cow-intro}). Detailed follow-up observations were made possible by AT2018cow's proximity (60\,Mpc) and the early discovery. The evolution of AT2018cow is described in more detail in Section~\ref{sec:observations}. Highlights include a fading optical light curve with no emerging supernova component (e.g., \cite{2019MNRAS.484.1031P}), sustained high blackbody temperatures (blue optical colors; e.g., \cite{2019ApJ...872...18M}), optical spectra resembling interacting supernovae (Type IIn/Ibn \cite{2019MNRAS.488.3772F}) or TDEs \cite{2019MNRAS.484.1031P}, X-rays with pronounced variability \cite{2018MNRAS.480L.146R,2019ApJ...872...18M,2019ApJ...871...73H} and a spectral ``hump'' \cite{2019ApJ...872...18M}, radio emission peaking at high frequencies (millimeter wavelengths) \cite{2019ApJ...871...73H}, 
strong optical polarization \cite{2023MNRAS.521.3323M}, and luminous ($10^{7}\,L_\odot$) UV emission detected by the Hubble Space Telescope years after the initial transient \cite{2022MNRAS.512L..66S, 2023ApJ...955...43C, 2023MNRAS.525.4042I}.


The discovery of AT2018cow was illustrative of a change in the field: 
wide-field, high-cadence surveys such as the Zwicky Transient Facility (ZTF \cite{2019PASP..131g8001G,2019PASP..131a8002B}) and ATLAS enabled FBOTs to be found in real time and at low redshifts, and followed to later phases when spectroscopic features emerged. Several spectroscopic classes were established, including helium-rich supernovae of Type~IIb and Type~Ibn \cite{2023ApJ...949..120H}, and a peculiar transient (AT2018kzr) attributed to a compact-object merger \cite{2019ApJ...885L..23M,2020MNRAS.497..246G}.  
In addition, multiwavelength observations of FBOTs made it clear that luminous X-ray and radio emission---as observed in AT2018cow---was uncommon  
(e.g., \cite{2019ApJ...887..169H, 2023ApJ...949..120H, 2025ApJ...988L..14E, 2025ApJ...988L..60S, 2026arXiv260420346S}). 
On the other hand, observations of several events with comparable optical luminosities to AT2018cow ($M\lesssim-20\,$mag) revealed similarly luminous radio and/or X-ray emission \cite{2020ApJ...895L..23C,2020ApJ...895...49H}. As a result, searches for AT2018cow analogs began to focus on the most luminous ($M\lesssim-20\,$mag) short-duration events, leading to the identification of over a dozen FBOTs with high optical and radio (or X-ray) luminosities \cite{2024MNRAS.527L..47C,2022ApJ...934..104Y,2021MNRAS.508.5138P,2023Natur.623..927H,2026MNRAS.549ag678P,2026ApJ..1007...38S,2026arXiv260813003F}. Such events have been referred to as ``luminous FBOTs'' (LFBOTs; e.g., \cite{2022ApJ...932...84M}) or ``AT2018cow-like'' transients (e.g., \cite{2023Natur.623..927H}).

\begin{figure}[!h]
    \centering
    \includegraphics[width=\linewidth]{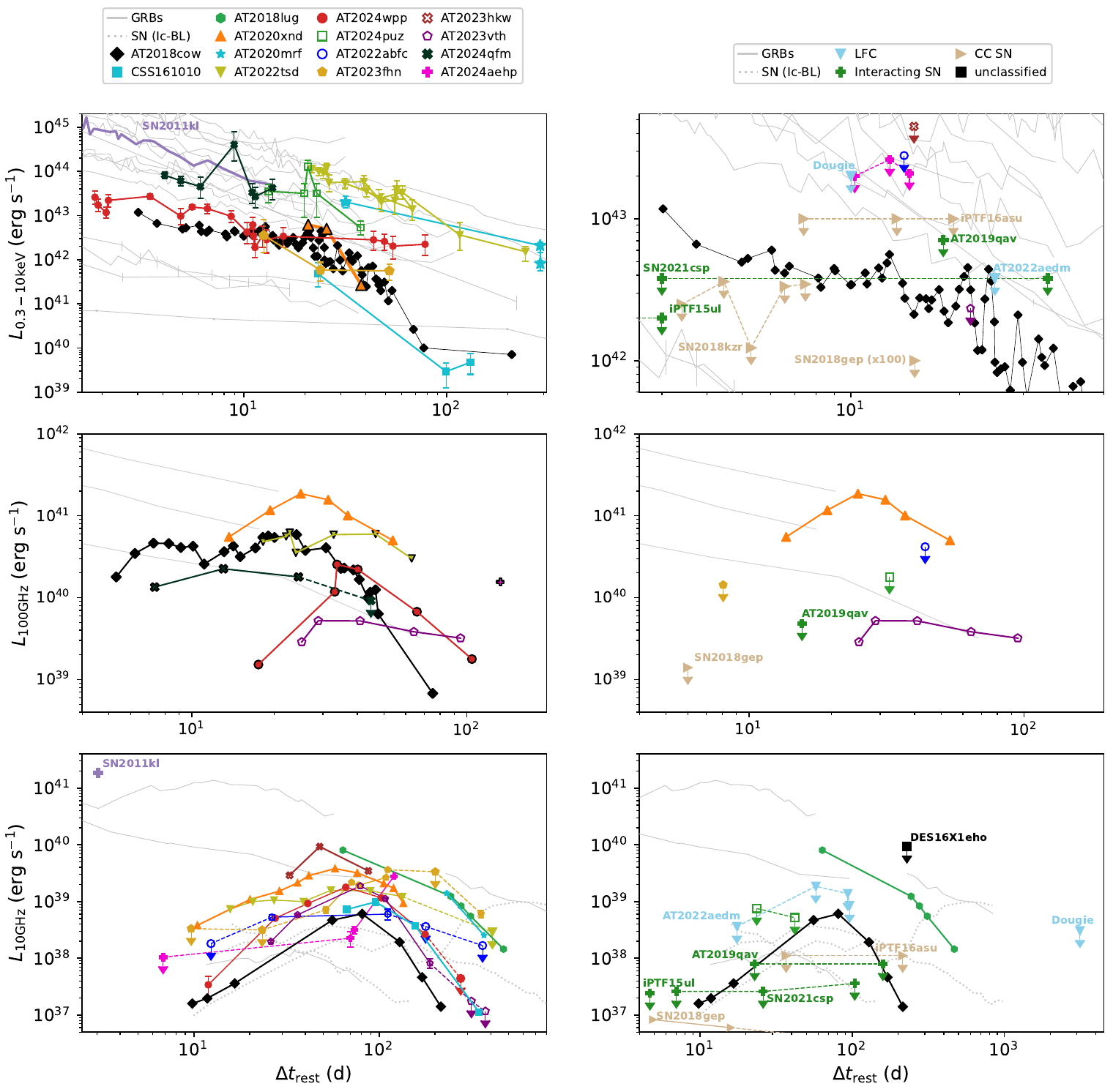}
    \vspace{-0.3in}
    \caption{Constraints on X-ray (top), mm-wave (100\,GHz; middle---except AT2018cow which is at 230\,GHz), and cm-wave (10\,GHz; bottom) observations of FBOTs. The light grey lines in the background show X-ray counterparts to long-duration gamma-ray bursts and energetic supernovae, for reference. 
    Left column shows LFBOTs discussed in this review and potentially related events with luminous X-ray and/or radio detections. Right column shows FBOTs with only non-detections at that wavelength, with some detected LFBOTs (same symbols as on the left)  for reference.}
    \label{fig:xray-radio-summary}
\end{figure}

This review focuses on LFBOTs. The term has been used inconsistently in the literature, sometimes phenomenologically and sometimes referring to a shared progenitor. We clarify our working usage in the ``Terminology'' inset on the next page, but emphasize that this is not intended to be a physical classification. 
Table~\ref{tab:summary} lists the LFBOTs emphasized in this review, and a small number of potentially related events\footnote{Two additional events, AT2026dbl \cite{2026TNSAN..36....1W,2026TNSAN..38....1J} and AT2026nik \cite{2026TNSAN.152....1W,2026GCN.44749....1Q}, have not been published at the time of writing this review.}. Most have been discovered in optical surveys, via searches for fast and luminous optical transients (e.g., \cite{2020ApJ...895...49H,2021MNRAS.508.5138P,2026arXiv260813003F}) or for persistently blue transients offset from galaxy nuclei \cite{2025ApJ...995..228S}. An LFBOT was also identified \cite{2022ApJ...934..104Y} in eROSITA X-ray telescope \cite{2021A&A...647A...1P} data, and the transient SN2011kl (which has been argued to be related to LFBOTs \cite{2026arXiv260707819V}) was discovered as ultra-long GRB\,111209A \cite{2013ApJ...766...30G,2014ApJ...781...13L}. 
For completeness, Figure~\ref{fig:xray-radio-summary} shows the X-ray and radio constraints on all FBOTs. We review the observational properties of LFBOTs in Section~\ref{sec:observations} and the proposed progenitor models in Section~\ref{sec:theories}. In Section~\ref{sec:future} we discuss how connections to other phenomena could be established, and other avenues for investigation.

\clearpage

\begin{terminologybox}

The terms FBOT and LFBOT have been used in different ways in the literature. In this review, we use them as descriptive working terms rather than imposing hard boundaries.

\smallskip

\textbf{Fast blue optical transient (FBOT).}
Short-duration transients with blue colors at peak light whose combination of peak luminosity and duration cannot be explained by radioactive $^{56}$Ni decay. Searches typically impose a duration cut of $\approx10$--15\,d in order to reduce false positives (ordinary supernovae). 
Spectroscopy has revealed diverse origins: established core-collapse supernova classes in which shock interaction and/or post-shock cooling is an important power source (IIb, IIn, Ibn, Ic-BL), and peculiar events of uncertain origin (e.g., Dougie, AT2018kzr).

\smallskip

\textbf{Luminous fast blue optical transient (LFBOT).}
An FBOT with luminous X-ray and/or radio emission that is not a conventional supernova (spectrum does not develop broad absorption features even after peak). The most commonly regarded examples have $M_g\lesssim-20\,$mag, durations $\lesssim10\,$d, and luminous X-ray and/or radio emission. Most are off-nuclear. A well-observed subset closely resemble AT2018cow in their multiwavelength behavior; the resemblance across independent emission components in these ``AT2018cow-like events'' suggests a common origin. 

\end{terminologybox}

\begin{landscape}
\begin{table*}[h]
\centering
\setlength{\tabcolsep}{0.25em}
\caption{Summary of the luminous fast blue optical transients that are the focus of this review. 
Values are reported in as close to the rest frame as possible. See text for references.  
}
\label{tab:summary}
\begin{tabular}{ccccccccccc}
\hline\hline
Name & $z$ & Nuclear & $t_{1/2,g}$ & $M_{\mathrm{pk},g}$ & $L_\mathrm{X}$ & $\nu L_\mathrm{\nu,10\,\mathrm{GHz}}$ & $\nu L_\mathrm{\nu,100\,\mathrm{GHz}}$ & Balmer emission lines & $M_\mathrm{Ni}$  & Notable behavior \\ 
 &  & offset &  &  & at 14--24\,d & at 60--90\,d & at 24--35\,d & width/first detected/shift & limit &  \\
 &  & kpc ($^{\prime\prime}$) & (d) & (mag) & (erg\,s$^{-1}$) & (erg\,s$^{-1}$) & (erg\,s$^{-1}$) & (km\,s$^{-1}$/d/[blue/red]) & ($M_\odot$) \\ 
\hline
\multicolumn{11}{c}{LFBOTs discussed in this review}\\
CSS161010 & 0.033 & 0.3 (0.4) & $4.3\pm0.4$ & $-20.66\pm0.07$ & $\lesssim 5 \times 10^{41}$ & $7\times10^{38}$ & -- & 30,000/21/blue & $<0.07$ & \\
& & & & & & & & &  
\\
AT2018cow & 0.014 & 1.7--1.8 & $3.06\pm0.12$ & $-20.87\pm0.05$ & 1--5$ \times 10^{42}$ & $6\times10^{38}$ & $4\times10^{40}$ & 1000--4000/15/red & $<0.18$ & UV/optical \\ 
& & (5.7--6.0) & & & & & (230\,GHz) & & & plateau (yrs) \\
AT2018lug & 0.271 & 0.6--3.0 & $4.0\pm0.1$ & $-21.17\pm0.05$ & -- & $8\times10^{39}$ & -- & -- & -- & \\ 
 &  & (0.15--0.7) & & &  & &  &  & & \\ 
AT2020mrf & 0.135 & 1.0--1.5 & $7.2\pm0.2$ & $-20.0\pm0.1$ & $2\times10^{43}$  & $1\times10^{39}$  & -- & -- & -- & Late (1\,yr) X-ray \\
 & & (0.4--0.6) &  &  & (30\,d) & (230\,d) &  &  &  & variability \\
AT2020xnd & 0.243 & 0.16--2.0 & 4--7 & $-21.03\pm0.04$ & $6\times10^{42}$ & $3\times10^{39}$ & $2\times10^{41}$ & -- & -- & \\
 & & (0.04--0.5) & & & & & & & & \\
AT2022tsd & 0.256 & 5.7--6.5 & 5.1--9.1 & $-20.64\pm0.13$ & $1\times10^{44}$ & $2\times10^{39}$ & $6\times10^{40}$ & -- & -- & Late (30--120\,d) \\
 & & (1.4--1.6) & & & & & & & & optical flares \\
AT2022abfc & 0.212 & 2.0 (0.6) & 4.7--5.7 & $-20.57\pm0.12$ & $<3\times10^{43}$ & $6\times10^{38}$ & $<4\times10^{40}$ & -- & -- & \\
 &  &  &  & & & & (44\,d) & -- & -- & \\
AT2023fhn & 0.24 & 5.35 or 16.5 & $\gtrsim4.2$ & $-21.16\pm0.07$ & $6\times10^{41}$ & $2\times10^{39}$ & $<1\times10^{40}$ & -- & -- & -- \\
 & & (1.4 or 4.2) & & & & & (8\,d)& & &  \\
AT2023hkw & 0.339 & 12--15.5 & $6.9\pm1.3$ & $-21.16\pm0.25$ & $<4.5\times10^{43}$ & $4\times10^{39}$ & -- & -- & -- & -- \\
 &  & (2.4--3.1) &  &  & & &   &  &  &  \\
AT2023vth & 0.0747 & 2.8--3.0 & $7.7\pm2.8$ & $-21.38\pm0.08$ & $<2\times10^{42}$ & $2\times10^{39}$ & $4\times10^{39}$ & -- & -- & -- \\
 & & (1.9--2.1) &  &  &  &  &  &  &  &  \\
AT2024puz & 0.356 & 5 (1) & $10.38\pm0.53$ & $-21.74\pm0.05$ & $3\times10^{43}$-- & $<5\times10^{38}$ & $<2\times10^{40} $ & -- & $<3.1$ & -- \\
 &  & &  &  & $1\times10^{44}$ & (42\,d) & &  &  &  \\
AT2024qfm & 0.2270 & 3--5 & $>4.2$ & $-20.92\pm0.11$ & $4\times10^{43}$ & -- & $2\times10^{40}$ & -- & -- & -- \\
 &  & (0.8--1.4) &  &  & (14\,d) &  & &  &  &  \\
AT2024wpp & 0.0868 & 5.0--6.3 & $6.71\pm0.63$ & $-21.82\pm0.05$ & $3\times10^{42}$ & $2\times10^{39}$ & $2\times10^{40}$ & 3000/36/blue& $<0.29$ & $E_\mathrm{rad,UVOIR}\approx 10^{51}\,$erg \\ 
& & (3.0--3.8) & & & & & & \emph{and} at $v=0$ & & X-ray hardening at 50\,d \\
\hline
\hline
\multicolumn{11}{c}{Potentially related phenomena}\\
AT2024aehp & 0.1715 & $<0.5$ & $\gtrsim12.0$ & $-20.08\pm0.09$ & $<2\times10^{43}$ & $3\times10^{38}$ & $2\times10^{40}$ & -- & -- & Late ($>100$\,d)  \\ 
&  & ($<0.2$) &  & & & & (130\,d) &  &  & radio turn-on \\ 
SN2011kl & 0.677 & $<0.250$ & $22.7\pm5.9$ & $-20.31\pm0.13$ & $\lesssim5\times10^{43}$ & -- & -- & -- & -- & Ultra-long GRB \\
 &  & ($0.011\pm0.038$) &  &  & & & &  & &  \\
\hline
\end{tabular}
\vspace{2.5mm}
\end{table*}
\end{landscape}

\vspace{-0.2in}

\section{Observational Properties}
\label{sec:observations}


\subsection{UVOIR light curves and nickel mass constraints}
\label{sec:uvoir-lc}

\begin{figure}[!b]
\vspace{0pt}
    \centering
    \includegraphics[width=\linewidth]{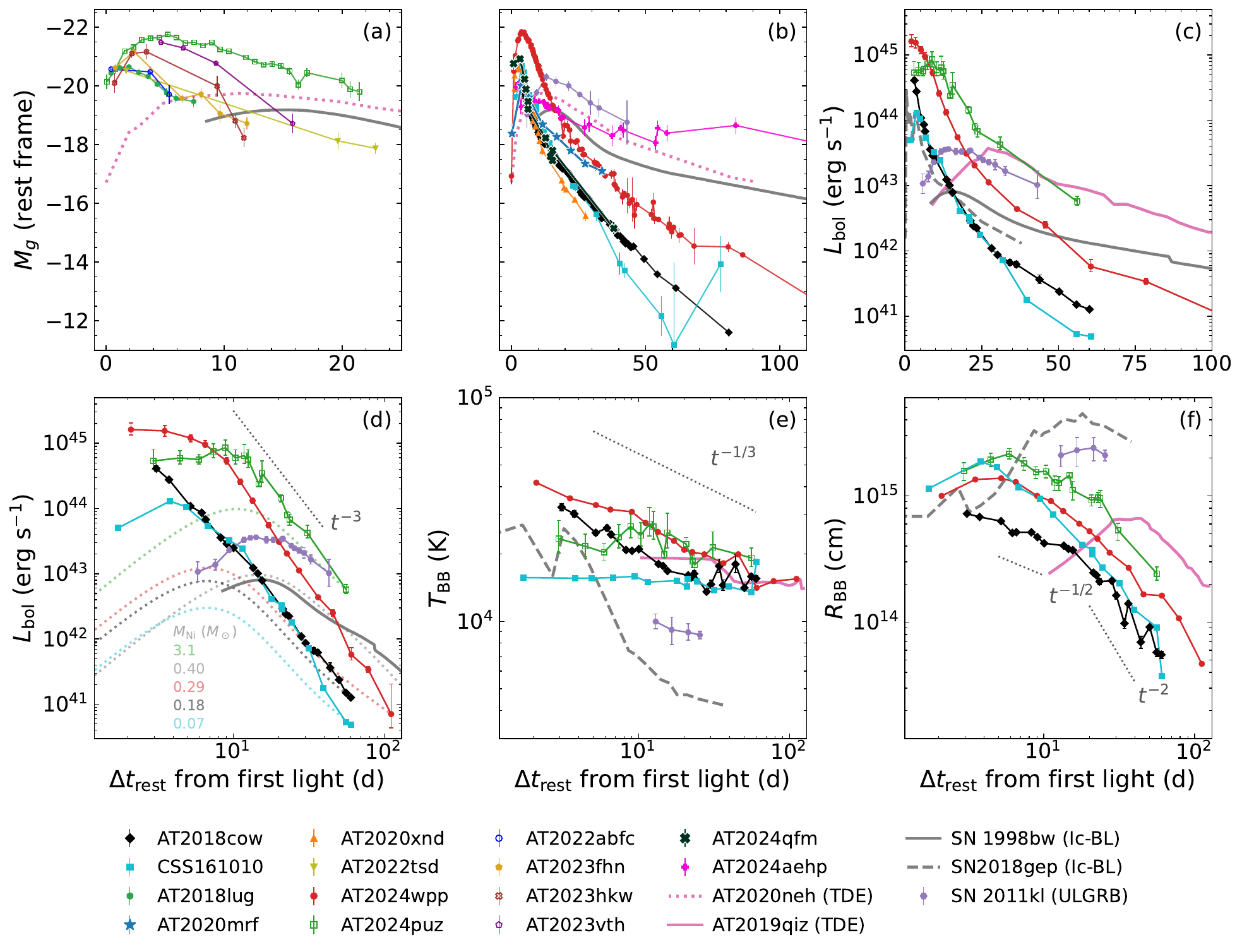}
    \vspace{-0.35in}
    \caption{
    (a) and (b): $g$-band light curves of LFBOTs and potentially related transients, for events whose monitoring (a) stopped $<25\,$d and (b) continued $>25$\,d. (c)--(f): Bolometric light curve in linear (c) and log (d) space, and the evolution of the blackbody temperature (e) and radius (f). For comparison, we show light curves of Ic-BL supernovae SN\,1998bw \cite{2011AJ....141..163C} and SN\,2018gep \cite{2019ApJ...887..169H} and the TDEs AT2020neh \cite{2022NatAs...6.1452A} and AT2019qiz \cite{2020MNRAS.499..482N}. In (d)--(f) we show power laws for reference. In addition, in (d) we show radioactive decay curves and upper limits on $M_\mathrm{Ni}$ for each of the objects with bolometric light curves. For AT2024puz $M_\mathrm{Ni}=M_\mathrm{ej}$, for SN\,1998bw $M_\mathrm{ej}=7\,M_\odot$ \cite{2013MNRAS.434.1098C}, and for the other events $M_\mathrm{ej}=1\,M_\odot$.}   
    \label{fig:optical-photometry}
\end{figure}

Optical coverage of LFBOTs ranges widely in time and frequency (Figure~\ref{fig:optical-photometry}, panels a--c). The three low-redshift ($z<0.1$) events discovered and confirmed within days of first light---AT2018cow, CSS161010, and AT2024wpp---as well as AT2024puz had multi-band observations spanning months, and showed very similar UVOIR light-curve evolution  \cite{2019ApJ...872...18M, 2019MNRAS.484.1031P, 2024ApJ...977..162G, 2025ApJ...995..228S, 2026ApJ...997L..10L, 2026MNRAS.549ag678P}. The light curves rose quickly to a high luminosity ($L_\mathrm{bol}\gtrsim10^{44}\,$erg\,s$^{-1}$), then declined rapidly ($L\sim t^{-3.5}$ for AT2024wpp \cite{2026MNRAS.549ag678P, 2026ApJ...997L..10L}, $L\sim t^{-2.8}$ for CSS161010 \cite{2024ApJ...977..162G}, and $L\sim t^{-2.5}$ for AT2018cow \cite{2019MNRAS.484.1031P, 2019ApJ...872...18M}), with no observed second peak from radioactive decay (panel d). AT2024wpp had the highest peak luminosity \cite{2026MNRAS.549ag678P, 2026ApJ...997L..10L}, $4.5\times$ more UV-luminous than AT2018cow \cite{2026ApJ...997L..10L}, similar only to MUSSES2020J \cite{2022ApJ...933L..36J}. In fact, AT2024wpp had the highest peak bolometric luminosity of any thermal transient for 20 days, and its integrated radiated UVOIR energy was $10^{51}\,$erg \cite{2026MNRAS.549ag678P, 2026ApJ...997L..10L} (higher than $7\times10^{49}$\,erg for CSS161010 \cite{2024ApJ...977..162G} or $10^{50}\,$erg for AT2018cow \cite{2019ApJ...872...18M}). The color temperature is roughly constant, in contrast to the declining temperatures observed in typical supernovae and more similar to what is observed in TDEs (panel e). Finally, the photospheric radius rapidly expands early on ($v\approx0.1c$), reaches a maximum of $\approx10^{15}$\,cm and then contracts (panel f)  \cite{2026MNRAS.549ag678P,2019MNRAS.484.1031P, 2019ApJ...872...18M, 2024ApJ...977..162G}. 

In Table~\ref{tab:summary} we report the rest-frame time above half-maximum light, as well as the peak absolute magnitude, in $g$-band for all objects (panels a and b of Figure~\ref{fig:optical-photometry}). The values for the objects through AT2022tsd were obtained from the literature \cite{2020ApJ...895...49H, 2023ApJ...949..120H, 2023Natur.623..927H,2024ApJ...977..162G}, with the exception of the duration for CSS161010, which was calculated for this work. For the remaining objects we calculated the values for this work in a consistent way using the published light curves. For comparison, we show the potentially related AT2024aehp and SN2011kl, whose optical light curves evolve substantially more slowly \cite{2019A&A...624A.143K,2026ApJ..1007...38S}. 

Although nickel decay cannot power the peak of LFBOT light curves, it could contribute to powering the late-time light curves, and there have been several approaches to constraining $M_\mathrm{Ni}$. Using the early (first month) AT2018cow light curve, assuming typical supernova properties, it was found that $M_\mathrm{Ni}<0.05$--0.06\,$M_\odot$ \cite{2019MNRAS.484.1031P,2019ApJ...872...18M}. The upper limit is less constraining if the ejecta mass is lower than that in a typical supernova, e.g., $M_\mathrm{Ni}<0.2$--0.4\,$M_\odot$---but then it was argued that Fe blanketing should have been observed in the spectra, which were instead dominated by a blue continuum, so a more reasonable estimate is $M_\mathrm{Ni}<0.1\,M_\odot$ \cite{2019ApJ...872...18M}. On the basis of bolometric data out to 60\,d, a limit of $M_\mathrm{Ni}<0.013$--0.019\,$M_\odot$ was placed assuming full $\gamma$-ray trapping and $M_\mathrm{ej}=0.1$--10\,$M_\odot$. Again, allowing leakage softens the constraint but was argued to conflict with the lack of observed line blanketing \cite{2023ApJ...955...42C}. There are no published limits for CSS161010, AT2024puz, or AT2024wpp---the other events with bolometric light curves. For AT2020xnd, limits were placed at 20\,d on the basis of the $r$-band light curve, finding $M_{\rm Ni}<0.02\,M_\odot$ assuming $M_\mathrm{ej}>1\,M_\odot$ \cite{2021MNRAS.508.5138P}. Similar estimates performed for AT2020mrf and AT2018lug found $M_\mathrm{Ni}<0.4\,M_\odot$ \cite{2022ApJ...934..104Y,2020ApJ...895...49H}.

Given the variety of approaches, for this work we perform our own $M_\mathrm{Ni}$ constraints (Figure~\ref{fig:optical-photometry}, panel d). After the ejecta diffusion timescale of a few days, say $t\gtrsim 20\,$d, the bolometric light curve is set by Ni/Co decay as well as a gamma-ray trapping efficiency factor and positron kinetic energy deposition. We use an Arnett-like model \cite{1982ApJ...253..785A} to calculate the bolometric light curve, assuming a uniform ejecta density profile (which causes order unity uncertainties). We use an effective opacity $\kappa_\mathrm{\gamma}=0.03\,$cm$^{2}$\,g$^{-1}$ for the gamma-ray energy trapping and vary $M_\mathrm{Ni}$. The limiting models (chosen so that they intersect the latest point in the bolometric light curve) are shown in panel (d) of Figure~\ref{fig:optical-photometry} and the resulting upper limits are given in Table~\ref{tab:summary}. We confirm that our model reproduces the $M_\mathrm{Ni}=0.4\,M_\odot$ for $M_\mathrm{ej}=7\,M_\odot$ of SN\,1998bw (also shown in the figure panel).  
We emphasize that these are conservative upper limits. 

\subsection{UVOIR spectra}
\label{sec:uvoir-spectra}

We begin by describing the spectra of AT2018cow, CSS161010, and AT2024wpp, which have the most detailed data due to their low redshifts and early discovery. We then describe generic features of the remaining LFBOTs. We emphasize that the velocity-offset structure of the lines observed in CSS161010 and AT2024wpp has not, to our knowledge, been observed in any other transient. 

\subsubsection{AT2018cow}

At $<2\,$d, the spectra were entirely featureless (except for host galaxy and Milky Way interstellar medium absorption lines) \cite{2019MNRAS.484.1031P}. From 4--8\,d, the spectra showed a very broad absorption feature\footnote{This feature led to early suggestions of a Ic-BL supernova classification.} implying high velocities of $v\sim0.1c$ \cite{2019MNRAS.484.1031P,2019ApJ...872...18M}. The broad absorption feature disappeared, and at $>10$\,d intermediate-width emission features of hydrogen and helium appeared. Specifically, \ion{He}{2} 4686 was visible on days 9, 11, 12, and 14---then faded \cite{2019MNRAS.484.1031P}. At 15\,d, \ion{He}{1} 5876 and 5015 as well as H$\alpha$ appeared with asymmetric line profiles, redshifted by 3000--4000\,km\,s$^{-1}$ \cite{2018ApJ...865L...3P, 2019MNRAS.484.1031P,2019ApJ...872...18M}: over the following 10--20\,d, the profiles evolved blueward and developed a ``wedge'' shape. At $>30$\,d, weak \ion{Ca}{2} and possibly \ion{O}{1} appeared, in addition to an upturn between 8000--9000\,\AA\ (also reported as an excess in the $z$-band photometry) \cite{2019MNRAS.484.1031P}. A near-infrared (NIR) spectrum obtained at 17\,d showed redshifted \ion{He}{1} with a similar velocity and profile to the lines in the optical spectrum \cite{2019ApJ...872...18M}. Swift UV grism spectra (170--430\,nm) were featureless from 5\,d, and at 13\,d showed a weak feature near \ion{He}{2} 2511\,\AA, and a broad feature near 2710\,\AA\ that could be \ion{He}{1} or \ion{He}{2} \cite{2019MNRAS.487.2505K}. 

\begin{figure}[tb]
\centering
\begin{minipage}[b]{0.48\linewidth}
\vspace{0pt}
  \centering
  \includegraphics[width=0.95\linewidth]{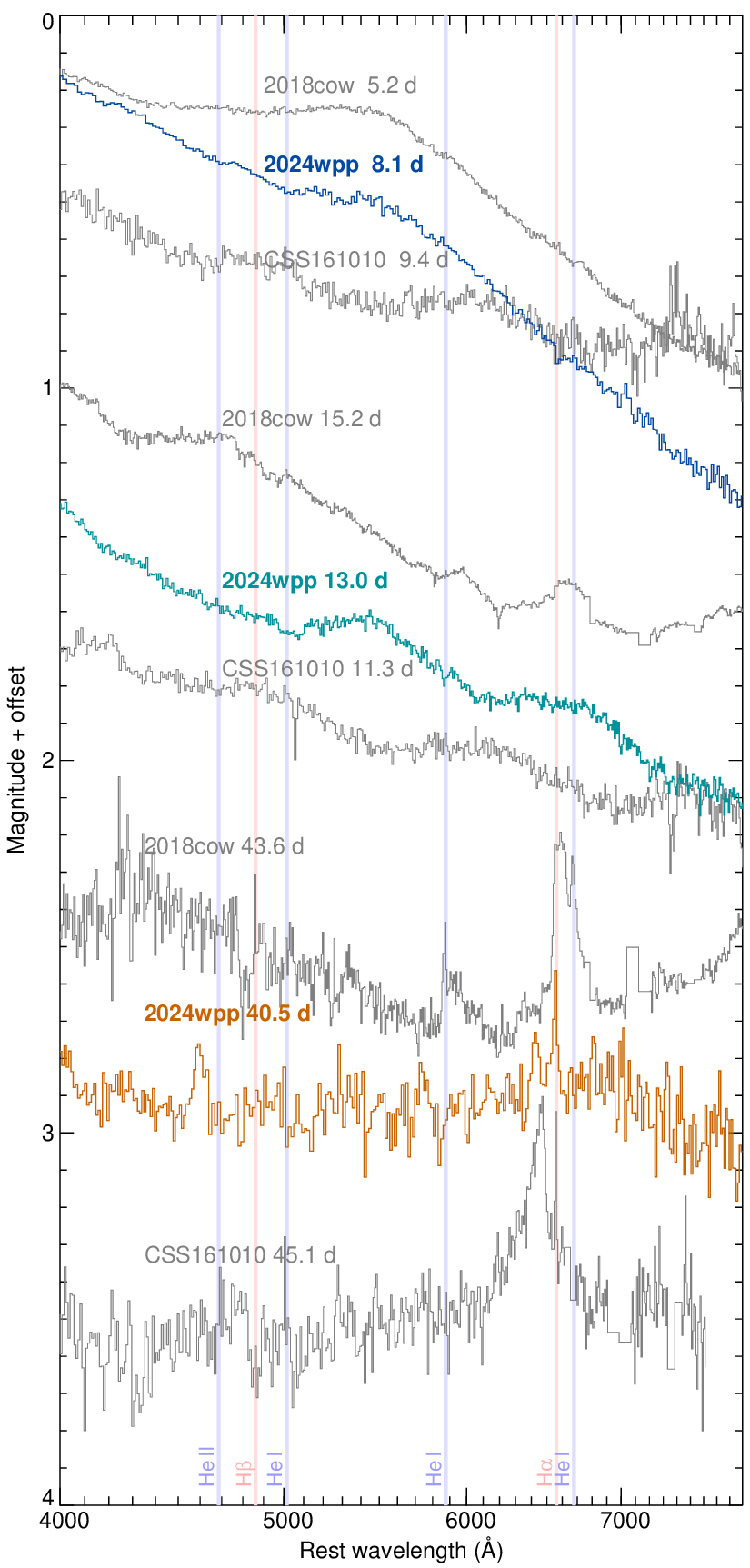}
\end{minipage}
\hfill
\begin{minipage}[b]{0.48\linewidth}
\vspace{0pt}
  \centering
  \includegraphics[width=\linewidth]{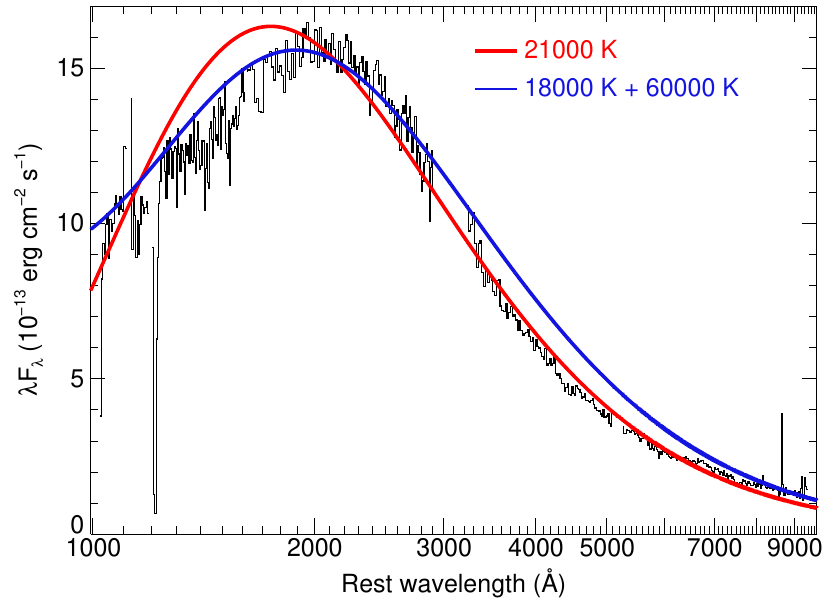}
  \includegraphics[width=\linewidth]{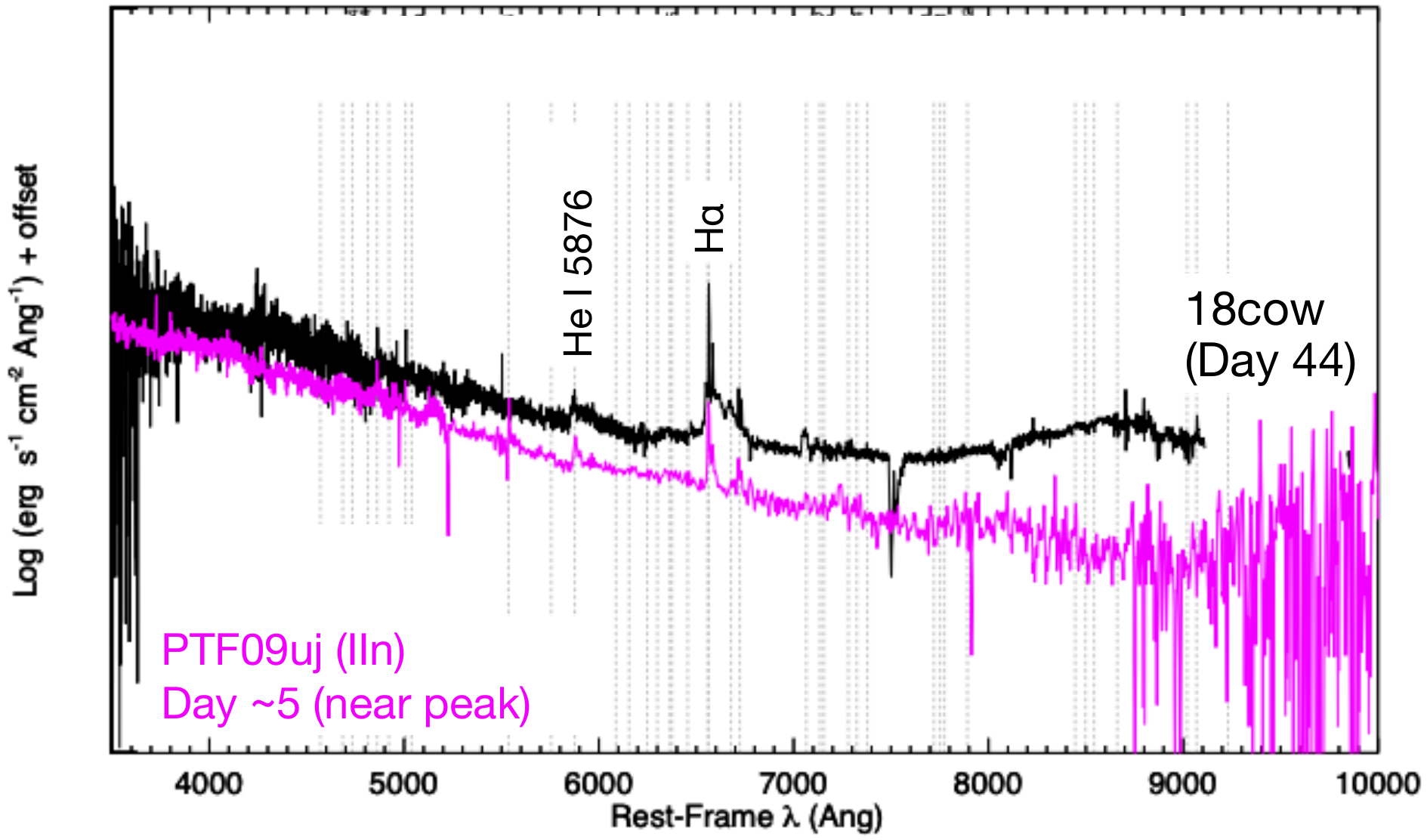}
  \vspace{0.5em}
  \includegraphics[width=\linewidth]{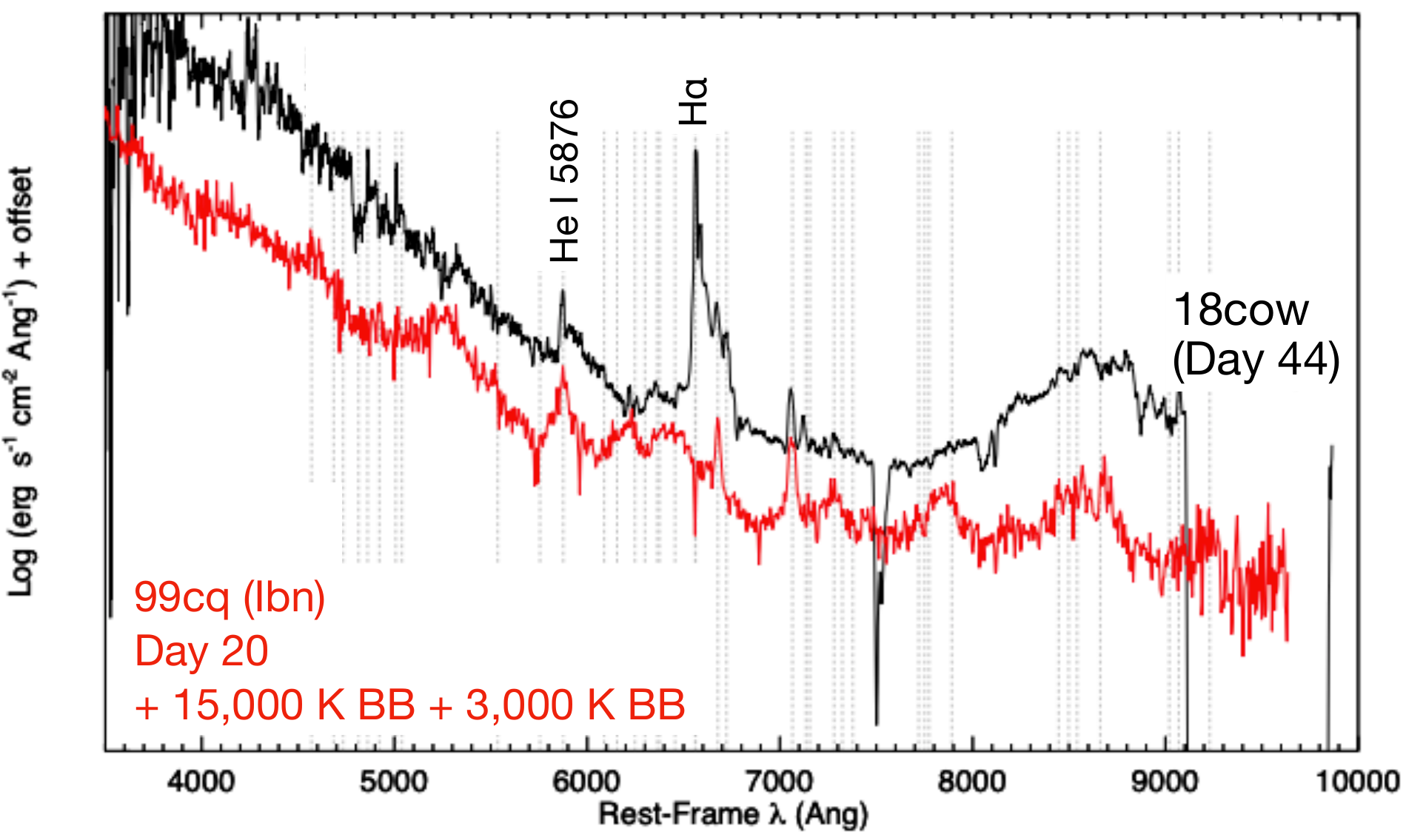}
\end{minipage}
\vspace{-0.2in}
\caption{\emph{Left:} Select optical spectra from AT2018cow, AT2024wpp, and CSS161010. Figure modified from Perley et al. (2026) \cite{2026MNRAS.549ag678P}, with permission. \emph{Top right:} Combined Hubble Space Telescope UV and optical spectra of AT2024wpp at 20\,d. Figure reproduced from Perley et al. (2026) \cite{2026MNRAS.549ag678P}, with permission. \emph{Center right and bottom right:}
Comparison of the Day 44 optical spectrum of AT2018cow to an early spectrum of two fast and luminous transients: PTF09uj (Type~IIn; center right) and SN\,1999cq (Type~Ibn; bottom right). The hydrogen emission line was argued to be similar to that observed in PTF09uj, while the helium features and upturn in the 8000--9000\,\AA\ range were argued to be similar to SN\,1999cq. Figures modified from Fox \& Smith (2019) \cite{2019MNRAS.488.3772F}, with permission.}
\label{fig:spec-evolution}
\end{figure}

It has been argued that features in the spectra of AT2018cow are similar to what has been observed in interacting supernovae of Types IIn and Ibn \cite{2019MNRAS.488.3772F,2021ApJ...910...42X} (Figure~\ref{fig:spec-evolution}, right). In interacting supernovae, these features are typically attributed to accelerated circumstellar material swept up in the shock. It was pointed out \cite{2019MNRAS.488.3772F} that the Ibn SN\,2006jc had spectra dominated by intermediate-width lines (1000--4000\,km\,s$^{-1}$) \cite{2007ApJ...657L.105F, 2007Natur.447..829P}, a similar velocity to what was observed in AT2018cow, and that velocities of 4000\,km\,s$^{-1}$ are often observed in the post-shocked circumstellar material of Type~IIn SNe \cite{2012ApJ...744...10K,2017MNRAS.466.3021S}. It was also pointed out \cite{2019MNRAS.488.3772F} that the line types, widths, and strengths are particularly similar to FBOT Type~Ibn SN\,1999cq \cite{2000AJ....119.2303M}, including the upturn longward of \ion{Ca}{2}; and that the hydrogen emission feature is similar to that of the FBOT Type~IIn PTF09uj \cite{2010ApJ...724.1396O}. In addition, it was shown that AT2018cow may have also had narrow (few hundred km\,s$^{-1}$) lines from dense pre-existing material, such as narrow \ion{He}{1} lines in the Day 44 spectrum \cite{2019MNRAS.488.3772F}. 

\subsubsection{CSS161010}

\begin{figure}[tb]
    \centering
    \includegraphics[width=0.535\linewidth]{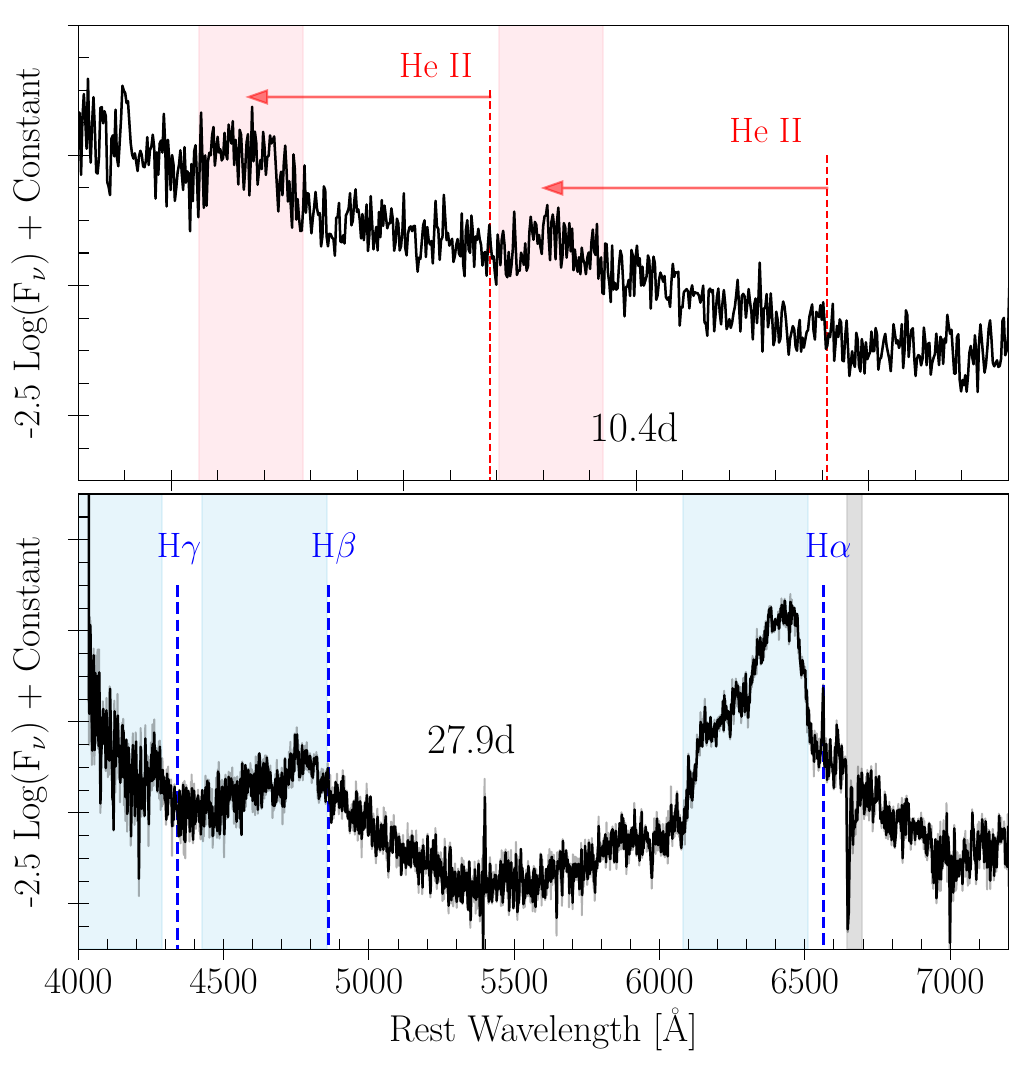}
    \includegraphics[width=0.455\linewidth]{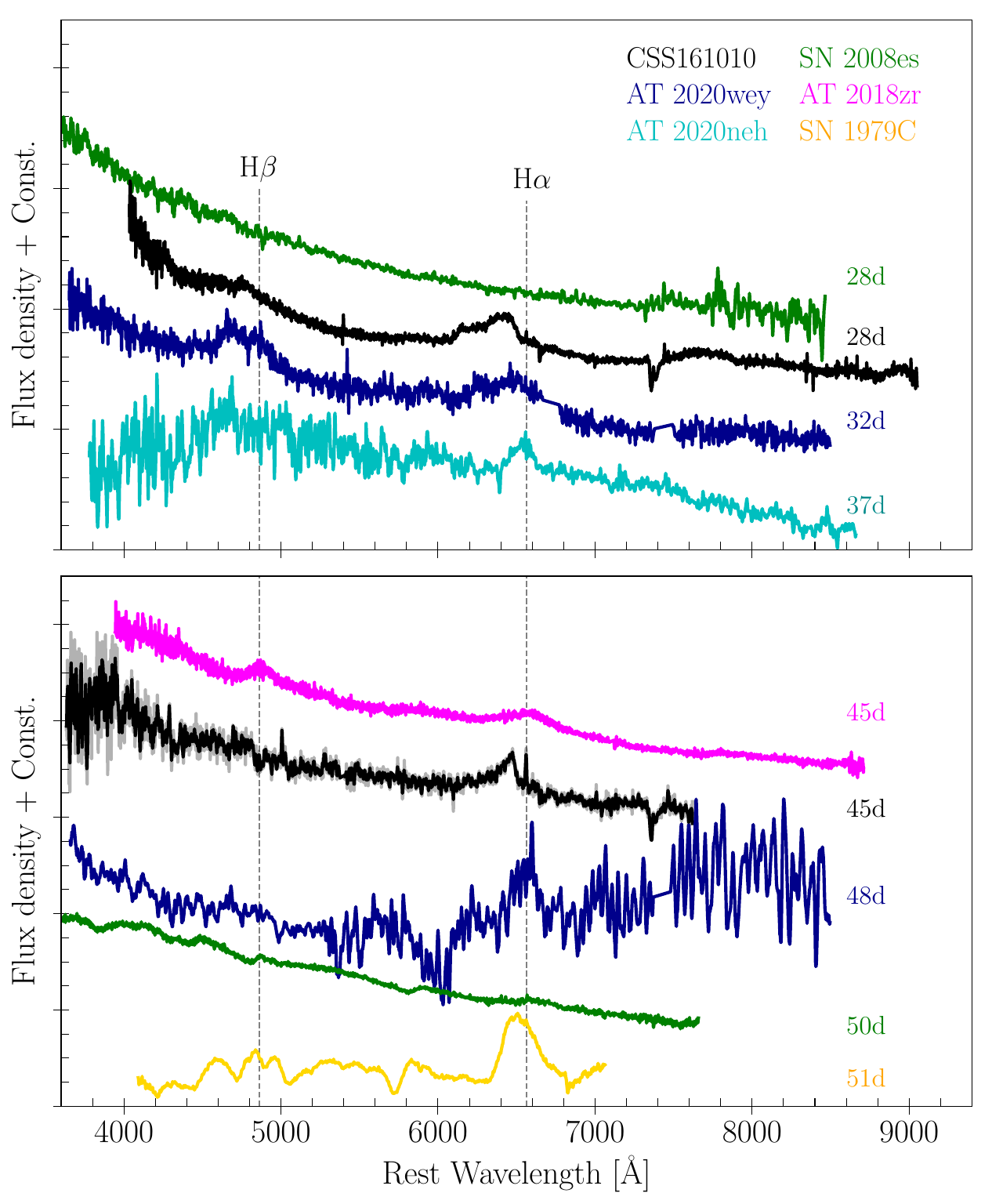}
    \vspace{-0.35in}
    \caption{Optical spectra of CSS161010. Figures modified from Gutiérrez et al. (2024) \cite{2024ApJ...977..162G}, with permission. \emph{Left:} Spectra at two select epochs. \emph{Right:} Spectra at two epochs (black) compared to supernovae (green, yellow) and tidal disruption events (other colors).}
    \label{fig:css-spectra}
\end{figure}

Like AT2018cow, CSS161010 had a featureless blue spectrum at early times ($<10\,$d) and exhibited transient \ion{He}{2} features: in a spectrum at 10.4\,d (Figure~\ref{fig:css-spectra}, top left), \ion{He}{2} 4686 and 5411 emission appeared blueshifted at 33,000 km\,s$^{-1}$, and disappeared by 21\,d. After 21\,d, the evolution was quite different from AT2018cow: two broad features attributed to H$\alpha$ and H$\beta$ appeared (Figure~\ref{fig:css-spectra}, bottom left), with no absorption, evolved to a complex flat-topped profile, then developed a narrower asymmetric shape at $>40\,$d. The velocities at the maxima of the profiles declined from 10,000 km\,s$^{-1}$ (21\,d) to 4000 km\,s$^{-1}$ (58\,d), while the velocities at the bluest parts of the profiles declined from 33,000 km\,s$^{-1}$ to 10,000 km\,s$^{-1}$ at those same dates. So, CSS161010 had a spectrum dominated by hydrogen from 20\,d onwards, while for AT2018cow it was dominated by \ion{He}{1}. The most similar spectral evolution in the literature to CSS161010 is that observed in TDEs (Figure~\ref{fig:css-spectra}, right) \cite{2024ApJ...977..162G}.

\subsubsection{AT2024wpp}

Like AT2018cow, the spectra were featureless in the first 6 days \cite{2026MNRAS.549ag678P}; a broad feature was detected at early times (8--13\,d) \cite{2025MNRAS.537.3298P,2026MNRAS.549ag678P}; and asymmetric, intermediate-width (full width $\delta v\sim3000\,$km\,s$^{-1}$) emission lines from H and He (H$\alpha$, \ion{He}{1} 5876, \ion{He}{2} 4686) emerged after one month and showed limited evolution\footnote{The full moon precluded detailed spectra from 11--36\,d.} \cite{2026MNRAS.549ag678P,2026ApJ...997L..10L}. Peculiarly, the H$\alpha$ line showed a prominent double-peaked structure, one peak at $v=0$ and the other blueshifted by 6400--6600\,km\,s$^{-1}$, with a gap of $\delta v\sim2000$\,km\,s$^{-1}$ and the flux returning to the continuum level in between \cite{2026MNRAS.549ag678P,2026ApJ...997L..10L}. A Hubble Space Telescope (HST) far-UV (FUV) spectrum was obtained at 20\,d  (Figure~\ref{fig:spec-evolution}, top right): from 1,000--10,000\,\AA\ the spectrum is reasonably well described by a featureless 21,000\,K blackbody \cite{2026MNRAS.549ag678P}. As discussed in Section~\ref{sec:nir-excess}, a NIR excess emerged at 20--30\,d (also observed in AT2018cow) \cite{2025MNRAS.537.3298P,2026ApJ...997L..10L}. Unlike the AT2018cow NIR spectrum, no He features were detected at 24\,d \cite{2026ApJ...997L..10L}. 

\subsubsection{Connection to X-ray--UVOIR ratio}

In AT2018cow and AT2024wpp, the emergence of spectral features with velocities $\approx10^{3}\,$km\,s$^{-1}$ appears to coincide with the ratio of the X-ray to UVOIR luminosity reaching unity: the ratio of the X-ray luminosity $L_X$ to the UVOIR luminosity $L_\mathrm{UVOIR}$ began at $\approx10^{-2}$ and eventually reached $\approx1$ \cite{2019ApJ...872...18M, 2026ApJ...997L..10L}. In AT2018cow, the transition to $L_X/L_\mathrm{UVOIR}\approx1$ occurred at $\approx20$--30\,d \cite{2019MNRAS.484.1031P, 2019ApJ...872...18M, 2026ApJ...997L..10L} while in AT2024wpp it occurred somewhat later \cite{2026ApJ...997L..10L}. 
CSS161010 did not have sufficiently detailed X-ray data for a similar analysis. 

\subsubsection{Other LFBOTs}

The spectra of LFBOTs are all dominated by a hot blackbody (e.g.,  \cite{2019MNRAS.484.1031P, 2022ApJ...934..104Y, 2023Natur.623..927H, 2015Natur.523..189G,2024MNRAS.527L..47C, 2024ApJ...977..162G,2026arXiv260813003F}). An early broad absorption feature, similar to that observed in AT2018cow, was also observed in AT2020xnd \cite{2021MNRAS.508.5138P} and AT2020mrf \cite{2022ApJ...934..104Y}. In addition to a broad absorption feature, a spectrum of AT2020xnd at $\Delta t\approx10\,$d showed a weak, narrow emission feature attributed to H$\alpha$, likely from the host galaxy \cite{2021MNRAS.508.5138P}. The spectrum of the potentially related SN\,2011kl was blue and approximately featureless redward of $\sim3600\,\AA$, but with a pronounced rest-frame UV downturn interpreted as blended broad features from high-velocity superluminous supernova-like ejecta \cite{2015Natur.523..189G}. Irrespective of the interpretation, the downturn differs from the best-observed LFBOTs. 




\subsection{X-ray emission}

The X-ray luminosities of LFBOTs are significantly higher than supernovae, and more similar to GRBs in the local universe. Some events have only X-ray non-detections, but those are not deep enough to rule out emission identical to AT2018cow (Figure~\ref{fig:xray-radio-summary}). We begin by summarizing the X-ray behavior of AT2018cow.

\subsubsection{AT2018cow}

\begin{figure}[tb]
    \centering
    \includegraphics[width=0.54\linewidth]{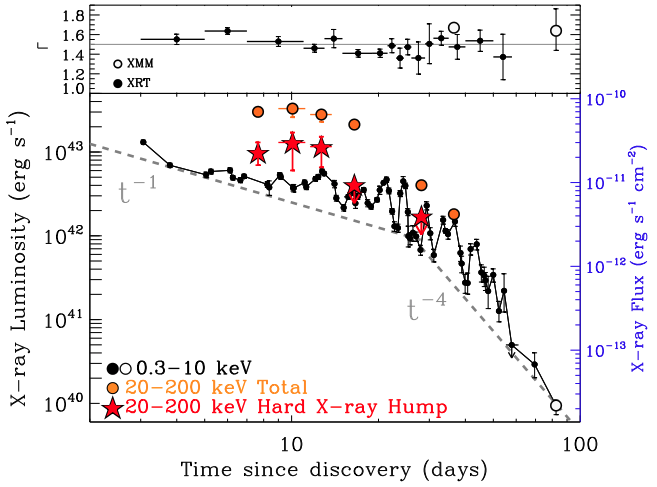}
    \includegraphics[width=0.45\linewidth]{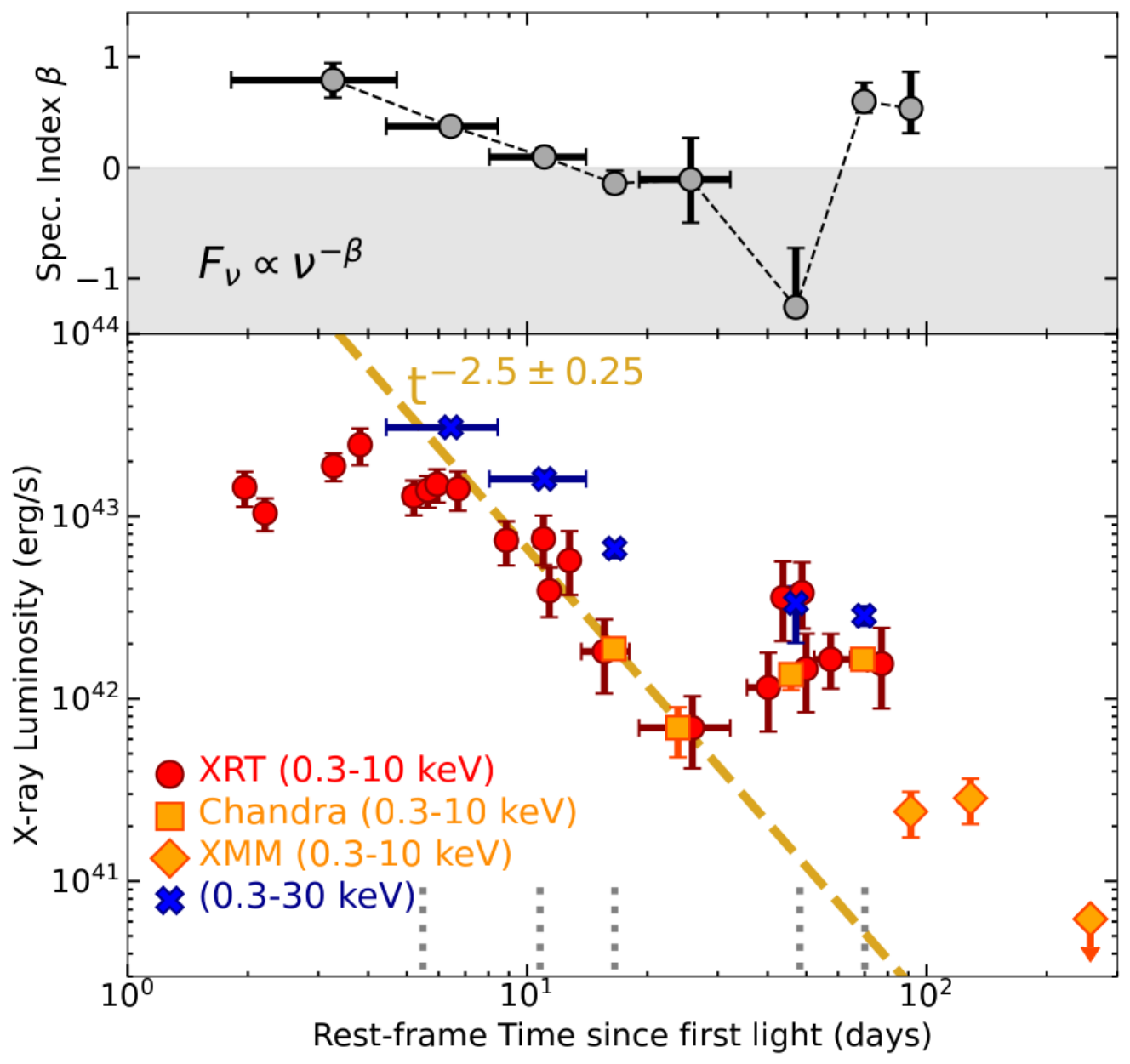}
    \vspace{-0.3in}
    \caption{X-ray light curve and photon (spectral) index for AT2018cow (left, reproduced from Margutti et al. 2019 \cite{2019ApJ...872...18M} with permission) and AT2024wpp (right, reproduced from Nayana AJ et al. 2025 \cite{2025ApJ...993L...6N} with permission). }
    \label{fig:xray-variability}
\end{figure}

The X-rays faded over 3--60\,d and showed significant variability on few-day timescales (Figure~\ref{fig:xray-variability}; \cite{2018MNRAS.480L.146R, 2019ApJ...871...73H, 2019ApJ...872...18M, 2019MNRAS.487.2505K}). The variability has been interpreted as superimposed on a smoothly varying component \cite{2018MNRAS.480L.146R, 2019ApJ...872...18M}, or as continuous flaring with some evolution with no underlying smoothly varying component \cite{2019MNRAS.487.2505K}. Regardless, the evolution appears to show two phases \cite{2019ApJ...871...73H, 2019ApJ...872...18M, 2019MNRAS.487.2505K}: a shallower decay until $\approx20$ days (during which $7\times10^{48}$\,erg was radiated in the Swift XRT band alone \cite{2019ApJ...871...73H}), then a steeper decay. The temporal decay index depends on how the flaring is treated in the fitting: the index ranges from $t^{-0.55}$ \cite{2018MNRAS.480L.146R} to $t^{-1}$ in the early phase \cite{2019ApJ...872...18M, 2019MNRAS.487.2505K}, and from $t^{-3}$ \cite{2019MNRAS.487.2505K} to $t^{-4}$ \cite{2019ApJ...872...18M} in the late phase. The earlier phase showed variability at the level of 50\% \cite{2018MNRAS.480L.146R, 2019ApJ...872...18M}, while in the later phase it increased to an order of magnitude \cite{2019ApJ...871...73H}. 

Results from variability analyses have been mixed. One analysis found significant power on a timescale of 4\,d in the first 40\,d, along with a high degree of correlation with the UV light curve \cite{2019ApJ...872...18M}. Another analysis found no significant timescales, concluding that the variability was ``burst-like'' rather than periodic \cite{2019MNRAS.487.2505K}. A 3.7$\sigma$ quasiperiodic oscillation was claimed from the entire 60 days of soft X-ray data from NICER, at 224\,Hz (4.4\,ms) \cite{2022NatAs...6..249P} while a $>3\sigma$ quasiperiodic oscillation with period 250\,s was claimed from an analysis of both XMM-Newton data and \emph{Swift}/XRT \cite{2022RAA....22l5016Z}. 

\begin{figure}[tb]
\centering
\includegraphics[width=\linewidth]{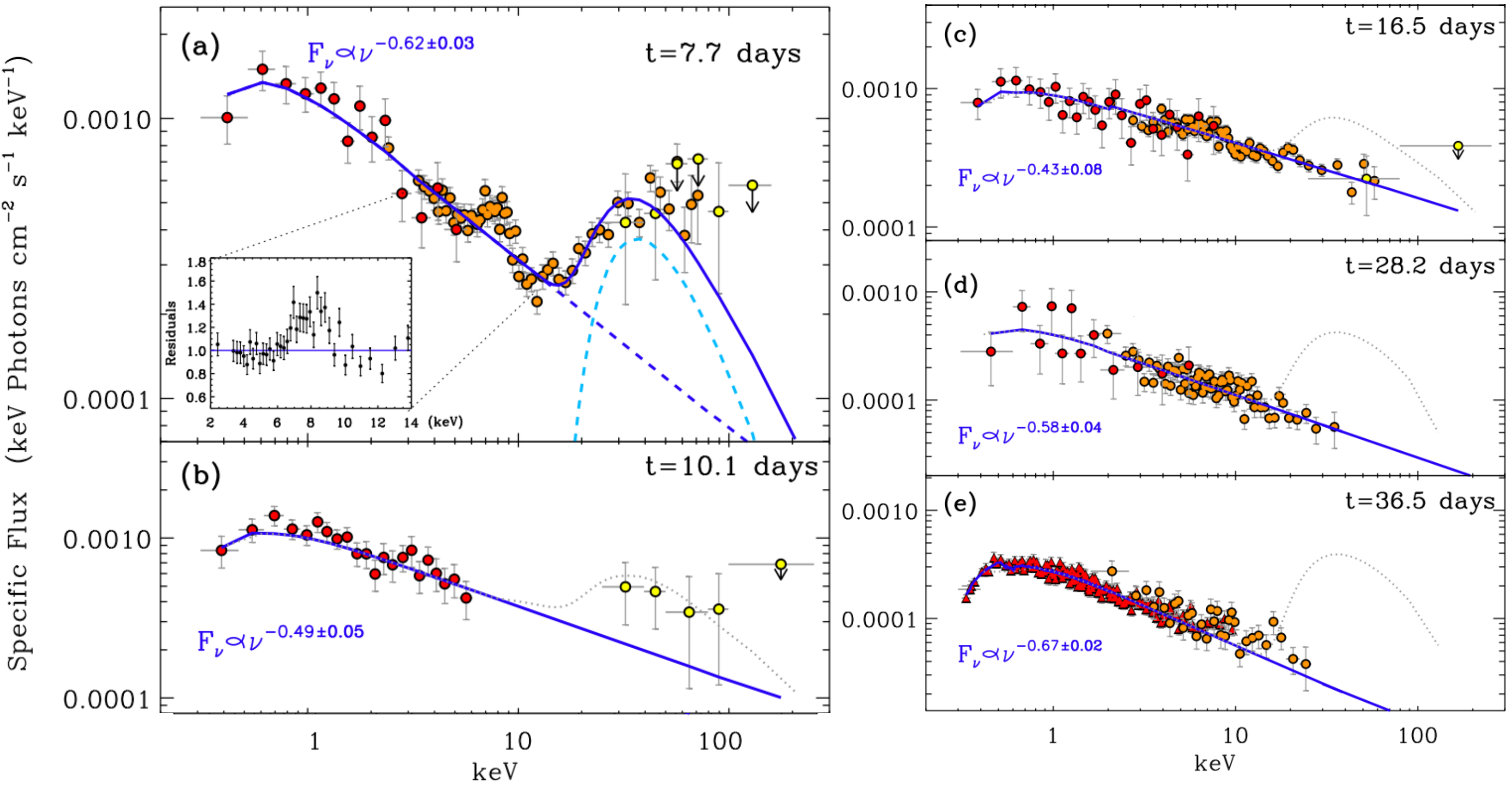}
\vspace{-8mm}
\caption{The evolution of the X-ray spectrum of AT2018cow, reproduced from Margutti et al. (2019) \cite{2019ApJ...872...18M} with permission. The early spectra showed two clear components, a soft power-law component with a harder ``bump''. In the later spectra, the hard bump disappeared, and the entire spectrum was well described by a single power law. In addition, the first spectrum showed a residual centered at 8\,keV that has been modeled as a Fe emission line.}
\label{fig:18cow-xray}
\end{figure}

It is generally agreed that the two phases in the light-curve evolution  correspond to phases in the spectral behavior (Figure~\ref{fig:18cow-xray}). Although no spectral evolution was observed in the XRT band (photon index $\Gamma=$1.5--1.7  \cite{2018MNRAS.480L.146R, 2019ApJ...872...18M}), there was a high-energy component detected up to 100\,keV present only in the early phase \cite{2019ApJ...871...73H, 2019ApJ...872...18M}. In the later phase, the spectrum was well described by a single power law extending up to high energies.  In addition to the transient hard component, at 7.7\,d the X-ray spectrum showed a feature centered at 8\,keV with a width of 1\,keV. Modeled as Fe K$\alpha$ emission, it implies a blueshift and Doppler broadening of the order $0.1c$ \cite{2019ApJ...872...18M}. 

Finally, the X-ray spectra showed no evidence for intrinsic neutral hydrogen absorption. An XMM-Newton observation constrained the intrinsic neutral hydrogen density at 36.5\,d to be $N_\mathrm{H,int}<0.05\times10^{22}\,$cm$^{-2}$. In total, the radiated energy in X-rays from 3--60\,d was $10^{49}\,$erg in soft X-rays and a similar amount in the hard X-ray component, not corrected for beaming \cite{2019ApJ...872...18M}.   

\subsubsection{Other LFBOTs}

X-ray variability has been observed in several other events, including AT2024wpp \cite{2025ApJ...993L...6N, 2026MNRAS.549ag678P}, AT2024puz (factor of 4 on 3\,d timescales) \cite{2025ApJ...995..228S}. In AT2020mrf, variability with a factor of $6$ over a day was observed at $\approx1\,$month and with a factor of $3\times$ over a day at $\approx1\,$yr \cite{2022ApJ...934..104Y}. Chandra observations of AT2022tsd revealed flux variations of factors of a few on timescales of tens of minutes (at the $3\sigma$ level), although no clear high-amplitude flares \cite{2023Natur.623..927H}. AT2024qfm exhibited an order-of-magnitude jump in X-ray luminosity over two days, much more pronounced than the variability observed in other events \cite{2026ApJ..1007...38S}. Rapid X-ray variability was also observed in the potentially related SN\,2011kl (factors of 10--100 on 100\,s timescales \cite{2014ApJ...781...13L}).

AT2020xnd ($\Gamma=1.4$) and AT2022tsd ($\Gamma=1.9$) \cite{2022ApJ...926..112B,2023RNAAS...7..126M, 2023Natur.623..927H} had no evidence for spectral evolution. For AT2020xnd, a NuSTAR non-detection was not sensitive enough to rule out a transient hard X-ray component identical to that in AT2018cow \cite{2022ApJ...926..112B}. Like AT2018cow, AT2024wpp has been modeled as having two distinct spectral components: a soft component and a transient hard component that peaked around 50\,d (later than the 8\,d in AT2018cow) \cite{2025ApJ...993L...6N}. Spectra of AT2020mrf and AT2024puz at 1\,month showed no statistical preference between a power law and thermal spectrum, with tentative evidence for spectral hardening at later times \cite{2022ApJ...934..104Y,2025ApJ...995..228S}. SN\,2011kl had complex spectra with departures from a single power-law fit: it is unclear if these were due to rapid temporal variation or multiple components \cite{2014ApJ...781...13L}. 

As shown in Figure~\ref{fig:xray-radio-summary}, the X-ray luminosities vary by orders of magnitude at late times. The most luminous events at $>100\,$d are AT2020mrf and AT2022tsd. Chandra observations of AT2020mrf at $\Delta t \approx 1$ year revealed a luminous detection, $200\times$ that of AT2018cow at 200\,d and $300\times$ that of CSS161010 at 300\,d \cite{2022ApJ...934..104Y}. 
AT2022tsd had the highest X-ray luminosity overall, at $L_X=10^{44}\,$erg\,s$^{-1}$ at 23\,d. Its light curve during this period could be fit by $t^{-2}$ with variability superimposed or alternatively a broken power law with similar pre- and post-break indices to AT2018cow \cite{2023RNAAS...7..126M}. However, fitting the observations out to 300\,d favors a single power-law decline ($t^{-1.81\pm0.13}$) with variability, with the total radiated energy in X-rays exceeding $10^{50}\,$erg \cite{2023Natur.623..927H}. 

Finally, we note slightly different measurements in the literature. CSS161010 had three \emph{Swift}/XRT observations from 24--32\,d, the first of which had been reported as a non-detection \cite{2024ApJ...977..162G} or alternatively a very marginal (3.3$\sigma$) detection \cite{2026ApJ..1007...38S}. AT2023fhn was observed by Chandra \cite{2024MNRAS.527L..47C} in four epochs spanning 10--200\,d \cite{2024A&A...691A.329C}. A detection was initially reported in the first epoch only \cite{2024A&A...691A.329C}, but an independent analysis of the data found detections in the first three epochs, with luminosities close to AT2018cow at the same epochs \cite{2025ApJ...993L...6N}. 





\subsection{Radio emission}

Among the LFBOTs discussed here, all except AT2024puz have luminous radio detections; the available upper limits for AT2024puz do not exclude AT2018cow-like radio emission. We begin by summarizing the radio behavior of AT2018cow and how it compares to more commonly observed radio transients, such as GRBs and supernovae. We then describe the emission observed in other events. 

\subsubsection{AT2018cow}

\begin{figure}[tb]
    \centering
    \vspace{0pt}
    \includegraphics[width=0.59\linewidth]{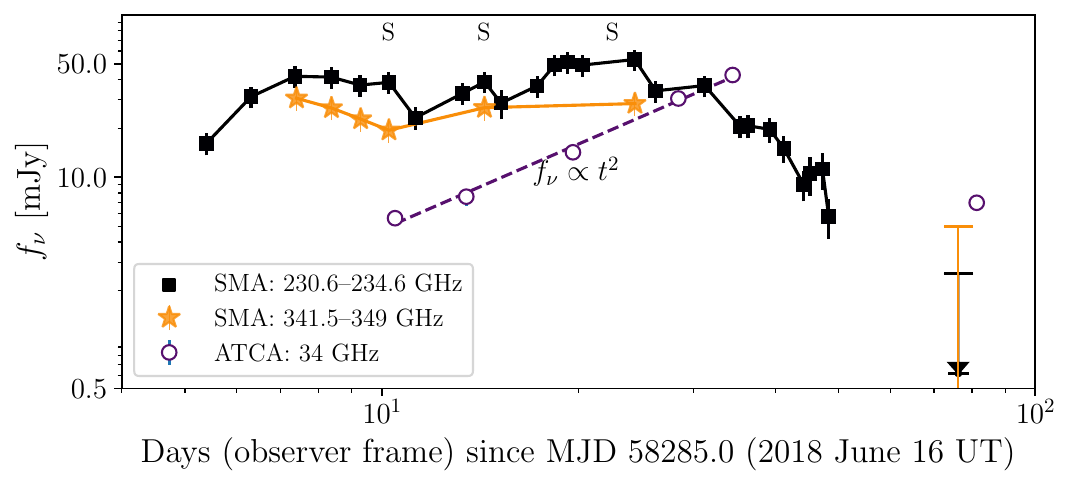}
    \includegraphics[width=0.4\linewidth]{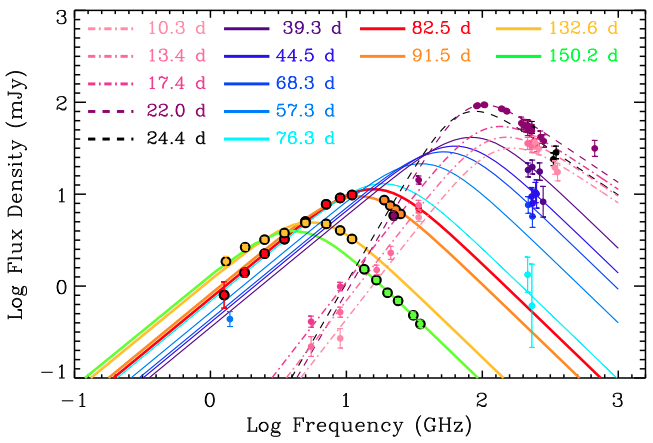}
        \vspace{0.1em} 
    \includegraphics[width=\linewidth]{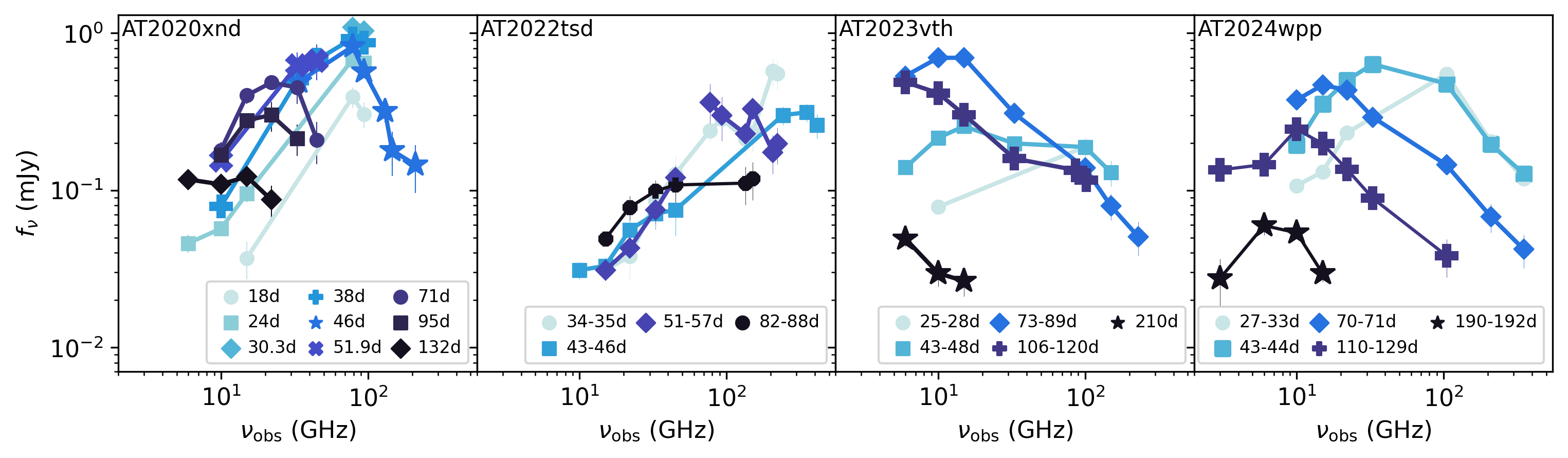}
    \vspace{-0.35in}
    \caption{Radio evolution. \emph{Top left:} AT2018cow radio light curves at cm (34\,GHz) and mm wavelengths (230\,GHz and 340\,GHz) showed smooth rising behavior at low frequencies and variability at high frequencies. Figure modified from Ho et al. (2019) \cite{2019ApJ...871...73H} with permission. \emph{Top right:} The radio SED evolution of AT2018cow: after $\approx10\,$d the evolution was reasonably well described by a broken power law with peak frequency and peak flux density declining as power laws in time. However, this behavior underpredicts the early-time high-frequency emission. Figure reproduced from Margutti et al. (2019) \cite{2019ApJ...872...18M} with permission.
    \emph{Bottom row:} Radio SED evolution of AT2020xnd \cite{2022ApJ...932..116H}, AT2022tsd \cite{2023Natur.623..927H}, AT2023vth \cite{2026ApJ..1007...38S}, and AT2024wpp \cite{2026MNRAS.549ag678P}. 
    }
    \label{fig:cow-radio}
\end{figure}

In supernovae and GRBs, radio emission is typically synchrotron radiation that follows a broken power law spectral energy distribution (SED). Light curves often show a frequency-dependent power-law rise to peak, then decay (e.g., \cite{2021ApJ...908...75B}). Radio observations of AT2018cow---spanning 5--570\,d and frequencies from 250\,MHz to nearly 1\,THz---showed this behavior to some extent (Figure~\ref{fig:cow-radio}, top right), but there were also significant differences. At high frequencies (230\,GHz; mm wavelengths) the luminosity was rivaled only by GRBs \cite{2019ApJ...871...73H}, while at cm wavelengths the luminosity was that of very luminous supernovae \cite{2019ApJ...871...73H, 2019ApJ...872...18M, 2021ApJ...912L...9N} (Figure~\ref{fig:xray-radio-summary}). The SED peak was at mm wavelengths for tens of days (Figure~\ref{fig:cow-radio}, top right), while in most cosmic explosions the peak shifts down to cm wavelengths after only a few days. The unusual SED evolution is also reflected in the light curves: the 34\,GHz light curve rose as a smooth $t^{2}$ power law for 40\,d, while the 230\,GHz emission initially rose (likely when it was below the peak frequency), flattened out and showed significant variability, then diminished around 50\,d \cite{2019ApJ...871...73H} (Figure~\ref{fig:cow-radio}, top left). 
Also, while the SED evolution after 10\,d was reasonably well described by a smoothed broken power law, with peak flux density $F_{\mathrm{pk}}\propto t^{-1.7}$ and peak frequency $\nu_\mathrm{pk}\propto t^{-2.2}$ \cite{2019ApJ...872...18M}, this evolution significantly underpredicts the early-time high-frequency emission (Figure~\ref{fig:cow-radio}, top right). Finally, at low frequencies, the spectral index flattened out and was significantly shallower than the $\nu^{5/2}$ expected from synchrotron self-absorption \cite{2019ApJ...871...73H, 2021ApJ...912L...9N}. 

Very Long Baseline Interferometry (VLBI) observations did not resolve the source. Assuming circular symmetry, the average apparent expansion velocity was constrained to $<0.49c$ by 98\,d (3$\sigma$)---but one-sided expansion or elongation along certain directions could allow for speeds up to a factor of 2 higher. A 3$\sigma$ limit on the proper motion yielded $<0.51c$, independent of the assumption of circular symmetry. The conclusion was that a sustained relativistic jet was unlikely to have been present \cite{2020MNRAS.491.4735B}. Observations with the European VLBI network spanning 260\,d constrained the proper motion to $<0.14c$ \cite{2020ApJ...888L..24M}. We summarize constraints from radio polarization in Section~\ref{sec:obs-polarization}.

Modeling the radio emission from AT2018cow  implied a fast but non-relativistic shock speed of $v=0.1c$ \cite{2019ApJ...871...73H, 2019ApJ...872...18M} or $v=0.2c$ \cite{2021ApJ...912L...9N}, and relatively high CSM densities of $10^{5}\,$cm$^{-3}$ \cite{2019ApJ...871...73H} that dropped over time at a rate steeper than expected for a stellar wind \cite{2019ApJ...872...18M, 2021ApJ...912L...9N}. We discuss the theory of radio modeling in more detail in Section~\ref{sec:theory-radio}. 

\subsubsection{Other LFBOTs}

\begin{figure}[tb]
\centering
\begin{minipage}[b]{0.51\linewidth}
\vspace{0pt}
  \centering
  \includegraphics[width=\linewidth]{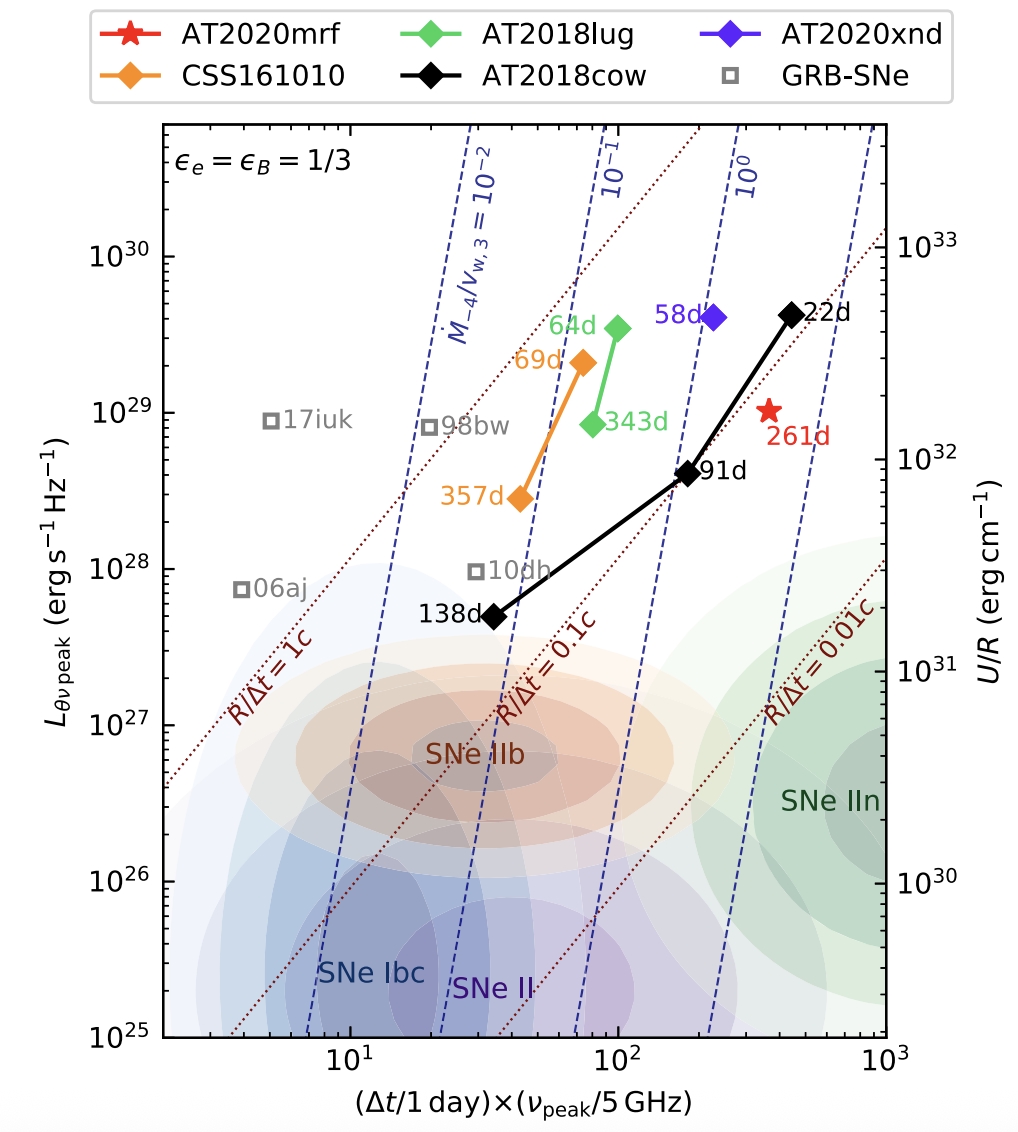}
\end{minipage}
\hfill
\begin{minipage}[b]{0.48\linewidth}
\vspace{0pt}
  \centering
  \includegraphics[width=\linewidth]{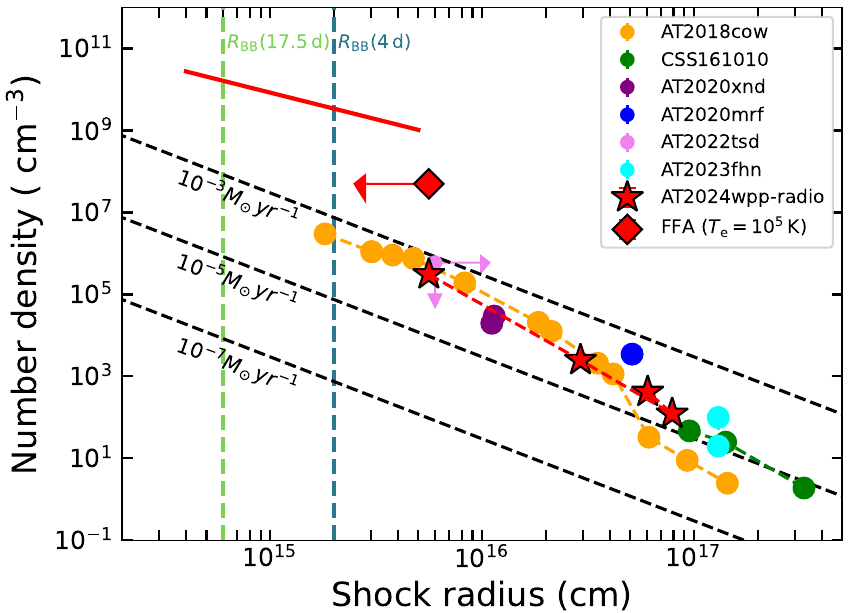}
\end{minipage}
\vspace{-0.3in}
\caption{\emph{Left:} Peak frequency vs. peak luminosity of radio transients including different types of core-collapse supernovae (SNe II, SNe IIn, SNe Ibc, and SNe IIb) as well as supernovae associated with gamma-ray bursts (GRBs). Individual objects are shown as colored points. Diagonal lines show contours of constant shock speed (grey dotted) and constant mass-loss rate (ambient density; blue dashed). Figure reproduced from Yao et al. (2022) \cite{2022ApJ...934..104Y} with permission. 
\emph{Right:} Density profile of several observed events, reproduced from Nayana A. J. et al. (2025) \cite{2025ApJ...993L...6N} with permission. The measured densities are quite similar at fixed radii, interpreted as an ambient medium produced in a consistent way prior to the terminal event. The mass-loss rates corresponds to an (arbitrarily chosen) outflow velocity of $v_{\rm csm}=1000\rm\, km\,s^{-1}$.
}
\label{fig:radio-modeling}
\end{figure}

The radio SEDs consistently show a high peak frequency for the time of observation (Figure~\ref{fig:radio-modeling}, left), which drops over time (Figure~\ref{fig:cow-radio}, bottom). In CSS161010, for example, the peak frequency and flux density evolved as $\nu_p \propto t^{-1.3}$ and $F_p \propto t^{-1.8}$ \cite{2020ApJ...895L..23C}. The SED seems to evolve in two stages: an overall brightening phase, before a cascade down to lower frequencies and flux, with the transition occurring from tens of days to over a hundred days \cite{2022ApJ...926..112B, 2022ApJ...932..116H, 2024A&A...691A.329C, 2025ApJ...993L...6N, 2026MNRAS.549ag678P, 2026ApJ..1007...38S}. The cm-wave radio light curves of several events exhibited very steep declines after the peak, as steep as $f_\nu \propto t^{-7}$ at 6\,GHz (340--400\,d) in AT2018lug \cite{2020ApJ...895...49H} and $t^{-5}$ at 8\,GHz for CSS161010 at $>99$\,d \cite{2020ApJ...895L..23C}. The steep fade appears to be chromatic, occurring at later times at lower frequencies \cite{2020ApJ...895...49H, 2022ApJ...932..116H}. 

The below- and above-peak spectral indices vary between the different events. For CSS161010, the values were $f_\nu \propto \nu^{2}$ below the peak and $f_\nu \propto \nu^{-1.3}$ above the peak \cite{2020ApJ...895L..23C}. In AT2020xnd, the SED at 46\,d showed a very steep spectral index $f_\nu \propto \nu^{-2}$. One possible explanation is that the emission is dominated by a relativistic Maxwellian electron population, although adiabatic expansion of a shocked shell without continuous shock-acceleration can also result in steeply declining light curves \cite{2022ApJ...932..116H}. Radio observations of AT2020mrf at 261\,d and 417\,d showed shallow pre-peak spectral indices and flat-topped SEDs, motivating an inhomogeneous CSM model \cite{2022ApJ...934..104Y}. Observations of AT2022tsd also showed a very shallow $f_\nu \propto \nu^1$ SED, which peaked at hundreds of GHz for more than a month (Figure~\ref{fig:cow-radio}) \cite{2023Natur.623..927H}. In contrast to the varying shock speeds, the ambient density profile appears similar from event to event (Figure~\ref{fig:cow-radio}, bottom right). 

The implied shock speeds can be significantly higher than those observed in AT2018cow: AT2018lug had $\beta \gtrsim 0.3$ \cite{2020ApJ...895...49H} while CSS161010 had $\beta \gtrsim 0.55$ at 100\,d (with 0.01--0.1\,$M_\odot$ coupled to the fast ejecta) \cite{2020ApJ...895L..23C}. In AT2024wpp, a significantly increasing velocity (from the shock-radius evolution) was reported over 32--73\,d \cite{2025ApJ...993L...6N}, but independent observations over the same time period found a close to constant shock speed $v=0.15c$ \cite{2026MNRAS.549ag678P}. The inferred shock radius also appeared to ``flatten out'' at late times  \cite{2026MNRAS.549ag678P}. 



Radio luminosity values for a time window when most light curves peak are provided in Table~\ref{tab:summary}. The frequency is chosen to be as close to rest frame 10\,GHz as possible, but in practice the rest frame values range from 6.2\,GHz (CSS161010) to 18.8\,GHz (AT2022tsd). Similarly, the frequency for the mm emission is chosen to be as close to rest frame 100\,GHz as possible, but in practice ranges from 104--136\,GHz. However, given the expected spectral indices at these times, the difference in frequency should not have a significant impact on the spectral luminosity values. Among the potentially related events, SN2011kl had very limited radio coverage, while AT2024aehp exhibited a late ($<100\,$d) radio turn-on \cite{2026ApJ..1007...38S}. 

\subsection{Host galaxies and environment}

AT2018cow's proximity enabled certain observations  that have not been possible for other, higher-redshift LFBOTs. 
We begin by describing the environment of AT2018cow, then describe the bulk properties of the host galaxies of the other LFBOTs.

\subsubsection{AT2018cow}

\begin{figure}[tb]
    \centering
    \includegraphics[width=\linewidth]{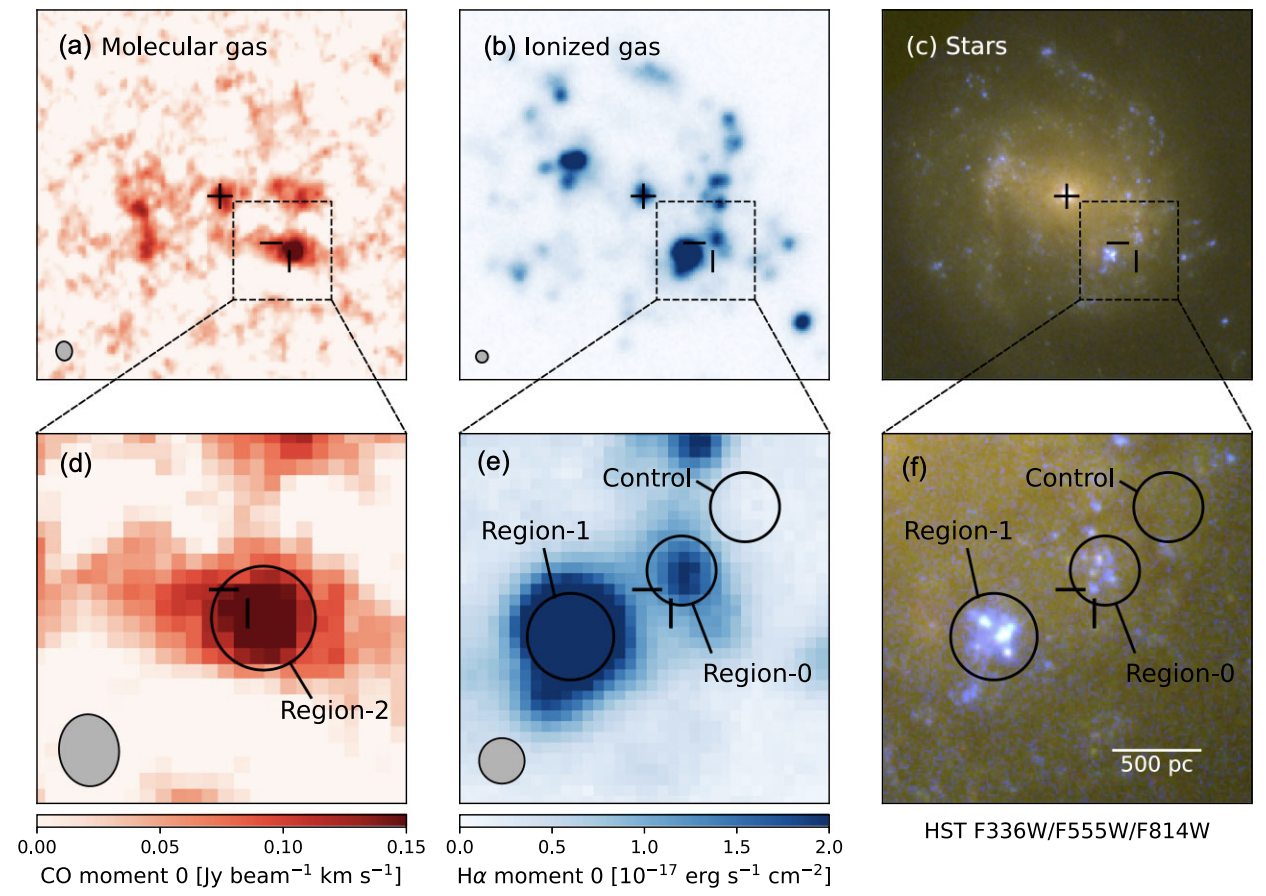}
    \vspace{-0.3in}
    \caption{\emph{Top:} The host galaxy of AT2018cow in (a) molecular (CO) gas, (b) ionized (H$\alpha$) gas, and (c) stars, with the galaxy nucleus marked with a cross and the position of AT2018cow marked with cross-hairs. \emph{Bottom:} The environment around AT2018cow, with several regions marked that were used for analysis. AT2018cow is coincident with a strong enhancement in molecular gas, and multiple star-formation regions. Figure reproduced from Sun et al. (2023) \cite{2023MNRAS.519.3785S} with permission.}
    \label{fig:at2018cow-host}
\end{figure}

AT2018cow occurred 1.7\,kpc from the nucleus of a star-forming galaxy \cite{2018ApJ...865L...3P}, coincident with a spiral arm \cite{2019ApJ...872...18M} (Figure~\ref{fig:at2018cow-host}). In addition to weak spiral features, the galaxy has a barred morphology, and likely a weak active galactic nucleus (AGN) \cite{2019MNRAS.484.1031P}. The stellar mass (1.4--1.7$\times10^{9}\,M_\odot$) and star-formation rate (0.19--0.22$\,M_\odot\,$yr$^{-1}$) make it a dwarf galaxy similar to the Large Magellanic Cloud (LMC) \cite{2019MNRAS.484.1031P, 2020MNRAS.495..992L}. Studies of the atomic (\ion{H}{1}) and molecular (CO) gas properties found characteristics typical of normal star-forming dwarf galaxies \cite{2019MNRAS.485L..93R, 2019A&A...627A.106M, 2019ApJ...879L..13M, 2023MNRAS.519.3785S}, except for an asymmetric ring of high \ion{H}{1} column density around the galaxy optical center. AT2018cow was coincident with this ring, which has been attributed to interaction with a companion galaxy, and used to argue for massive-star progenitors (as it would be an ideal environment for star formation \cite{2019MNRAS.485L..93R}). 

An \ion{H}{2} region was visible slightly southeast of AT2018cow's position, but no underlying point source was detected in optical pre-imaging \cite{2019MNRAS.484.1031P}. An ALMA image of cold molecular gas showed that AT2018cow was close to a peak in intensity (left panel of Figure~\ref{fig:at2018cow-host}), with a molecular gas surface density of $14\,M_\odot\,$pc$^{-2}$ \cite{2019ApJ...879L..13M}, again favoring a connection with star formation. IFU imaging found that the site was typical of core collapse supernovae, with a young stellar population age (few $\times$ 10\,Myr), slightly sub-solar metallicity, and proximity to strong star formation. Similarly to the \ion{H}{1} study, the H$\alpha$ distribution was found to be irregular and significantly asymmetric, weighted in the direction of a faint tidal tail, supporting a recent dynamical interaction. AT2018cow was found to be close to two bright compact star-formation regions, one 130\,pc away, the other (the brightest in the galaxy) 570\,pc away (central panel of Figure~\ref{fig:at2018cow-host}) \cite{2020MNRAS.495..992L}. 

In summary, AT2018cow was in the vicinity of a prominent molecular gas concentration, and two giant star-formation complexes. Although AT2018cow appears coincident with one of the complexes (Figure~\ref{fig:at2018cow-host}, right), it was likely in the foreground due to its much lower extinction; if so, then a non-detection of the associated stellar population would constrain its age to be $\gtrsim10\,$Myr. If AT2018cow occurred inside the region, its age would be very low, only a few Myr \cite{2023MNRAS.519.3785S}. 

\subsubsection{Other LFBOTs}

Following AT2018cow, the next few LFBOTs reported were located in dwarf galaxies. The host galaxy of AT2018lug had a notably high specific star-formation rate (sSFR = $1.4\times10^{-8}\,$yr$^{-1}$) making it a starburst galaxy \cite{2020ApJ...895...49H}, while CSS161010 had a particularly low host-galaxy mass ($10^{7}\,M_\odot$) \cite{2020ApJ...895L..23C}. AT2020xnd \cite{2021MNRAS.508.5138P} and AT2020mrf \cite{2022ApJ...934..104Y} were also in dwarf hosts. This led to suggestions that these transients are similar to long-duration GRBs (LGRBs) and Type~I superluminous supernovae (SLSN-I), which have a preference for low-mass and low-metallicity galaxies.

The picture changed somewhat after that. AT2022tsd was in a higher-mass host ($10^{10}\,M_\odot$) \cite{2023Natur.623..927H}, while AT2023fhn was $>3.5$ half-light radii from the two closest galaxies (16.5\,kpc from the larger galaxy and 5.35\,kpc offset from a dwarf satellite). Only 1\% of core collapse supernovae occur at such large offsets. The masses of several other events (e.g., AT2022abfc and AT2023hkw) were close to $10^{11}\,M_\odot$ \cite{2026ApJ..1007...38S}. Finally, AT2024puz was 5\,kpc offset from a $10^{8}\,M_\odot$ galaxy with a low to moderate star formation rate, but was selected by a strategy that requires a $>10^{\prime\prime}$ offset \cite{2025ApJ...995..228S}. 

Nuclear (galactocentric) offsets vary in the literature, in part due to different methods for measuring the transient position (e.g., ZTF public alerts \cite{2026ApJ..1006...75N} or VLA radio positions \cite{2026ApJ..1007...38S}) and the host-galaxy centroid (different optical surveys). So far there are two published compilations of physical offsets \cite{2026ApJ..1006...75N, 2026ApJ..1007...38S}. From individual papers and these compilations, 
CSS161010 has been reported to be coincident with the host-galaxy nucleus \cite{2020ApJ...895L..23C} but also to be offset by $\approx300\,$pc \cite{2024ApJ...977..162G, 2026ApJ..1007...38S, 2026ApJ..1006...75N}. AT2018lug has been reported to be offset by $0\farcs28\pm0\farcs13$ \cite{2020ApJ...895...49H} or $0\farcs57\pm0\farcs15$ \cite{2026ApJ..1006...75N}, corresponding to 0.9--3.0\,kpc\footnote{One offset value incorrectly used luminosity distance instead of angular diameter distance \cite{2020ApJ...895...49H}.}. Finally, AT2023fhn has two possible host galaxies: a more massive spiral and a dwarf satellite, with significantly different offsets \cite{2024MNRAS.527L..47C}. Some offsets presented in the literature assume that the progenitor originated in the massive spiral \cite{2026ApJ..1007...38S, 2026ApJ..1006...75N}. In Table~\ref{tab:summary} we use the HST measurements for AT2023fhn and otherwise present the range of values in the literature. 

Analysis of host-galaxy photometry and spectroscopy for 11 LFBOTs using \texttt{Prospector} found that all occurred in actively star-forming galaxies, for which the highest rate of star formation, or a burst of star formation, occurred within the past 100\,Myr. It was also found that the host galaxies are statistically distinct from those of SLSN-I (in that they have higher masses and lower specific star-formation rates), but that it is possible that they are similar to those of LGRBs. The sSFR was also found to be higher than that of core collapse supernovae hosts. 
The host stellar and gas-phase metallicities are sub-solar: the $12+\log(O/H)$ metallicity is statistically distinct (lower) than those of field galaxies and core collapse supernovae but higher than those of SLSN-I and LGRBs. They occupy similar regions of the BPT diagram to SLSN-I and LGRBs. Finally, their offset distributions are similar to core-collapse supernovae, and they seem to occur in fainter pixels of their host galaxies, unlike LGRBs and core collapse supernovae \cite{2026ApJ..1006...75N}. However, as most of the host galaxies are not well resolved from the ground, such a study ultimately requires HST. 

\begin{figure}[b]
\centering
\vspace{0pt}
  \centering
  \includegraphics[width=0.59\linewidth]{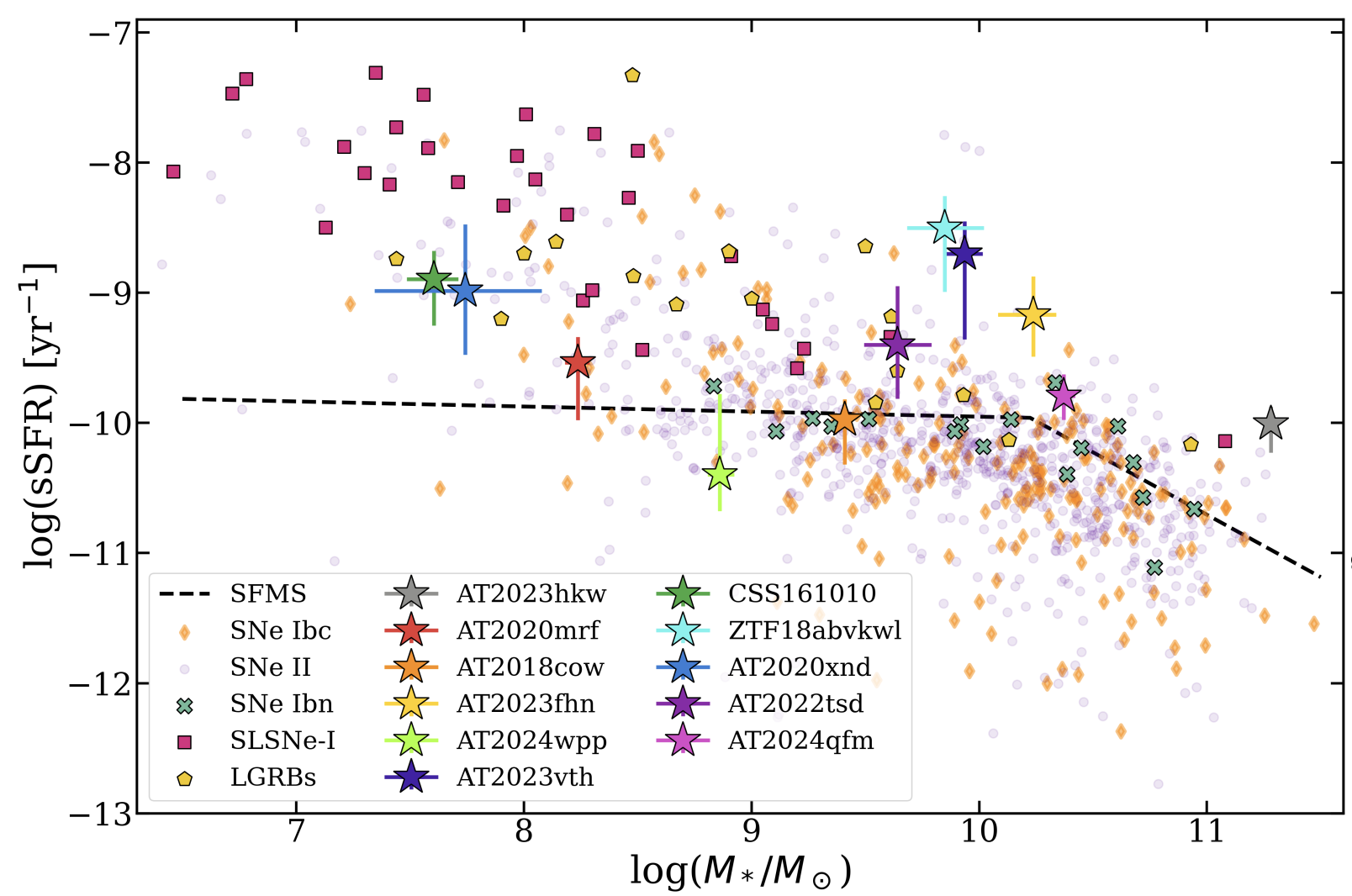}
  \includegraphics[width=0.4\linewidth]{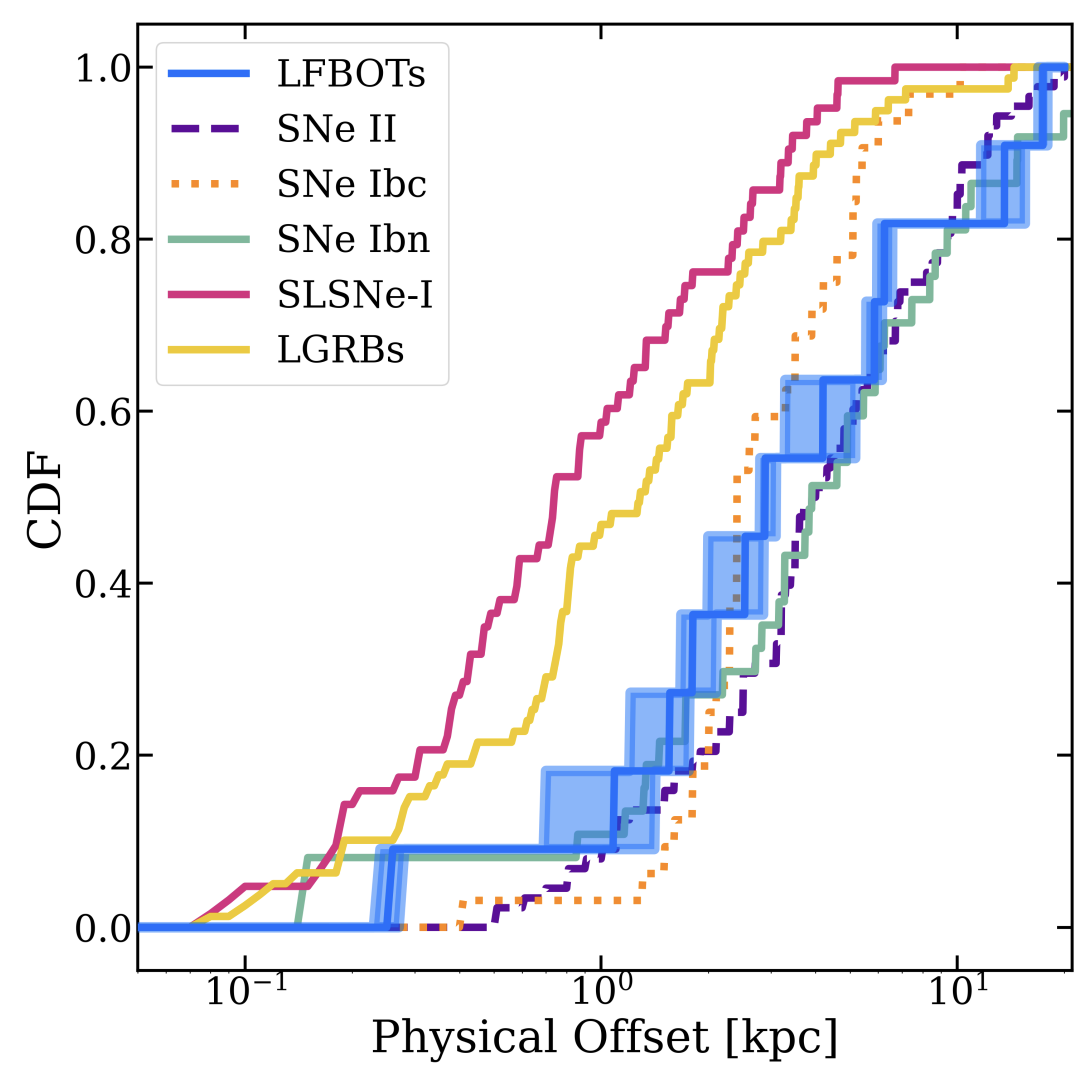}
  \vspace{-0.3in}
\caption{\emph{Left:} Stellar mass vs. star formation rate of host galaxies for most of the transients in Table~\ref{tab:summary}, compared to other classes of extragalactic transients. \emph{Right:}  Offset distribution of the transients compared to other classes of extragalactic transientns in the literature. The offset distribution so far is similar to that of core collapse supernovae. Both figures reproduced from Nugent et al. (2026) \cite{2026ApJ..1006...75N} with permission.  
}
\label{fig:hosts}
\end{figure}

An independent analysis of the host galaxy properties accounted for the selection bias in discovering these objects: that a host association is required, which preferentially selects for more massive galaxies. Even accounting for this selection bias, the mass distribution of the host galaxies appears to be higher than that of SLSN-I, and more similar to those of core collapse supernovae \cite{2026ApJ..1007...38S}. 
A summary of the host-galaxy properties is shown in Figure~\ref{fig:hosts}. 

\subsection{Constraints on optical and radio polarization}
\label{sec:obs-polarization}

Two LFBOTs (AT2018cow and AT2024wpp) had optical polarization observations. AT2018cow showed strong time- and wavelength-dependent polarization, reaching 7\% at red wavelengths at 5.7\,d, declining below the sensitivity threshold, with a second ``bump'' at blue wavelengths at 12\,d (reaching 2\%). This polarization was the highest recorded for a non-relativistic transient and exceeds the theoretical maximum for a thermal electron-scattering photosphere \cite{2023MNRAS.521.3323M}. Polarimetric observations of AT2024wpp showed no significant polarization ($<0.5\%$) from 6.1--14.4\,d \cite{2025MNRAS.537.3298P}. 

Two epochs of radio polarimetry were obtained of AT2018cow by ALMA, at 11\,d and 17\,d, at 100\,GHz and 230\,GHz. The level of polarization was constrained to $<0.2\%$ at both epochs, implying internal Faraday depolarization from a high ambient density \cite{2019ApJ...878L..25H}. 

\subsection{Optical flares}

\begin{figure}[tb]
\centering
\includegraphics[width=\linewidth]{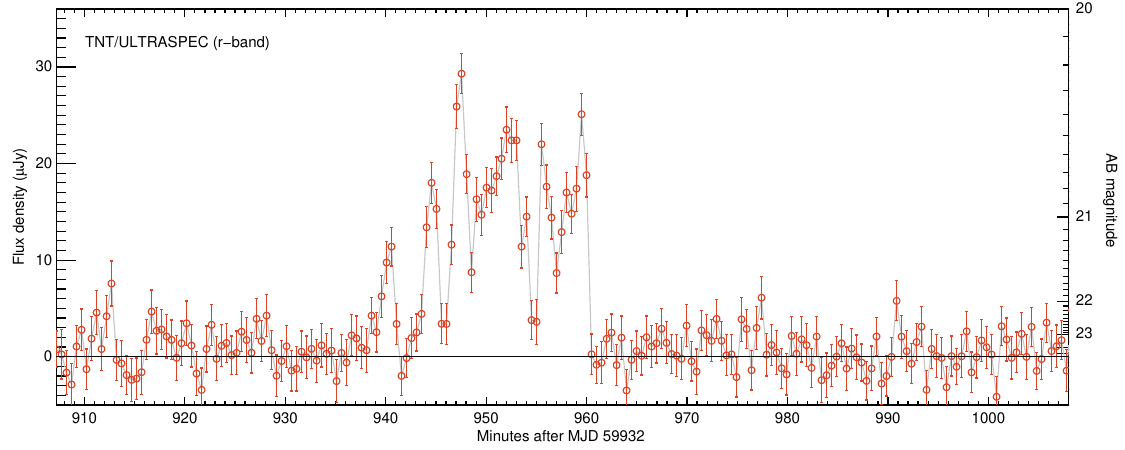}
\vspace{-7mm}
\caption{A flare from AT2022tsd observed by ULTRASPEC, an instrument mounted on the 2.4\,m Thai National Telescope. Light curve was measured at 30\,s cadence. Figure modified from Ho et al. (2023) \cite{2023Natur.623..927H} with permission.}
\label{fig:flare}
\end{figure}

A routine optical imaging sequence of AT2022tsd at 100\,d resulted in a surprisingly bright detection, several magnitudes in excess of the extrapolation of the steadily fading light curve. Visual inspection revealed that the  ``detection'' was a flare lasting 20~minutes. A retrospective search of survey data from ZTF and Pan-STARRS revealed other flare detections at the position of AT2022tsd as early as 30 days after the discovery. Over the following weeks, 60 hours of observations of AT2022tsd with 8 different telescopes resulted in the detection of 14 flares (e.g., Figure~\ref{fig:flare}), as late as 120\,d after the initial transient discovery. 
Such energetic ($10^{44}\,$erg\,s$^{-1}$ or $10^{47}\,$erg in each flare, uncorrected for beaming), long-lasting flaring was unprecedented for cosmic transients. The red flare colors imply non-thermal emission and near-relativistic outflow speeds. No periodicity was detected between or within flares \cite{2023Natur.623..927H}. A search for similar flaring in the aftermath of AT2024wpp ruled out behavior identical to that in AT2022tsd \cite{2025ApJ...993...76O}. A search in Transiting Exoplanet Survey Satellite (TESS) data found no flares at the positions of any LFBOTs, implying that the flaring in AT2022tsd declined in brightness or shut off after hundreds of days---and that flaring at the same luminosity cannot be present in all events on the same timescales \cite{2026arXiv260617129J}. It has been suggested that the flares arise from late-time accretion onto a compact object from a binary companion \cite{2024ApJ...972L..17L}; we discuss this and other possible sources of late-time accretion in Section~\ref{sec:theory-late-accretion}.

\subsection{A NIR excess}
\label{sec:nir-excess}

Although the UV/optical emission from AT2018cow was generally well modeled by a blackbody, an excess above this blackbody was observed in the $R$, $I$, and near-infrared bands, and attributed to a separate component. This excess was observed throughout the first 45 days \cite{2019MNRAS.484.1031P, 2019ApJ...872...18M}, beginning at 3.4\,d (essentially, when a blackbody fit to the SEDs began to underpredict the NIR flux) \cite{2019MNRAS.484.1031P}. The red excess could also be observed in the spectra after 10\,d, with the component appearing to peak around 10,000\AA\ \cite{2019MNRAS.484.1031P}. The component was reasonably well described by both a cool (3000\,K) blackbody and a power law with $F_\nu \propto \nu^{-0.75}$---although the power law index is poorly constrained.\footnote{In most epochs it cannot be meaningfully constrained; for the best epochs the uncertainty is $\pm0.3$ (1$\sigma$ confidence; Daniel Perley, private communication).}

In addition, the NIR bands showed minor but significant variability on timescales of 2--3\,d \cite{2019MNRAS.484.1031P}. It has been speculated that this NIR excess could be connected to the observed sub-mm emission \cite{2019MNRAS.484.1031P, 2019ApJ...872...18M, 2019ApJ...871...73H} although the variability is not correlated in an obvious way. It is not connected to an extrapolation of the radio observations at $\nu<100\,$GHz, nor could it be an extrapolation of the X-ray component to the NIR band \cite{2019ApJ...872...18M}. It has also been proposed to come from dust formation, as in interacting supernovae \cite{2019MNRAS.488.3772F}. 

A NIR excess was also observed in AT2024puz (31\,d and 56\,d \cite{2025ApJ...995..228S}) and AT2024wpp  (appearing around 20\,d \cite{2025MNRAS.537.3298P, 2026ApJ...997L..10L}). The power-law slope measured from the NIR photometry was $F_\nu \propto \nu^{-0.3}$, while fitting the overall spectrum with a blackbody + power law model gave $F_\nu \propto \nu^{-3}$ for the power law component. The ratio of the soft X-ray luminosity to the luminosity of the NIR component (integrated over the power law) is $\approx0.5$, similar to AT2018cow. In addition, as in AT2018cow extrapolating the radio emission underpredicts the NIR emission \cite{2026ApJ...997L..10L}. 

\subsection{A late-time UV, and possibly X-ray, plateau}

\begin{figure}[tb]
    \centering
    \vspace{-5mm}
    \includegraphics[width=0.46\linewidth]{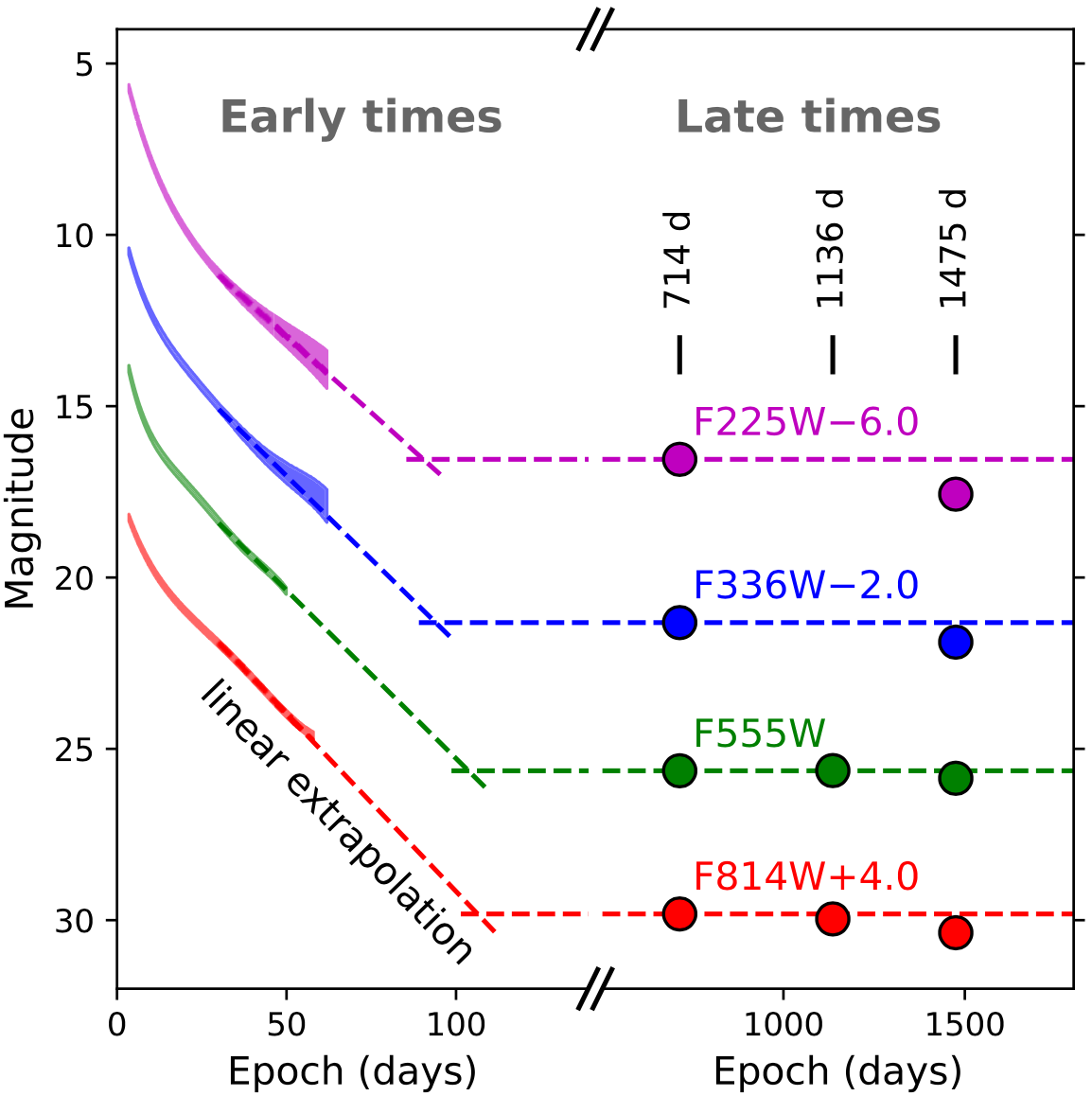}
    \includegraphics[width=0.52\linewidth]{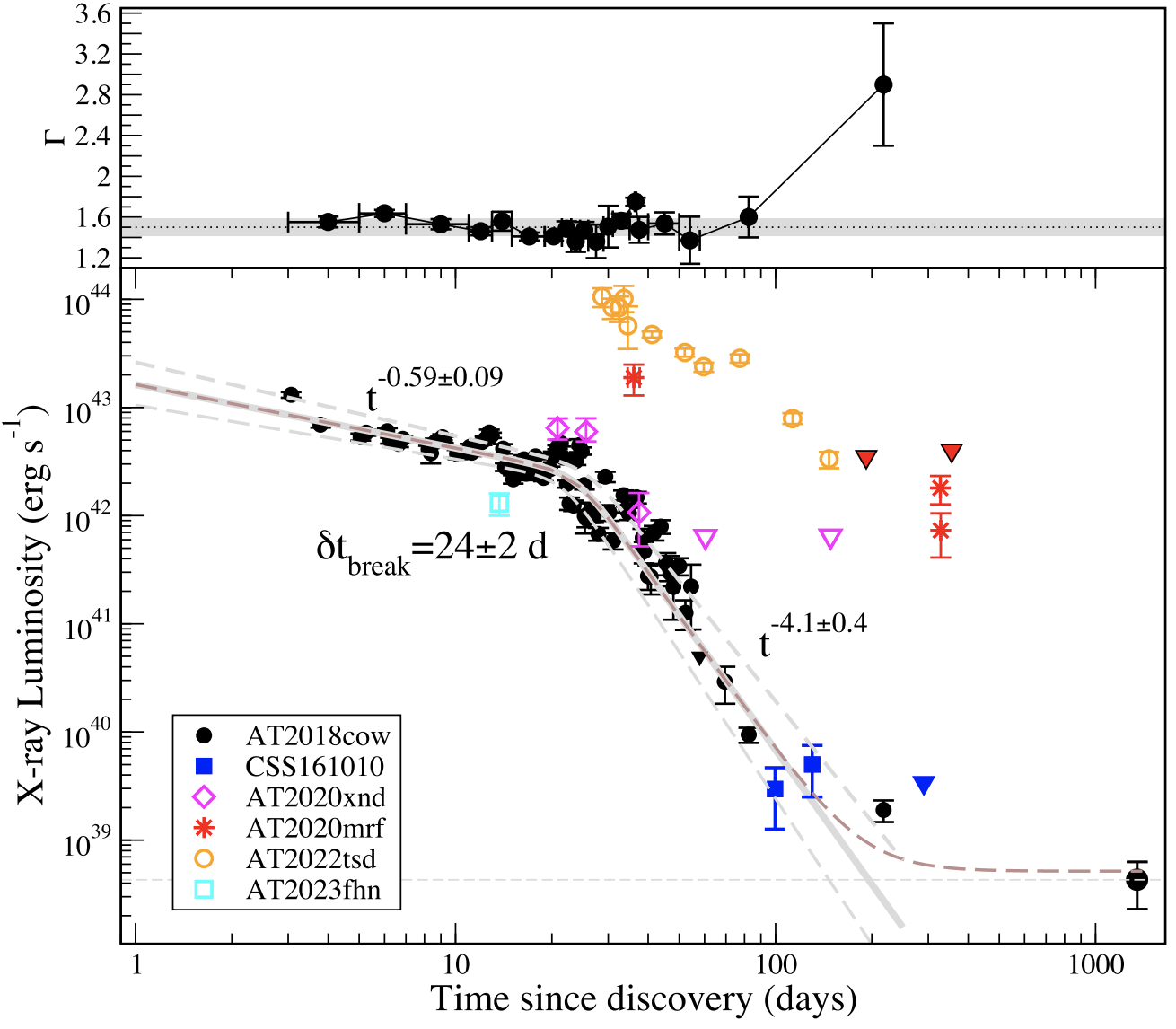}
    \vspace{-0.1in}
    \caption{\emph{Left:} UV-optical light curve of AT2018cow, reproduced from Sun et al. (2023) \cite{2023MNRAS.519.3785S} with permission. The initially rapidly fading light curve plateaued. \emph{Right:} The X-ray light curve and photon index $\Gamma$ of AT2018cow, reproduced from Migliori et al. (2024) \cite{2024ApJ...963L..24M} with permission. The source was robustly detected at 200\,d with a much softer spectral index. The late-time detection after 1000\,d has some unknown contribution from the underlying host galaxy so should be regarded as an upper limit.}
    \label{fig:plateau}
\end{figure}

HST observations of AT2018cow at 714\,d and 1136\,d detected a surprisingly bright blue unresolved ($<20\,$pc) source at the position of the transient, significantly exceeding predictions from extrapolating the earlier-time light curve (Figure~\ref{fig:plateau}, left). The SED was consistent with being on the RJ tail of a blackbody ($L_\nu \propto \nu^2$), constraining $T\gtrsim 10^{4}$--$10^{5}$\,K, and $\log(L/L_\odot) \approx 7$, with $R\sim$ tens of $R_\odot$. The source was significantly bluer than AT2018cow was at early times \cite{2022MNRAS.512L..66S, 2023ApJ...955...43C, 2023MNRAS.525.4042I}. One analysis identified prominent H$\alpha$ from the late-time source \cite{2022MNRAS.512L..66S}, but an independent analysis of the same data found no significant H$\alpha$ detection at the transient site \cite{2023ApJ...955...43C}. A third epoch at 1475\,d revealed that the emission had faded significantly in the UV filters, and that the SED had shifted downward, implying a lower temperature and lower luminosity \cite{2023MNRAS.519.3785S}. Some analyses found that the source faded only in the UV filters, and remained stable in the optical filters \cite{2023MNRAS.519.3785S, 2023ApJ...955...43C}, while another found that there was also significant fading in the optical band \cite{2023MNRAS.525.4042I}. In a final epoch at $\approx2000\,$d, the fading was found to be only marginal, with significant fading only in one UV filter \cite{2025MNRAS.544L.108I}. In general, measurements of the UV and optical photometry have significant discrepancies, resulting in different inferred blackbody parameters (but no significantly different conclusions)---this is likely in large part due to approaches to the photometry, particularly in subtracting the complex diffuse background \cite{2023MNRAS.525.4042I}. All analyses concluded that the source was likely associated with the transient, because of the fading in brightness over time, and the fact that the colors were significantly bluer than the colors of compact star-forming regions in the host galaxy \cite{2023MNRAS.519.3785S, 2023ApJ...955...43C, 2023MNRAS.525.4042I}. 

In addition to a UV plateau, late-time X-ray observations ($\approx200\,$d) showed a flattening out (Figure~\ref{fig:plateau}, right) and significant softening in the X-ray spectrum. An additional detection at $>1000\,$d has an unknown contribution from the underlying host galaxy environment, so it cannot be definitely determined whether the emission at that late stage arises from the transient \cite{2024ApJ...963L..24M}. 


\subsection{Limits on prompt gamma-ray emission}


No associated prompt GRBs have been detected. For AT2018cow, the interplanetary network rules out at 97\% confidence GRBs with peak luminosity $>10^{47}\,$erg\,s$^{-1}$ \cite{2019ApJ...872...18M}.  

\subsection{Volumetric rate}

From the ZTF (optical) flux-limited survey, the volumetric rate of events with similarly fast and luminous optical light curves is 0.9--12.5\,Gpc$^{-3}$\,yr$^{-1}$ (95\% confidence), corresponding to 0.001--0.01\% of the core-collapse supernova rate \cite{2026MNRAS.549ag678P}. The very low rate is consistent with searches in VLA Sky Survey (radio) data, which found that the rate of transients with radio light curves similar to AT2018cow is $\lesssim 0.34\%$ of the local core-collapse supernova rate \cite{2025PASP..137h4102S}. 

\section{Progenitor Models}
\label{sec:theories}

Many progenitor models have been proposed for LFBOTs. In this section, we summarize the models proposed in the literature and critically examine their strengths and weaknesses. We start by considering the basic constraints from individual emission components (optical, X-rays, radio, NIR, and late-time UV plateau). Then, we discuss how these emission components may be realized in global progenitor models.

\subsection{Basic considerations: key ingredients}\label{sec:basic_constraints}

Before reviewing the global models, we consider the empirical constraints based on the energy budget, mass budget, timescales, composition, and spectra. Models that do not immediately satisfy some of these constraints would need additional work to justify their relevance. We defer the discussion of the host environment to \S \ref{sec:proposed_models}. A summary of the principal physical ingredients can be found in Table \ref{tab:physical_requirements} at the end of this section.

\subsubsection{A powerful central engine with peak duration of hours to days}\label{eq:engine_power_constraints}
\leavevmode\\

\noindent The peak bolometric luminosity $L_{\rm pk}\gtrsim \mbox{a few}\times 10^{44}\rm\, erg\,s^{-1}$ and rapid evolution $t_{\rm 1/2}\sim \,$a few days  suggest a central engine that is capable of delivering $E_{\rm eng} \gtrsim E_{\rm rad}\gtrsim 10^{50}\rm\, erg$ on the timescale of a few days or shorter. In the case of AT2024wpp \cite{2026ApJ...997L..10L, 2026MNRAS.549ag678P}, the total \textit{radiated energy} is $E_{\rm rad}\simeq 1\times 10^{51}\rm\, erg$ (the typical ejecta energy of core-collapse supernovae). This favors a powerful central engine, because a standard $10^{51}\rm\, erg$ SN ejecta interacting with a CSM would require fine-tuning to match the radiated energy, peak luminosity, and spectra of LFBOTs.

Another argument for the existence of a central engine comes from the fast ejecta expansion speed ($\gtrsim 0.1c$) inferred from either the expanding optical photosphere or the radio emission. Such expansion speeds suggest that the engine is powered by either a neutron star (NS) or a black hole (BH). Indeed, all progenitor models proposed in the literature share this property.

Two different power sources have been considered: rotational energy of a strongly magnetized NS and gravitational energy from accretion onto either NS or BH. We note that these two cases operate rather similarly in terms of their time-dependent engine power evolution, which can be roughly described by a broken power-law
\begin{equation}
    L_{\rm eng}(t) \simeq {E_{\rm eng}\over t_0} (1 + t/t_0)^{-\beta},
\end{equation}
where $t_0$ is the \textit{initial} energy release timescale and $\beta$ is the temporal decay index. For the NS rotationally powered case, we have $\beta=2$ for standard magnetic dipole spindown power. For the disk accretion-powered case, $4/3\lesssim \beta \lesssim 8/3$ for a self-similar temporal evolution, depending on the angular momentum loss carried by the disk wind (e.g., \cite{2008MNRAS.390..781M}), and numerical simulations of long-term disk evolution support $\beta\approx 2$ \cite{2013MNRAS.435..502F}. We generally require $t_0 \lesssim t_{\rm 1/2}$, meaning that the initial engine timescale is shorter than or comparable to $t_{\rm 1/2}\sim\,$a few days. 

We disfavor the case of $t_0\ll t_{\rm 1/2}$ based on the considration of radiative efficiency, because this would mean that most of the engine's energy is injected at $t\sim t_0$ before radiation can escape from the system. In such a situation, most of the energy $E_{\rm eng}$ will be in the form of kinetic energy of the ejecta, whereas only the energy injection on the timescale of $t_{\rm 1/2}$ may be efficiently radiated from the system, so we require
\begin{equation}
    t_{1/2} L_{\rm eng}(t_{1/2}) \sim E_{\rm eng} (t_0/t_{\rm 1/2})^{\beta-1} \sim \eta_{\rm rad}^{-1} E_{\rm rad},
\end{equation}
where $\eta_{\rm rad}<1$ is the radiative efficiency. Thus, we generally require
\begin{equation}
    E_{\rm eng} \sim (t_0/t_{1/2})^{1-\beta} \eta_{\rm rad}^{-1} E_{\rm rad} \gtrsim (t_0/t_{1/2})^{1-\beta} E_{\rm rad}.
\end{equation}
For $\beta\approx 2$, we find $E_{\rm eng}\gtrsim (t_0/t_{1/2})^{-1} E_{\rm rad}$. This rules out very short-lived central engines with $t_0\lesssim 1\rm\, hr$ (such as those powering GRB jets) because that would require $E_{\rm eng}\gtrsim 10^2 E_{\rm rad}$ and leads to extremely large ejecta kinetic energy not seen in LFBOTs. We conclude that the central engines of LFBOTs most likely have
\begin{equation}
    \mbox{hours}\lesssim t_0\lesssim \mbox{days}, \ \ E_{\rm eng}\gtrsim 10^{50}\rm\, erg.
\end{equation}
Purely from the radiative efficiency point of view, the most ``energetically efficient'' engines would be those with $t_0\sim t_{\rm 1/2}$.

In the following, we discuss the constraints based on the energy budget and evolutionary timescale of the engine in NS rotation- and accretion disk-powered cases.

In the case of rotational energy $E_{\rm rot}$ from a strongly magnetized NS, the energy budget constrains the NS spin period
\begin{equation}\label{eq:Pns}
    P_{\rm ns} = \lrb{2\pi^2 I_{\rm ns}\over E_{\rm rot}}^{1/2} = 14\mr{\,ms}\, \lrb{I_{\rm ns}\over 10^{45}\mr{\,g\,cm^2}}^{1/2} \lrb{E_{\rm rot}\over 10^{50}\rm\, erg}^{-1/2},
\end{equation}
where $I_{\rm ns}$ is the moment of inertia of a NS. Moreover, the rotational energy is released on a spindown timescale due to magnetic dipole emission \cite{2006ApJ...648L..51S}
\begin{equation}
\begin{split}
    t_{\rm sd} & = {E_{\rm rot}\over L_{\rm sd}} = 
    {I_{\rm ns}^2 c^3\over 4(1+\sin^2\chi) \mu_{\rm B}^2 E_{\rm rot}}\\
    &=0.52\mr{\,d}\, \lrb{1+\sin^2\chi\over 1.5}^{-1} \lrb{I_{\rm ns}\over 10^{45}\mr{\,g\,cm^2}}^2 \lrb{\mu_{\rm B}\over 10^{33}\mr{\,G\,cm^3}}^{-2} \lrb{E_{\rm rot}\over 10^{50}\rm\, erg}^{-1},
\end{split}
\end{equation}
where $\mu_{\rm B}=B_{\rm ns} R_{\rm ns}^3$ is the magnetic dipole moment, $B_{\rm ns}$ is the surface dipolar B-field strength at the magnetic equator (half of that at the magnetic poles), $R_{\rm ns}$ is the neutron star radius, and $\chi$ is the inclination angle between the magnetic and spin axes (hereafter taken as $\sin^2\chi\simeq 0.5$ for simplicity). By considering a spin-down timescale of a few days, we can then obtain a constraint on the magnetic moment
\begin{equation}
    \mu_{\rm B} = 4.2\times10^{32}\mr{\,G\,cm^3} \lrb{t_{\rm sd}\over 3\mr{\,d}}^{-1/2} {I_{\rm ns}\over 10^{45}\mr{\,g\,cm^2}} \lrb{E_{\rm rot}\over 10^{50}\rm\, erg}^{-1/2}.
\end{equation}
For $R_{\rm ns}\simeq 10\rm\, km$, $I_{\rm ns}\simeq 10^{45}\rm\, g\,cm^2$, and $E_{\rm rot}\sim E_{\rm eng}$, we find the surface dipolar B-field strength
\begin{equation}
    B_{\rm ns} \simeq 4\times10^{14}\mr{\,G} \lrb{t_{\rm 1/2}\over 3\mr{\,d}}^{-1/2} \lrb{E_{\rm eng}\over 10^{50}\mr{\,erg}}^{-1/2}.
\end{equation}
The constraints on $P_{\rm ns}$ and $B_{\rm ns}$ are broadly consistent with those obtained by Prentice et al. (2018) \cite{2018ApJ...865L...3P}, Margutti et al. (2019) \cite{2019ApJ...872...18M}, and L. Li et al. (2024) \cite{2024ApJ...963L..13L}. We see that a rapidly spinning magnetar is required in this case. Since our constraints on the engine energy $E_{\rm eng}$ are based on the radiative output on the timescale of $t_{\rm 1/2}\sim\,$a few days, the consideration is that the spindown time $t_{\rm sd}$ near the peak of the optical light curve is comparable to $t_{\rm 1/2}$. The \textit{initial} spindown time $t_0$ may be shorter than $t_{1/2}$, so our $P_{\rm ns}$ in eq. (\ref{eq:Pns}) is strictly an upper limit for the initial spin period, but the constraint on $\mu_{\rm B}$ remains the same because the initial rotational energy is correspondingly higher.

On the other hand, the engine power may be supplied by accretion onto a NS or BH. In this case, it is possible to constrain the mass and angular momentum of the accretion disk based on $E_{\rm eng}$ and $t_{1/2}$. For an accretion efficiency $\eta_{\rm acc}\sim 0.1$ near the NS surface or the BH's innermost stable circular orbit (ISCO), one can estimate the accreted mass onto the compact object
\begin{equation}
    M_{\rm acc} = {E_{\rm eng}\over \eta_{\rm acc}c^2} = 5.6\times10^{-4}\mr{\,M_\odot}\, {E_{\rm eng}\over 10^{50}\rm\, erg}.
\end{equation}
This is an estimate of the mass gained by the compact object. However, since the accretion power needed for LFBOTs is highly super-Eddington (by a factor of $10^5$ or more for a stellar-mass compact object) and the gas in the accretion flow does not cool efficiently via neutrino emission (which requires an accretion rate of the order $10^{-3}\,M_\odot\rm \,s^{-1}$ or higher), it is likely that most of the mass in the outer regions of the accretion disk is lost by disk wind as is the case in Advection-Dominated Accretion Flow (ADAF, \cite{1995ApJ...444..231N, 1999MNRAS.310.1002S, 1999MNRAS.303L...1B, 2014ARA&A..52..529Y}). Recent simulations \cite{2024ApJ...977..200C, 2024ApJ...973..141G} show that, as a result of disk wind, the accretion rate in ADAF decreases as the gas approaches the inner edge of the accretion disk roughly as a power-law
\begin{equation}\label{eq:ADAF_Mdot_r}
    \dot{M} \propto r^{p}, \ \ p\simeq 0.5.
\end{equation}
In the case where the mass inflow is supplied at the disk outer radius $r_{\rm d}$ (typically $\gtrsim R_\odot$) which is many orders of magnitude larger than the inner boundary of the disk $r_{\rm in}$ (near the innermost stable circular orbit, ISCO), such a power-law solution is the natural outcome of the self-similarity of the system \cite{1999MNRAS.303L...1B}, as there is no physical scale in between $r_{\rm in}$ and $r_{\rm d}$. In such a self-similar solution, the total disk mass is much higher than $M_{\rm acc}$, i.e.,
\begin{equation}\label{eq:total_disk_mass_from_energy}
    M_{\rm d} \simeq M_{\rm acc} (r_{\rm d}/r_{\rm in})^p \sim 0.1M_\odot \lrb{r_{\rm d}\over R_\odot}^{1/2} \lrb{r_{\rm in}\over 10\mr{\,km}}^{-1/2} {E_{\rm eng}\over 10^{50}\mr{\,erg}},
\end{equation}
where we have adopted a fiducial $p=0.5$. In this picture, despite the fact that most of the accreted mass is lost from the outer regions of the disk (as $\dot{M}\propto r^{0.5}$), most of the accretion power is generated near the inner edge of the disk (as $GM\dot{M}/r\propto r^{-0.5}$).

It should be noted that the result of $p\simeq 0.5$ from numerical simulations of ADAFs \cite{2024ApJ...977..200C, 2024ApJ...973..141G} has not been observationally tested in the super-Eddington regime, so the power-law index $p$ could in principle be anywhere between 0 (no wind) and 1 (the heaviest possible mass loss allowed by accretion energy). Turning the argument around, if the central engines in LFBOTs are indeed powered by accretion (which is tentatively supported by the observations of late-time UV plateau, see below), one can place stringent constraints on the power-law index $p$ --- LFBOTs then become a natural laboratory to study fundamental accretion physics.

On the other hand, the rapid evolution of LFBOTs places an interesting constraint on the radius of the outer disk $r_{\rm d}$ based on the timescale for viscous accretion
\begin{equation}
    t_{\rm vis} \simeq \alpha^{-1} \lrb{H\over r_{\rm d}}^{-2} \sqrt{r_{\rm d}^3\over GM},
\end{equation}
where $\alpha$ is the Shakura-Sunyaev viscosity parameter \cite{1973A&A....24..337S}, $H$ is the vertical pressure scale-height of the outer disk, and $M$ is the mass of the accreting compact object. By requiring $t_{\rm vis}\simeq t_{1/2}$ and $H/r_{\rm d}\simeq 0.3$ (a weakly bound, geometrically thick disk), it is possible to place a constraint on the outer disk radius near the peak time,
\begin{equation}\label{eq:disk_radius_from_viscous_time}
    r_{\rm d} \simeq 2.8R_\odot\, \lrb{t_{1/2}\over 3\mr{\,d}}^{2/3} \lrb{M\over 10M_\odot}^{1/3} \lrb{\alpha\over 0.1}^{2/3} \lrb{H/r_{\rm d}\over 0.3}^{4/3}.
\end{equation}
Note that the requirement of $t_{\rm vis}\simeq t_{1/2}$ is based on the assumption that the emission near the peak time is directly powered by the \textit{current} accretion. It is possible that the conversion between accretion power to the UV/optical emission has some delay (due to e.g., dynamical expansion or photon diffusion), and in that case the above constraint on the disk radius is an upper limit. For this reason, and because of possible viscous expansion of the disk before $t_{\rm 1/2}$, the \textit{initial} disk radius may be smaller than the above constraint. One can use the constraint on the initial viscous time $t_{\rm vis,0} \gtrsim 1\rm\, hr$ (as argued above) to place a lower limit on the initial disk radius
\begin{equation}
    r_{\rm d,0}\gtrsim 10^{10}\mr{\,cm}\, \lrb{M\over 10M_\odot}^{1/3} \lrb{\alpha\over 0.1}^{2/3} \lrb{H/r_{\rm d}\over 0.3}^{4/3}.
\end{equation}
This shows that an \textit{initial} disk radius much smaller than $R_\odot$ (as in the case of failed supernovae) is in principle allowed. In the late-time UV plateau phase (a few years after the optical peak), due to further viscous expansion, the disk outer radius is likely much larger than the above constraint.

Motivated by the models proposed by Kremer et al. (2019, 2021) \cite{2019ApJ...881...75K, 2021ApJ...911..104K}, Metzger (2022) \cite{2022ApJ...932...84M}, Tsuna \& Lu (2025) \cite{2025ApJ...986...84T}, and Klencki \& Metzger (2025) \cite{2026ApJ..1005....2K} which are based on a merger between a star and a compact object or a tidal disruption event (TDE) of a star by a compact object, it is useful to compare the above disk radius constraint to the tidal disruption radius of a star of mass $M_*$ and radius $R_*$,
\begin{equation}
    r_{\rm t} \simeq R_*(M/M_*)^{1/3} = 2.1 R_\odot {R_*\over R_\odot} \lrb{M_*\over M_\odot}^{-1/3} \lrb{M\over 10M_\odot}^{1/3}. \ \mbox{ (for TDEs)}
\end{equation}
Note that the above tidal disruption radius is only valid in the limit $M\gg M_*$, which is not the case for a NS. However, for stellar-mass compact objects and for typical stellar mass $M_* \lesssim 10M$, we expect the disk radius to be of the order the stellar radius anyway (as given by the orbital angular momentum of the  encounter \cite{2023MNRAS.524.6358K}), so the above expression approximately holds.

The comparison between $r_{\rm t}$ and the above constraint on $r_{\rm d}$ shows that the central engines of LFBOTs are consistent with a TDE of main-sequence or Wolf-Rayet stars ($R_*$ not much larger than a few $R_\odot$), but the compact object mass is not constrained by this argument alone as both radii are proportional to $M^{1/3}$. However, as we show below (\S \ref{sec:IMBH_TDE}), the physics of gas circularization/disk formation limits the compact object to be stellar mass ($M\lesssim 100M_\odot$) --- TDEs of main-sequence stars by intermediate-mass or supermassive BHs ($M\gtrsim 100M_\odot$) are disfavored because the stellar debris may not efficiently circularize to form an accretion disk on a timescale of a few days. Another constraint is that the late-time disk accretion signatures from TDEs by such high-mass BHs are inconsistent with X-ray observations \cite{2024ApJ...963L..24M}.

If the emission near the peak time is directly powered by the \textit{current} accretion (assuming nearly instantaneous conversion between accretion power to the UV/optical emission), then one can plug the disk radius $r_{\rm d}$ (eq. \ref{eq:disk_radius_from_viscous_time}) into the power-law (eq. \ref{eq:total_disk_mass_from_energy}) and obtain
\begin{equation}
    M_{\rm d} \simeq 0.25\,M_\odot\, \lrb{t_{1/2}\over 3\mr{\,d}}^{1/3} \lrb{M\over 10M_\odot}^{1/6} \lrb{\alpha\over 0.1}^{1/3} \lrb{H/r_{\rm d}\over 0.3}^{2/3}\lrb{r_{\rm in}\over 10\mr{\,km}}^{-1/2} {E_{\rm eng}\over 10^{50}\mr{\,erg}}.
\end{equation}
This again shows that a stellar-mass disk is plausible for LFBOTs.

One caveat in the accretion-powered picture is that, even in the ADAF regime (i.e., inefficient radiative cooling), the structure of the accretion flow onto a NS is likely quite different from that for a BH, because of the different inner boundary conditions. For an accreting BH, the boundary condition is set by the general relativistic spacetime without ambiguity (for a given BH mass and spin). However, a NS has a ``hard surface'', and gas must lose its internal energy before settling onto the NS surface. The energy loss is primarily due to neutrino emission \cite{1993ApJ...411L..33C}. Inefficient cooling leads to the formation of an extended envelope (radius $R_{\rm env}\gg R_{\rm ns}$) surrounding the NS \cite{2025ApJ...987...71C}, which may reduce the accretion efficiency $\eta_{\rm acc}$ by a factor of $R_{\rm ns}/R_{\rm env}$. However, NS accretion depends strongly on the (highly uncertain) magnetospheric field strength and magnetosphere-disk interaction physics (e.g., \cite{1976MNRAS.175..395B, 1979ApJ...232..259G, 2004ApJ...616L.151R, 2015MNRAS.447.1847M, 2016ApJ...822...33P, 2017ApJ...851L..34P}). Future work should address the question of whether a NS accretor is physically plausible for LFBOTs (they are energetically allowed in the current analysis).

\subsubsection{UVOIR emission: low radiating mass}\label{sec:low_ejecta_mass}
\leavevmode\\

\noindent
The photon diffusion time for a homologously expanding ejecta is given by
\begin{equation}\label{eq:diffusion_time_homologous}
\begin{split}
    t_{\rm diff}^{(1)} \simeq 0.2\, \lrb{\kappa M_{\rm ej}^{3/2} \over E_{\rm ej}^{1/2} c}^{1/2}
    \simeq 10\mr{\,d} \lrb{\kappa\over 0.1\mr{\,cm^2/g}}^{1/2} \lrb{M_{\rm ej}\over M_\odot}^{3/4} E_{\rm ej,51}^{-1/4},
\end{split}
\end{equation}
where the prefactor of $0.2$ weakly depends on the density profile of the ejecta \cite{1982ApJ...253..785A}, $\kappa$ is the (Rosseland-mean) opacity of the ejecta for optical photons, and our fiducial value is equal to the electron scattering opacity of singly ionized helium (appropriate for temperature $T\sim 3\times10^4\rm\,K$ near peak light). We see that, if the ejecta expands homologously, the fast evolution of LFBOTs ($t_{1/2}\sim\,$a few days) requires a low ejecta mass of the order $M_{\rm ej}\sim 1M_\odot$ or less (unless the ejecta opacity is extremely low $\ll 0.1\rm\,cm^2\,g^{-1}$, see \cite{2015MNRAS.450.1295W}).

Another possibility is that the system is not evolving homologously. For instance, the ejecta (or pre-existing CSM) may have already expanded to a large radius by the time the central engine turns on. If the outer edge of the ejecta (or CSM) is located at $R_{\rm ej}$, then the \textit{current} photon diffusion time is given by 
\begin{equation}\label{eq:diffusion_time_one_snapshot}
    t_{\rm diff}^{(2)} \simeq {\kappa M_{\rm ej}\over 4\pi c R_{\rm ej}} \simeq 6.1\mr{\,d}\, {\kappa \over 0.1\mr{\,cm^2/g}} {M_{\rm ej}\over M_\odot} \lrb{R_{\rm ej}\over 10^{15}\mr{\,cm}}^{-1}.
\end{equation}
The diffusion time $t_{\rm diff}^{(2)}$ applies if the ejecta (or CSM) is optically thick and the radiation is produced in its deep interior. If we adopt $R_{\rm ej}$ to be comparable to the photospheric radii $R_{\rm ph}\sim 10^{15}\rm\, cm$ inferred from the luminosity and color temperature near the peak, we again conclude that the ejecta mass is likely $M_{\rm ej}\lesssim 0.5M_\odot$, as has been pointed out by Margutti et al. (2019) \cite{2019ApJ...872...18M}.

A caveat of the above analysis is that, by ejecta mass, we mean the mass of the gas that participates in producing the observed optical emission and the radiative transfer. It is possible that the system has some ``dark mass'' (e.g., cold, non-emitting gas) that is unconstrained by current observations.

\subsubsection{X-ray emission: jet internal dissipation or shock-powered?}\label{sec:X-ray_emission}
\leavevmode\\

\noindent
A minimum model for the X-ray emission needs to explain (1) significant variability on timescales of a few days (Figure~\ref{fig:xray-variability}); (2) the soft X-ray continuum shape (well described by a power-law $L_\nu\propto \nu^{-\beta}$ with $\beta\simeq 0.6$), (3) the early-time (first $\sim\!10$ days in AT2018cow) hard X-ray ``hump'' peaking near $40\rm\, keV$ together with a blue-shifted Fe K$\alpha$ line peaking near $8\rm\,keV$ (Figure~\ref{fig:18cow-xray}).

Besides the three main properties above, another intriguing behavior seen in AT2018cow and AT2024wpp is the coordinated evolution of the X-ray and UVOIR emission (e.g., \cite{2019ApJ...872...18M, 2026ApJ...997L..10L}). The ratio $L_{\rm X}/L_{\rm UVOIR}$ quasi-monotonically increased over time, from $\sim10^{-2}$ in the first few days to $\sim 1$ in the first few weeks, and then ``saturated'' around unity subsequently --- meaning that the X-ray and UVOIR emissions reached ``equipartition''. This coordinated behavior may suggest that UVOIR is likely the reprocessed emission from an \textit{embedded} X-ray source whose intrinsic luminosity was much higher than the observed $L_{\rm X}$ in the first few weeks \cite{2025ApJ...993L...6N}. Another notable behavior is that the ``equipartition time'' $t_{\rm eq}$ (when $L_{\rm X}/L_{\rm UVOIR}$ first reached unity) nearly coincided with the time when the intermediate-width emission lines (e.g., H$\alpha$) first appear, despite the fact that $t_{\rm eq}$ is earlier for AT2018cow ($\sim$15--20 d) than AT2024wpp ($\sim$35 d). Lebaron et al. (2026) \cite{2026ApJ...997L..10L} argued that that the coincidental timing between $L_{\rm X}/L_{\rm UVOIR}$ and the appearance of intermediate-width emission lines is due to the optical depth of the reprocessing layer dropping below unity (either due to ejecta expansion or X-ray ionization breakout), which caused the embedded X-ray source and the line emission region to be fully exposed. In this picture, the H-rich gas responsible for the intermediate-width emission lines is embedded within the UVOIR emitting gas (see \S \ref{sec:Balmer_lines}). For future events with as abundant data as in AT2018cow and AT2024wpp, it remains to be tested if such temporal evolution of $L_{\rm X}/L_{\rm UVOIR}$ and the emission line appearance timing are just coincidental or universal among LFBOTs.

In the following, we focus on the implications of the three main properties: variability, soft X-ray spectrum, and hard X-ray ``hump''.

If the variability timescale of $t_{\rm var}\sim{\rm few\ days}$ is associated with the local dynamical expansion time, $t_{\rm dy}=R/v$, then for $v\sim0.1c$ one obtains $R\sim vt_{\rm var}\sim10^{15}\ {\rm cm}$, comparable to the optical photospheric radius. This exercise, again, suggests that the X-ray emission is produced at radii $R\lesssim10^{15}\ {\rm cm}$, near or below the optical photosphere $R_{\rm ph}$. This is the strongest argument against the classical CSM-interaction picture in which the observed X-rays are produced exterior to the optical photosphere.

Most models place the X-ray source within or beneath the optically thick material responsible for the UVOIR emission near peak luminosity. We refer to this optically active material as the ``ejecta'', with mass $M_{\rm ej}$, following the operational definition in Section~\ref{sec:low_ejecta_mass}. Depending on the progenitor model,
it may consist of explosion ejecta, compact pre-existing CSM, engine-driven outflows, or a mixture of them.

In an aspherical system, the column density of this material may vary strongly with direction. The observed X-rays escape through a relatively low-column density, highly ionized region in which the opacity is dominated by electron scattering. We refer to this region as the \textit{scattering layer}, and denote its characteristic radius,
scattering optical depth, and \textit{isotropic-equivalent} mass by $R_{\rm sca}$, $\tau_{\rm s}$, and $M_{\rm sca}$, respectively. Denser ejecta at other solid angles may absorb a larger fraction of the high-energy radiation and reprocess it into UVOIR emission.

In the following, we first discuss possible emission mechanisms for the soft X-rays and then move on to the origin of the hard X-ray hump and constraints on the properties of the scattering layer.

The origin of the power-law spectrum ($L_\nu\propto \nu^{-\beta}$, $\beta\simeq 0.6$) in the soft X-ray band remains debated. Models in the literature fall into two categories: non-thermal emission (synchrotron or inverse-Compton) and thermal emission (free-free + Comptonization).

Synchrotron or inverse-Compton emission from a non-thermal (power-law) population of electrons requires a Lorentz factor distribution with $\d N/\d \gamma\propto \gamma^{-p}$ with $p\simeq 2.2$. This has been considered in the context of AT2018cow by L. Li et al. \cite{2024ApJ...963L..13L} who proposed that the X-rays may be produced by synchrotron emission from non-thermal electrons accelerated at the wind termination shock powered by a spinning down magnetar. Such an electron power-law index is seemingly consistent with the expectation from diffusive shock acceleration (e.g., \cite{1983RPPh...46..973D, 1987PhR...154....1B}). However, one can show that the emitting electrons in this picture must be undergoing fast cooling as a result of very high radiation and magnetic energy densities in their environment. The radiation energy density near radius $r\sim R_{\rm ph}\sim 10^{15}\rm\, cm$ can be estimated by
$$U_{\rm rad} \simeq \tau L/(4\pi r^2 c) \sim 2.6\times10^2\mr{\,erg\,cm^{-3}} \tau L_{44} r_{15}^{-2},$$
where $\tau\gtrsim 1$ is the Rosseland-mean optical depth of the emitting region. Under the conservative limit of $\tau\sim 1$ and defining the ratio between the magnetic and radiation energy densities in the emitting region $\xi = U_{\rm B}/U_{\rm rad}$, one can infer the magnetic field strength as $B\simeq 80\mr{\,G}\, \xi^{1/2} L_{44}^{1/2} r_{15}^{-1}$. The electron Lorentz factor corresponding to synchrotron emission in the X-ray band ($\sim10^{17}\rm\, Hz$) can then be inferred to be $\gamma \simeq 1.7\times10^4 \nu_{17}^{1/2} \xi^{-1/4} L_{44}^{-1/4} r_{15}^{1/2}$. Thus, we can estimate the synchrotron plus inverse-Compton cooling timescale as
\begin{equation}
    t_{\rm cool} \simeq {3\me c\over 4\gamma \sigma_{\rm T} (U_{\rm rad} + U_{\rm B})} \simeq 7\mr{\,s}\, {\xi^{1/4} \over 1+\xi} {r_{15}^{5/2}\over \nu_{17}^{1/2}L_{44}^{3/4}},
\end{equation}
where we have ignored Klein-Nishina suppression of the inverse-Compton cooling (but it is unlikely to change our conclusion). We conclude that, for physically plausible parameters, $t_{\rm cool}$ is much shorter than the dynamical timescale of the system near peak luminosity (few days). Thus, fast cooling requires that the electrons are accelerated with a much shallower (harder) Lorentz factor distribution with power-law index $p-1\simeq 1.2$, which is inconsistent with diffusive shock acceleration (typically 2--2.3).

We conclude that a simple fast-cooling synchrotron or inverse-Compton interpretation of the X-ray emission faces major difficulties, as has been pointed out by Margutti et al. (2019) \cite{2019ApJ...872...18M} and Ho et al. (2019) \cite{2019ApJ...871...73H}, unless the emitting band lies in a different spectral segment or the acceleration/cooling physics is non-standard. One remaining possibility is that the X-ray emission may come from internal dissipation of a relativistic jet launched from e.g., a spinning BH, where electrons may be \textit{continuously} accelerated while they undergo rapid cooling. The non-thermal X-ray emission from the jet may interact with the optically thick gas surrounding or ahead of the jet, and it may be possible to produce the ``Compton hump'' feature in the hard X-rays \cite{2014MNRAS.437..703M}. However, The dynamics of such a relativistic jet interacting with the surrounding gas as well as the detailed particle acceleration and emission mechanisms are still highly uncertain (e.g., \cite{2015PhR...561....1K}). 

In the following, we discuss the other class of models based on free-free emission and Comptonization by shock-heated thermal electrons.
Standard optically thin free-free emission from a single-temperature plasma produces a spectrum
\begin{equation}
    j_\nu\propto n^2 T^{-1/2} g_{\rm ff}(\nu,T)\, \mr{e}^{-h\nu/\kB T}, \ \ g_{\rm ff}\approx \ln\lrsb{\mr{e} + \exp\lrb{1.68 - {\sqrt{3}\over \pi} \ln \lrb{\nu_{17}T_8^{-1}}}},
\end{equation}
where $n$ is the gas particle number density, $g_{\rm ff}$ is the Gaunt factor\footnote{This expression for the Gaunt factor $g_{\rm ff}$ follows the functional form provided by Draine (2011) \cite{2011piim.book.....D} and is valid for $\kB T\gg Z \mr{Ry}$, where $\mr{Ry}=13.6\rm\, eV$ is the Rydberg energy and $Z$ is the charge number of the dominant ion species.} for non-relativistic electron temperatures $T$. Such an optically thin free-free spectrum is either too shallow (at low frequencies $h\nu\ll \kB T$) or too steep (at high frequencies $h\nu \gg \kB T$) as compared to the observed soft X-ray spectrum, except for a fine-tuned electron temperature.

One attractive possibility is that the system consists of plasmas at a wide range of electron temperatures. This is the case for the CSM interaction in SN1987A where the interaction between the SN ejecta with the clumpy equatorial ring produces plasmas with a broad distribution of temperatures (e.g., \cite{2021ApJ...916...76A, 2025ApJ...981...26S}) --- lower (higher) temperatures when the shock propagates through denser (less dense) regions. If we ignore the Gaunt factor (as we are mainly focusing on the emission near $h\nu \sim \kB T$) and assume the system to be optically thin, the total specific luminosity from the system is given by
\begin{equation}
    L_\nu\propto \int T^{-1/2} \mr{e}^{-h\nu/\kB T} n^2 \d V \sim \int_{T_{\rm min}(\nu)}^\infty {n^2\d V\over \d T} T^{-1/2} \d T,
\end{equation}
where the minimum temperature is given by $T_{\rm min}(\nu)\sim h\nu/\kB$ (as the plasmas at $T\ll T_{\rm min}$ do not contribute significantly at the given photon energy $h\nu$). Let us consider a system with a power-law distribution of electron temperatures
\begin{equation}
    {n^2\d V\over \d T} = {n\d N\over \d T} \propto T^{-\delta} \ \ \Rightarrow \ \  L_\nu\propto T_{\rm min}^{1/2-\delta} \propto \nu^{1/2-\delta},
\end{equation}
where $\d N = n \d V$ is the differential gas particle number. We see that a spectral index of $\beta\simeq 0.6$ can be reproduced if $\delta = 1/2+\beta \simeq 1.1$. However, in this picture, it is not clear how the plasma temperature distribution of ${n\d N/\d T} \propto T^{-1.1}$ should be physically interpreted, as this distribution depends on the spatial density distribution of the pre-existing CSM as well as free-free (and inverse-Compton) cooling of the shock-heated gas.

Another physically motivated model for plasmas with a broad temperature distribution is provided by Govreen-Segal et al. \cite{2026arXiv260118887G} (see also \cite{2025ApJ...993...46W}) based on free-free emission from electrons undergoing rapid cooling. The key idea is that as a population of electrons cool to lower temperatures, they emit at lower frequencies. In a steady state, the total emission spectrum from electrons at different temperatures is identical to the cumulative emission spectrum of a plasma over its entire cooling history. As the plasma cools from temperature $T$ to $T-\Delta T$ (assuming free-free cooling dominates over e.g., inverse-Compton), the radiated energy per electron is $\Delta E = (3/2)\kB \Delta T$, which is distributed with an instantaneous free-free spectrum $j_\nu\propto g_{\rm ff}(\nu,T)\,\mr{e}^{-h\nu/\kB T}$ (including Gaunt factor), so we obtain
\begin{equation}
    {d (\Delta E)\over \d \nu} \approx {h\Delta E\over \kB T} g_{\rm ff}\,\mr{e}^{-h\nu/\kB T} = {3h\over 2I_{\rm norm}} {\Delta T\over T} g_{\rm ff}(\nu,T)\,\mr{e}^{-h\nu/\kB T},
\end{equation}
where $I_{\rm norm} = \int_0^\infty g_{\rm ff}(x)\,\mr{e}^{-x}\d x\approx 1.46$ is a normalization factor and $x \equiv h\nu/(\kB T)$. 
Over the entire cooling history starting from an initial temperature $T_0$ to a final temperature of $\sim 0$ (the exact final temperature is unimportant for X-ray emission), the cumulative emission spectrum near a given frequency $\nu$ is then given by
\begin{equation}\label{eq:ffcoolspec_numerical}
    {\d E\over \d \nu} = {3h\over 2I_{\rm norm}} \int_0^{T_0} T^{-1} g_{\rm ff}\, \mr{e}^{-h\nu/\kB T} \d T = {3h\over 2I_{\rm norm}}\int_{x_0}^{\infty} x^{-1} g_{\rm ff}(x)\, \mr{e}^{-x}\d x,
\end{equation}
where $x_0\equiv h\nu/(\kB T_0)$.

Let us consider the limit of $x_0\ll 1$ or $\kB T_0\gg h\nu$ (for a given frequency $\nu$). In this limit, it is easy to see that the contribution from $x\sim 1$ to $\infty$ is negligible and that most of the contribution is from the region of $x_0\ll x\ll 1$. Using the approximations of $g_{\rm ff}(x)\approx \sqrt{3}\pi^{-1} \ln (x^{-1})$ and $\mr{e}^{-x}\approx 1$ in the limit of $x\ll 1$, we find the following asymptotic result
\begin{equation}
    \lrb{\d E\over \d \nu}^{(1)} \approx {3\sqrt{3}h\over 2\pi I_{\rm norm}} \int_{x_0}^{\sim 1} (-\ln x) {\d x\over x} \approx {3\sqrt{3}h\over 4\pi I_{\rm norm}} \ln^2 (1/x_0),
\end{equation}
and the dimensionless spectrum is given by
\begin{equation}\label{eq:ffcoolspec_app1}
    {1\over \kB T_0} \lrb{\d E\over \d x_0}^{(1)} \approx {3\sqrt{3}\over 4\pi I_{\rm norm}} \ln^2 (1/x_0). \ \ (\mbox{for } x_0\ll 1)
\end{equation}
The spectral slope is then given by $$\beta(x_0) = -\d \ln L_\nu/\d \ln \nu \approx 2/\ln (1/x_0),$$ which is much steeper (softer) than the free-free spectrum from a single temperature plasma (e.g., $x_0=0.02$ corresponds to $\beta = 0.5$). Note that our treatment above is slightly different from that by Govreen-Segal et al. \cite{2026arXiv260118887G} who ignored the Gaunt factor, but the results are broadly similar. 

\begin{figure}[b]
    \centering
    \includegraphics[width=0.6\linewidth]{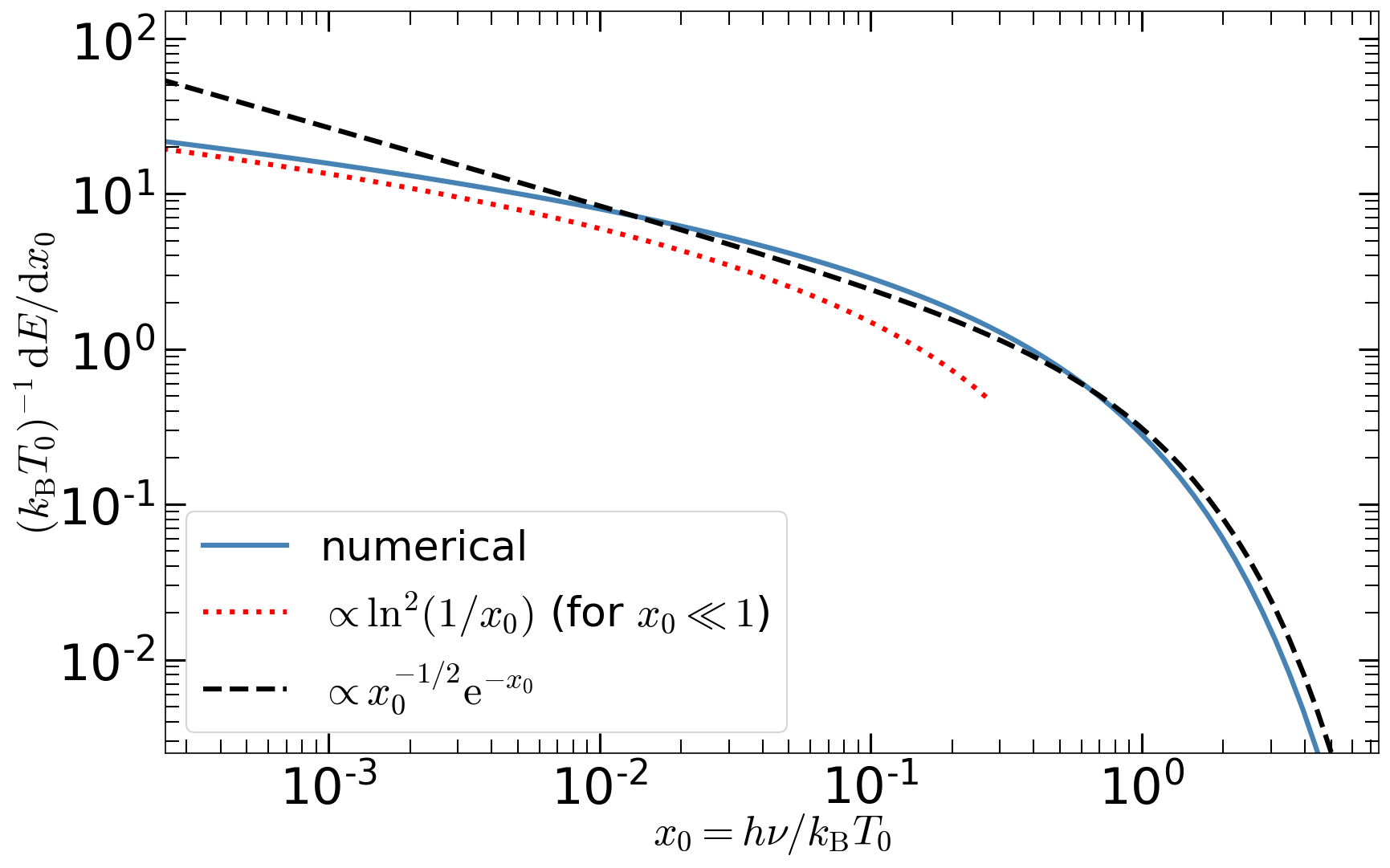}
        \caption{The cumulative emission spectrum of a plasma over the entire cooling history from an initial temperature $T_0$ to final temperatures $\ll h\nu/\kB$ (for a given frequency $\nu$). Here, we use the dimensionless quantity $x_0 = h\nu /(\kB T_0)$ to represent the frequency. The numerical result is shown as a solid blue line (eq. \ref{eq:ffcoolspec_numerical}), and two different approximations are shown as red dotted line (eq. \ref{eq:ffcoolspec_app1}) and black dashed line (eq. \ref{eq:ffcoolspec_app2}).
        }
    \label{fig:ffcoolspec_cumu}
\end{figure}

In fact, Govreen-Segal et al.  found that the cumulative spectrum from the cooling history is well described by the following analytical approximation
\begin{equation}\label{eq:ffcoolspec_app2}
    {1\over \kB T_0} \lrb{\d E\over \d x_0}^{(2)} \approx {3\over 2\sqrt{\pi}} x_0^{-1/2}\mr{e}^{-x_0}. \ \ (\mbox{for } x_0\gtrsim 10^{-2})
\end{equation}
A comparison between the numerical results and the analytical approximations (1) and (2) are shown in Figure~\ref{fig:ffcoolspec_cumu}.

This spectrum potentially applies when the central engine drives a radiative shock into earlier ejecta or CSM. An important assumption in the above calculation, however, is that the emitting system is effectively optically thin, such that the escaping spectrum is not substantially modified by absorption or Comptonization. In reality, the X-ray source may be embedded within the reprocessing layer introduced above. The observed X-ray spectrum may then be modified by Comptonization and bound-free absorption before escaping. More detailed modeling of these effects is needed. Along the solid angles through which X-rays escape in an aspherical system, the gas may be heated to sufficiently high temperatures that bound-free absorption is strongly suppressed. Another caveat regarding the electron temperature distribution is that the geometrically thin cool dense layer in the shock downstream is susceptible to instabilities in 3D, which may cause the shock front to be ``corrugated'' \cite{2018MNRAS.479..687S}. Such a structure may facilitate efficient turbulent mixing and lead to a different electron temperature distribution from that of a simplified 1D model.


Let us then discuss the possible origin(s) of the hard X-ray hump peaking near 40 keV.

It is possible to produce such a hump in the scattering layer, provided that the Compton-$y$ parameter is of order unity or somewhat larger,
\begin{equation}\label{eq:Compton_y}
    y = {4\kB T_{\rm C}\over \me c^2} \max(\tau_{\rm s}, \tau_{\rm s}^2) \sim 1,
\end{equation}
where $\tau_{\rm s}$ is the Thomson optical depth of the Comptonization layer and $T_{\rm C}$ is the electron/radiation temperature at Compton equilibrium. In this picture, from the peak temperature of the Compton hump in AT2018cow, $3\kB T_{\rm C}\simeq 40\rm\, keV$, we find $T_{\rm C}\simeq 1.5\times10^8\rm\, K$. Then, $y\sim 1$ requires a mildly optically thick layer with $\tau_{\rm s}\sim 3$. In this Comptonization picture, the equilibrium occurs at the Compton temperature \cite{2004MNRAS.347..144S}
\begin{equation}
    T_{\rm C} \approx {\me c^2\over 4\kB} {\int x^2(1-21x/5 + 147x^2/10)(\d n_{\rm ph}/\d x)\d x\over \int x(1-47x/8+567x^2/20)(\d n_{\rm ph}/\d x)\d x},
\end{equation}
which is valid as long as the maximum dimensionless photon energy $x_{\rm max}\ll 1$.
For a photon energy spectrum $\d n_{\rm ph}/\d x\propto x^{-1.6}$ (extrapolated from the soft X-ray power-law) and maximum energy $x_{\rm max}$, we obtain
\begin{equation}\label{eq:Compton_temperature}
    T_{\rm C}(x_{\rm max}) \approx  4.24\times 10^8\,{\rm K}\, x_{\max}\,
    \frac{ 1-2.45\,x_{\max}+6.05\,x_{\max}^2}{1-1.68\,x_{\max}+4.73\,x_{\max}^2}, \ (\mbox{for } x_{\rm max}\ll 1)
\end{equation}
From $T_{\rm C}\simeq 1.5\times10^8\rm\, K$, we infer $x_{\rm max}\simeq 0.4$ or $h\nu_{\rm max}\simeq 200\rm\, keV$. Since $x_{\rm max}$ is not very small, this estimate should be regarded as order-of-magnitude and ideally checked with the full Klein–Nishina energy-exchange rate. Nevertheless, the emission of such high energy photons requires mildly relativistic electrons that are accelerated by fast shocks (see eq. \ref{eq:thermal_electron_temperature}).

An alternative interpretation for the hard X-ray hump is that the soft X-rays are preferentially absorbed due to bound-free absorption in a layer between the X-ray emission region and the observer, as compared to hard X-rays which do not suffer from absorption, and this can potentially give rise to a positive spectral slope near frequencies where the effective optical depth $\tau_{\rm eff}\sim \sqrt{\tau_{\rm a}(\tau_{\rm a} + \tau_{\rm s})}$ is of order unity. This interpretation was proposed by Margutti et al. \cite{2019ApJ...872...18M, 2025ApJ...993L...6N}.

Recently, Govreen-Segal et al.~\cite{2026arXiv260118887G} proposed a more structured reprocessing geometry in which soft X-rays are ultimately absorbed by cold material at $T\sim10^4$--$10^5\ {\rm K}$ located \textit{interior} to the X-ray-emitting region. An optically thick, fully ionized scattering region outside the X-ray source reflects a fraction of the soft X-rays back toward this colder absorbing material.
If the scattering layer has optical depth $\tau_{\rm s}\gg 1$, then the transmission fraction for the soft X-rays will be of the order $\tau_{\rm s}^{-1}$ whereas nearly all the hard X-rays ($\gtrsim 10\rm\, keV$) will diffuse out as they do not suffer from significant absorption. However, we also require that the hard X-rays are not strongly Compton down-scattered, and this puts an upper limit on the scattering optical depth $\tau_{\rm s}\lesssim \sqrt{\me c^2/\epsilon}\lesssim 7 (\epsilon/10\rm{\rm \,keV})^{-1/2}$, where $\epsilon$ is the photon energy under consideration. 

In the above models for the hard X-ray hump, a common theme is that the X-rays are propagating through a scattering layer, which is likely heated up to the Compton temperature by the X-ray photons. Combining the conservative upper limit of $\tau_{\rm s}\lesssim10$ with the radius constraint from the variability timescale,
$R_{\rm sca}\lesssim10^{15}\ {\rm cm}$, gives an \textit{isotropic-equivalent}
mass along the X-ray escape directions,
\begin{equation}
M_{\rm sca}
=
\frac{4\pi R_{\rm sca}^{2}\tau_{\rm s}}
{\kappa_{\rm s}}
\simeq
0.3\,M_\odot
\frac{\tau_{\rm s}}{10}
\left(\frac{R_{\rm sca}}{10^{15}\ {\rm cm}}\right)^2
\left(\frac{\kappa_{\rm s}}
{0.2\ {\rm cm^2\ g^{-1}}}\right)^{-1}.
\label{eq:Msca}
\end{equation}
Here, $M_{\rm sca}$ is obtained by extending the column along the
observed X-ray escape directions over $4\pi$ steradians. It should not necessarily be identified with the UVOIR diffusion-inferred ejecta mass $M_{\rm ej}$ in an aspherical system. The two quantities are not separate masses to be added: $M_{\rm sca}$ characterizes a low-column region, whereas $M_{\rm ej}$ characterizes the mass participating in the UVOIR diffusion problem.



Nevertheless, the upper limit of $M_{\rm sca}\lesssim0.3\,M_\odot$
is consistent with the ejecta-mass constraint
$M_{\rm ej}\lesssim1\,M_\odot$ from the UVOIR diffusion-timescale argument.
As we show in the next subsection, an independent upper limit of a
similar magnitude follows from the requirement of X-ray ionization
breakout: soft X-rays cannot fully ionize a scattering layer with a
mass much greater than $\sim0.3\,M_\odot$. These arguments further
suggest that LFBOTs are likely produced by low-mass explosions.

\subsubsection{X-ray ionization breakout along low-column directions}\label{sec:Xray_ionization_breakout}
\leavevmode\\

\noindent The preceding discussion suggests that the X-ray source is embedded
beneath the optically thick ejecta responsible for the diffusion and
reprocessing of the UVOIR emission near peak luminosity. In an
aspherical system, however, the ejecta column density may vary strongly
with direction. The observed soft X-rays can escape along relatively
low-column directions where the gas is ionized to sufficiently high
charge states that bound-free absorption is strongly suppressed and
electron scattering dominates the opacity. We refer to the material
along such an X-ray escape direction as the \emph{scattering layer},
and denote its isotropic-equivalent mass by $M_{\rm sca}$.

Ionization breakout occurs when the ionizing photon supply is
sufficient to balance recombination throughout this scattering column.
The ionization balance is controlled primarily by soft X-rays with
$h\nu\sim1\ {\rm keV}$, which interact mainly with helium, carbon, and
oxygen. Our goal is to estimate the critical $M_{\rm sca}$ below which the
entire scattering layer can be maintained in a highly ionized,
density-bounded state, allowing the soft X-rays to escape.

Let us consider a hydrogen-poor scattering layer for which the
threshold for X-ray ionization breakout is controlled by oxygen; the
calculation can be generalized to other compositions. Near the
breakout threshold, the recombination rate in the X-ray-ionized region
is dominated by hydrogenic oxygen. At a given electron density
$n_{\rm e}$, the mass $M_{\rm ion}$ that can be maintained in a highly
ionized state is determined by balancing the ionization rate
$\dot{N}_{\rm ion}$ against the recombination rate
$\dot{N}_{\rm rec}$. The former is given by
\begin{equation}
    \dot{N}_{\rm ion}
    = {L_{\rm ion}\over \lara{\epsilon_{\rm ion}}},
\end{equation}
where $L_{\rm ion}$ is the luminosity of ionizing photons and
$\lara{\epsilon_{\rm ion}}\sim1\rm\,keV$ is the average energy cost per
ionization. The recombination rate is
\begin{equation}
    \dot{N}_{\rm rec}
    = {M_{\rm ion}X_{\rm O}\over A_{\rm O}\mproton}
      \alpha_{\rm B}n_{\rm e},
\end{equation}
where $X_{\rm O}$ is the oxygen mass fraction, $A_{\rm O}=16$ is its
mass number, $\mproton$ is the proton mass, and $\alpha_{\rm B}$ is the
Case-B recombination coefficient for a hydrogenic ion of nuclear charge
$Z$. We adopt \cite{2011piim.book.....D}
\begin{equation}
    \alpha_{\rm B}
    \simeq 1.4\times10^{-12}
    \lrb{Z/8}^{2.6}T_6^{-0.8}
    \rm\,cm^3\,s^{-1},
\end{equation}
where $T=10^6T_6\rm\,K$. The recombination timescale is very short,
\begin{equation}
    t_{\rm rec}
    =(\alpha_{\rm B}n_{\rm e})^{-1}
    \sim10\mr{\,s}\,
    \alpha_{\rm B,-12}^{-1}n_{\rm e,11}^{-1},
\end{equation}
so ionization--recombination equilibrium is justified. Equating
$\dot{N}_{\rm ion}$ and $\dot{N}_{\rm rec}$ gives
\begin{equation}\label{eq:Mion}
    M_{\rm ion}
    \simeq 0.06M_\odot
    {A_{\rm O}/16\over \lrb{X_{\rm O}/0.1} \lrb{\lara{\epsilon_{\rm ion}}/\mr{keV}}}
    \lrb{Z/8}^{-2.6}
    n_{\rm e,11}^{-1}
    L_{\rm ion,44}T_6^{0.8},
\end{equation}
where $L_{\rm ion,44}=L_{\rm ion}/10^{44}\rm\,erg\,s^{-1}$.

We next ask whether the entire scattering layer can be maintained in
this highly ionized state. We approximate the material along a
representative X-ray escape direction by a spherical-equivalent shell with characteristic radius $R_{\rm sca}$ and isotropic-equivalent
mass $M_{\rm sca}$. For a shell with thickness $\Delta R_{\rm sca}\sim R_{\rm sca}$, the electron density is related to this column mass by
\begin{equation}
    M_{\rm sca}
    \simeq 4\pi R_{\rm sca}^{3}
      n_{\rm e}\mu_{\rm e}\mproton,
\end{equation}
where we adopt $\mu_{\rm e}=2$.
This definition is equivalent to $M_{\rm sca}\simeq 4\pi R_{\rm sca}^{2}\tau_{\rm s}/\kappa_{\rm s}$ (eq. \ref{eq:Msca}). For a uniform sphere (instead of a shell), the actual enclosed mass would instead be $M_{\rm sca}/3$.

At the threshold for ionization breakout, the ionized mass equals the total scattering-layer mass, $M_{\rm ion}=M_{\rm sca}$. Eliminating $n_{\rm e}$ between this
condition and eq.~(\ref{eq:Mion}) gives the maximum scattering-layer mass that can be maintained in the required highly ionized state,
\begin{equation}\label{eq:Mionmax}
    M_{\rm ion,max}
    \simeq 0.3 M_\odot\,
    \lrsb{A_{\rm O}/16\over \lrb{X_{\rm O}/0.1} \lrb{\lara{\epsilon_{\rm ion}}/\mr{keV}}}^{1/2}
    \lrb{Z/8}^{-1.3}
    L_{\rm ion,44}^{1/2}
    R_{\rm sca,15}^{3/2}
    T_6^{0.4},
\end{equation}
where
$R_{\rm sca,15}=R_{\rm sca}/10^{15}\rm\,cm$.

The physical meaning of $M_{\rm ion,max}$ can be summarized by
considering two regimes. If $M_{\rm sca}<M_{\rm ion,max}$,
the ionizing photon supply is sufficient to maintain the entire
scattering layer in a highly ionized state. The system is then
density bounded, $M_{\rm ion}=M_{\rm sca}$, and bound-free absorption
is strongly suppressed. Soft X-rays can therefore escape, although
they may undergo multiple electron scatterings.

If instead $M_{\rm sca}>M_{\rm ion,max},$ the ionizing photons are exhausted before reaching the outer edge of the layer. The ionized region terminates at a ``Str\"omgren radius'' smaller than $R_{\rm sca}$, so that $M_{\rm ion}<M_{\rm sca}$, and a neutral or partially ionized column remains outside it. The system is then ionization bounded: soft X-rays are absorbed, and their energy is reradiated at lower frequencies
through recombination, line cooling, and continuum emission.

In the spherical-equivalent approximation, the condition
for $\sim 1 \rm \, keV$ soft X-rays to escape is
\begin{equation}\label{eq:Msca_ionization}
    M_{\rm sca}
    <M_{\rm ion,max}(R_{\rm sca})
    \simeq0.3M_\odot\,
    \lrb{X_{\rm O}/0.1}^{-1/2}\lrb{\lara{\epsilon_{\rm ion}}/\mr{keV}}^{-1/2}
    \lrb{Z/8}^{-1.3}
    L_{\rm ion,44}^{1/2}
    R_{\rm sca,15}^{3/2}
    T_6^{0.4}.
\end{equation}
We emphasize that the isotropic-equivalent mass $M_{\rm sca}$ along the X-ray escape directions should not necessarily be equal to the total ejecta mass $M_{\rm ej}$. This is because, in an aspherical system, soft X-rays can
escape along low-column directions for which
$M_{\rm sca}<M_{\rm ion,max}$, while denser directions remain
ionization bounded. High-energy radiation absorbed along these denser
directions can still be thermalized and reprocessed into the UVOIR
emission near peak luminosity. Thus, the detection of escaping X-rays
constrains the ejecta column along the low-column directions rather than
the total mass or angle-averaged density of the ejecta.

These ionization-breakout constraints apply to the material at
radii of order $10^{15}\rm\,cm$ or smaller along the directions from
which the X-rays escape. Independent evidence for an outer,
radio-emitting CSM at much larger radii,
$r\sim10^{16}$--$10^{17}\rm\,cm$, comes from the radio synchrotron
emission to be discussed next.

\subsubsection{Radio emission: fast outflow interacting with a dense ambient medium}\leavevmode\\
\label{sec:theory-radio}

\noindent Radio observations provide the strongest evidence for dense, radially confined CSM at larger radii. The radio synchrotron emission is produced by shock-accelerated relativistic electrons at distances $10^{16}\lesssim r\lesssim10^{17}\ {\rm cm}$ from the source. The hydrodynamics of an engine-driven fast outflow interacting with a power-law CSM density profile has been considered by Coughlin et al. \cite{2024ApJ...975L..14C, 2026PASA...43...96A}. As for the synchrotron emisison from the non-thermal electrons, the basic picture for understanding the radio SED is the ``equipartition model'' proposed by Chevalier \cite{1998ApJ...499..810C} and subsequently modified by Barniol Duran et al. \cite{2013ApJ...772...78B}. In the following, we first outline the model and its caveats, and then cautiously apply it to infer the CSM density profile.


In the simplest version of the equipartition model, the relativistic electron population at a given epoch is assumed to have a single power-law distribution in Lorentz factors $\d N/\d\gamma\propto \gamma^{-p}$. The expected radio SED is a broken power-law where the low frequency spectrum is self-absorbed $F_\nu\propto \nu^{5/2}$ and at high frequencies we have optically thin emission $F_\nu\propto \nu^{-(p-1)/2}$ from a power-law Lorentz factor distribution. The electron power-law index $p$ may have been affected by synchrotron and inverse-Compton cooling, which makes $p$ different from that expected from shock acceleration. The equipartition model outlined below is insensitive to the power-law index $p$.

At any given epoch, we identify the peak frequency as $\nu_{\rm a}$, where the synchrotron self-absorption optical depth $\tau(\nu_{\rm a})\simeq 1$, and the flux density $F_{\nu_{\rm a}}$ at the peak frequency. For a known source distance $D$, the specific luminosity at the peak frequency is then given by $L_{\nu_{\rm a}} = 4\pi D^2 F_{\nu_{\rm a}}$ (hereafter ignoring cosmological redshift factors). As we show below, it is possible to infer a characteristic radius $R$ of the radio emitting region using these two numbers: $\nu_{\rm a}$ and $L_{\nu_{\rm a}}$.

For simplicity, let us assume that the emitting plasma is spatially uniform and has a (randomly oriented) magnetic field strength of $B$ and that, at a given epoch, the plasma occupies a volume
\begin{equation}
    V = 4\pi R^3/3,
\end{equation}
where $R$ is the volume-equivalent radius. We ignore light-travel effects as the system is expanding at a non-relativistic speed (as inferred from the radio data self-consistently). Let $\gamma_{\rm a}$ be the Lorentz factor of electrons whose characteristic synchrotron frequency is equal to $\nu_{\rm a}$, and we can write
\begin{equation}
    \nu_{\rm a} = \gamma_{\rm a}^2{3 e B\over 4\pi \me c}.
\end{equation}
The emission spectrum of a mono-energetic population of electrons with Lorentz factor $\gamma_{\rm a}$ (with isotropic distribution of pitch angles) has the characteristic spectrum of $P_\nu \propto (\nu/\nu_{\rm a})^{1/3}$ at low frequencies and $P_\nu\propto \mr{e}^{-\nu/\nu_{\rm a}}$ at high frequencies. The peak spectral power near $\nu=\nu_{\rm a}$ is roughly given by
\begin{equation}
    P_{\nu_{\rm a}}\simeq {\sqrt{3}e^3B\over m_{\rm e} c^2}.
\end{equation}
Let us denote the number of electrons with Lorentz factors near $\gamma_{\rm a}$ as $N_{\gamma_{\rm a}}\simeq (\gamma \d N/\d\gamma)|_{\gamma_{\rm a}}$. Since the system is marginally optically thin near $\nu_{\rm a}$, the total spectral luminosity near $\nu_{\rm a}$ is roughly given by
\begin{equation}
    L_{\nu_{\rm a}} \simeq 4\pi j_{\nu_{\rm a}} V \simeq N_{\gamma_{\rm a}} P_{\nu_{\rm a}},
\end{equation}
where $j_{\nu_{\rm a}}$ is the specific emissivity at $\nu_{\rm a}$. On the other hand, the system is also marginally optically thick ($\tau_{\nu_{\rm a}}\simeq 1$), so the absorption coefficient at $\nu_{\rm a}$ is given by
\begin{equation}
    \alpha_{\nu_{\rm a}}\simeq 1/R,
\end{equation}
where we have assumed the typical path length for the emitted photons to be about $R$ (a geometric factor is ignored here for simplicity).

Importantly, $j_{\nu_{\rm a}}$ and $\alpha_{\nu_{\rm a}}$ are related to each other by Kirchhoff's law
\begin{equation}
    {j_{\nu_{\rm a}}\over \alpha_{\nu_{\rm a}}} = B_{\nu_{\rm a}}(T_{\rm exc})\simeq {2\over 3}\gamma_{\rm a} m_{\rm e} \nu_{\rm a}^2,
\end{equation}
where $B_\nu(T_{\rm exc})$ is the Planck function evaluated at the excitation temperature that is given by $3 \kB T_{\rm exc}\simeq \gamma_{\rm a} m_{\rm e} c^2$, and we have taken the RJ limit $h\nu_{\rm a}\ll \kB T_{\rm exc}$ ($h=\,$Planck constant) appropriate for radio emission. 

Counting the above equations, we find that we have 3 independent equations (based on two observables $\nu_{\rm a}$, $L_{\nu_{\rm a}}$ and one physical condition of $\tau(\nu_{\rm a})\simeq 1$) for 4 unknowns: $N_{\gamma_{\rm a}}$, $B$, $\gamma_{\rm a}$, and $R$. We then impose another (speculative) condition where the ratio between the energy carried by the emitting electrons and the energy carried by magnetic fields is given by a new parameter $\xi_{\rm eB}$, i.e.,
\begin{equation}\label{eq:xieB_definition}
    {N_{\gamma_{\rm a}} \gamma_{\rm a} m_{\rm e} c^2\over V B^2/8\pi} = \xi_{\rm eB}.
\end{equation}
Note that the above condition is slightly different from the original condition proposed by Chevalier \cite{1998ApJ...499..810C} and more in line with the model by Barniol Duran et al. \cite{2013ApJ...772...78B}. Here, the electron energy in the numerator only includes that carried by the \textit{observable} electrons with Lorentz factors near or above $\gamma_{\rm a}$; whereas the Chevalier model is based on the \textit{total} energy of all relativistic electrons down to the minimum Lorentz factor $\gamma_{\rm min}$. In the Chevalier model, it is assumed that electron Lorentz factor distribution follows a single power-law with $p\simeq 3$ and that the minimum Lorentz factor $\gamma_{\rm min}\sim 1$. However, these two assumptions may break down when (1) synchrotron or inverse-Compton cooling is important for electrons with $\gamma < \gamma_{\rm a}$ (see \cite{2022ApJ...932..116H} for a self-consistent treatment) and (2) when the shock velocity is sufficiently high $\gtrsim 0.1c$ such that $\gamma_{\rm m}$ may be significantly greater than unity. In the current approach (eq. \ref{eq:xieB_definition}) which is advocated by Barniol Duran et al., we do not make assumptions on the Lorentz factor distribution at $\gamma\ll \gamma_{\rm a}$, but the parameter $\xi_{\rm eB}$ depends not only on the microphysics of collisionless shocks but also on $\gamma_{\rm a}$. Nevertheless, similar in spirit to the original Chevalier model, our hope is that at least some of the final results depend sufficiently weakly on our assumption on the $\xi_{\rm eB}$ parameter and on the simplified treatment of the geometry of the system.

With the addition of the fourth equation based on $\xi_{\rm eB}$ (eq. \ref{eq:xieB_definition}), we obtain the solutions for all four unknowns
\begin{equation}
    \gamma_{\rm a} \simeq {\xi_{\rm eB}\over 4\sqrt{6}}^{2/17} \lrb{L_{\nu_{\rm a}}\over \me c^2}^{1/17} = 101\, \xi_{\rm eB}^{2/17} L_{\nu_{\rm a}, 30}^{1/17},
\end{equation}
\begin{equation}
    N_{\gamma_{\rm a}}\simeq {\sqrt{3}c\over 4\pi e^2} {L_{\nu_{\rm a}}\over \nu_{\rm a}} \gamma_{\rm a}^2 = 1.8\times10^{51}\, \xi_{\rm eB}^{4/17} L_{\nu_{\rm a}, 30}^{19/17} \nu_{\rm a,11}^{-1},
\end{equation}
\begin{equation}
    B\simeq {4\pi \me c\over 3e} {\nu_{\rm a}\over \gamma_{\rm a}^2} = 2.3\mr{\,G}\, \xi_{\rm eB}^{-4/17} L_{\nu_{\rm a}, 30}^{-2/17} \nu_{\rm a,11},
\end{equation}
\begin{equation}\label{eq:radius_from_equipartition}
    R\simeq {3\over 4\sqrt{2} \pi} \lrb{L_{\nu_{\rm a}}\over \gamma_{\rm a} \me}^{1/2} {1\over \nu_{\rm a}} = 5.5\times10^{15}\mr{\,cm}\, \xi_{\rm eB}^{-1/17} L_{\nu_{\rm a}, 30}^{8/17} \nu_{\rm a,11}^{-1},
\end{equation}
where $L_{\nu_{\rm a}, 30} = L_{\nu_{\rm a}}/10^{30}\rm\,erg\,s^{-1}\,Hz^{-1}$ and $\nu_{\rm a,11} = \nu_{\rm a}/10^{11}\rm\, Hz$.

With the above solutions, one can then calculate the sum of the energies carried by the emitting electrons (the observables ones with $\gamma\gtrsim \gamma_{\rm a}$) and magnetic fields,
\begin{equation}
    E_{\rm eB} \simeq N_{\gamma_{\rm a}}\gamma_{\rm a}\me c^2 + {R^3 B^2\over 6} = 1.5\times10^{47}\mr{\,erg}\, L_{\nu_{\rm a}, 30}^{20/17} \nu_{\rm a,11}^{-1}\lrsb{\xi_{\rm eB}^{6/17} + \xi_{\rm eB}^{-11/17}},
\end{equation}
where we have assumed that most of the observable electron energy is carried by those near $\gamma\sim\gamma_{\rm a}$ (reasonable as long as the electron index $p > 2$). We find that the energy $E_{\rm eB}$ reaches the minimum value $E_{\rm min}$ when $\xi_{\rm eB} = 11/6$ such that $\xi_{\rm eB}^{6/17} + \xi_{\rm eB}^{-11/17} = 1.91$.

The main advantage of the equipartition model is that the volume-equivalent radius $R$ is well constrained with minimum dependence on the unknown parameter $R\propto \xi_{\rm eB}^{-1/17}$. However, the other physical quantities $\gamma_{\rm a}$, $N_{\gamma_{\rm a}}$, $B$, and $E_{\rm eB}$ should be considered as more uncertain. The uncertainties are better understood if we express their dependence on $R$ instead of $\xi_{\rm eB}$,
\begin{equation}\label{eq:onezone_model_R_scalings}
\begin{split}
    \gamma_{\rm a} &\propto R^{-2} L_{\nu_{\rm a}} \nu_{\rm a}^{-2}, \\
    N_{\gamma_{\rm a}} &\propto R^{-4} L_{\nu_{\rm a}}^3 \nu_{\rm a}^{-5}, \\
    B &\propto R^4 L_{\nu_{\rm a}}^{-2} \nu_{\rm a}^{5}.
\end{split}
\end{equation}
Expressing these quantities in terms of $R$ (instead of the unconstrained parameter $\xi_{\rm eB}$) is motivated by the possibility of directly measuring $R$ by resolving the emitting region with radio VLBI observations. For fixed $L_{\nu_{\rm a}}$ and $\nu_{\rm a}$ (from observations), the total energy has the strongest dependence on $R$ as
\begin{equation}
    E_{\rm eB} = E_{\rm min} \lrsb{{11\over 17}(R/R_0)^{-6} + {6\over 17}(R/R_0)^{11}},
\end{equation}
where $E_{\rm min}$ and $R_0$ are the electron+magnetic field energy and radius for $\xi_{\rm eB} = 11/6$. 

\begin{figure}[b]
    \centering
    \includegraphics[width=0.6\linewidth]{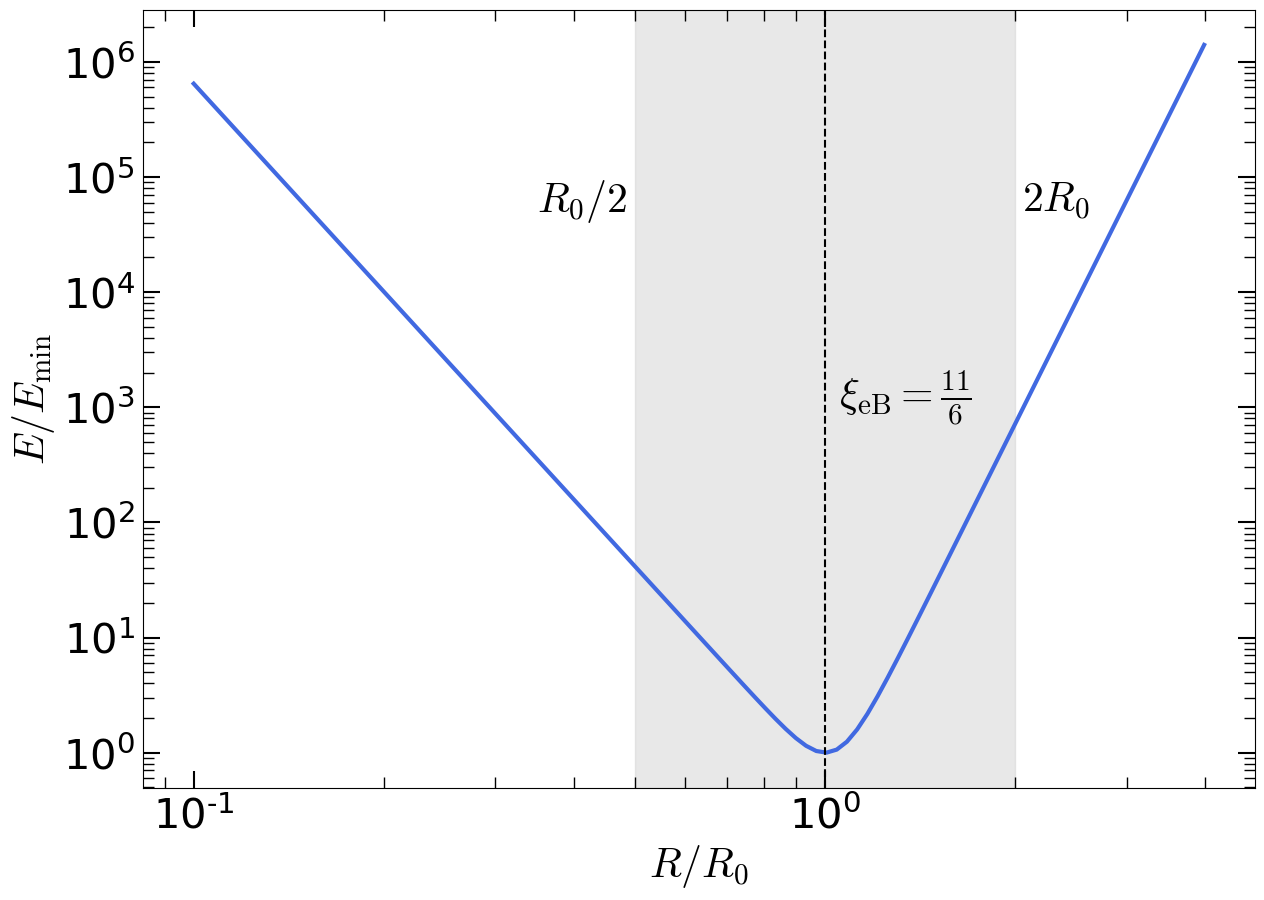}
        \caption{Inferred energy of emitting electrons and magnetic field $E$ vs. the volume equivalent radius $R$.
        }
    \label{fig:Emin_R}
\end{figure}

There are a few caveats in the above equipartition model. The first and main one is that the total energy $E_{\rm eB}$ depends strongly on the volume equivalent radius $R$ of the emitting region, as shown in Figure~\ref{fig:Emin_R}. The extremely strong dependence means that there are important \textit{geometrical factors} that can easily change $E_{\rm eB}$ by orders of magnitude. Even a variation in $R$ by a factor of $2$ gives rise to two to three orders of magnitude change in $E_{\rm eB}$. For this reason, the minimum energy obtained from the equipartition argument does not reflect the true energy of the system due to its large uncertainties. Turning the argument around, if the size of the emitting region can be well constrained by radio VLBI observations, then one can obtain a much better constraint on the total energy of the system.

The next issue of the equipartition model is that the electrons with Lorentz factors $\gamma\ll \gamma_{\rm a}$ are not directly constrained by observations, as their emission is self-absorbed. Due to various uncertainties regarding the Lorentz factor distribution at $\gamma\ll \gamma_{\rm a}$ (related to cosmic ray acceleration by collisionless shocks and synchrotron/inverse-Compton cooling), the true energy of all relativistic electrons is unknown. In fact, one can show that, in AT2018cow, the cooling timescale for electrons near $\gamma_{\rm a}$ is much shorter than the dynamical timescale in the earliest epochs of radio observations $t\lesssim 30\rm\, d$ (see e.g., \cite{2022ApJ...932..116H, 2025ApJ...993L...6N}). There is no compelling reason to believe $\xi_{\rm eB}$ will be order unity.

The final issue is that $E_{\rm eB}$ does not include the thermal energy carried by thermal population of electrons/ions and the kinetic energy carried by the bulk motion of the emitting gas. In the region heated by the forward shock (that propagates into the slowly moving CSM), we expect that about half of the energy is carried by thermal motions of ions/electrons and the other half is carried by the fluid's bulk motion (based on the Rankine-Hugoniot jump condition). Only a small fraction ($\sim 0.1$) of the total thermal energy is shared by non-thermal electrons and magnetic fields (e.g., \cite{2015PhRvL.114h5003P}). In fact, the thermal (Maxwell-J\"uttner) population of electrons may reach ultra-relativistic temperatures $\Theta\equiv \kB T/\me c^2\gg 1$ due to the fast shocks in LFBOTs \cite{2024ApJ...977..134M}. For a blast wave with velocity $v$, we expect the dimensionless electron temperature to be
\begin{equation}\label{eq:thermal_electron_temperature}
    \Theta \simeq 3 \, {T_{\rm e}/T_{\rm i}\over 0.3} (v/0.1c)^2,
\end{equation}
where $T_{\rm e}/T_{\rm i}$ describes the temperature ratio between the electrons and ions (as controlled by collisionless plasma processes, e.g., \cite{2015PhRvL.114h5003P}). The synchrotron spectrum by a Maxwellian population of electrons has been calculated by Mahadevan et al. \cite{1996ApJ...465..327M}. This has been used to fit the radio spectra of some LFBOTs in certain epochs where the high-frequency spectral slope is steeper than $F_\nu\propto \nu^{-1.5}$ (requiring electron index $p\gtrsim 4$), which is difficult to explain with power-law electrons accelerated from diffusive shock acceleration even with fast cooling \cite{2022ApJ...932..116H}.

We conclude that the most reliable constraint from the equipartition model is the size of the radio emitting region (eq. \ref{eq:radius_from_equipartition}). For AT2024wpp, the radio emitting region has a radius $R\sim 10^{16}\rm\, cm$ at $t\simeq 30\rm\, d$ and it later expands to $\sim 4\times10^{16}\rm\, cm$ at $t\simeq 100\rm\, d$. The radius change corresponds to an expansion rate of the order
\begin{equation}
    v \simeq {\d R\over \d t}\sim 0.1c.
\end{equation}

One can further try to infer the density profile of the CSM by invoking some \textit{additional} assumptions. The first assumption is that the emitting gas is undergoing homologous expansion via
\begin{equation}
    v \simeq {R\over t} \simeq 0.11c\, \xi_{\rm eB}^{-1/17} L_{\nu_{\rm a}, 30}^{8/17} \nu_{\rm a,11}^{-1} \lrb{t\over 20\mr{\,d}}^{-1},
\end{equation}
where $R(t)$ is radius inferred at time $t$ since the beginning of the explosion. This assumption of homologous expansion of the emitting region breaks down if the blast wave transitions from the initially coasting phase to the Sedov-Taylor phase or if the CSM density profile deviates from a single power-law, but this would only introduce an order unity correction to the inferred expansion speed (as the scaling of $v\propto R/t$ still holds). Suppose the unperturbed CSM density at radius $r=R$ from the center of explosion is $\rho_{\rm csm}(r= R)$, then the thermal energy density in the shock-heated region is given by $2\rho_{\rm csm} v^2$ (from the adiabatic shock jump condition). If one further assumes that the magnetic fields share a fraction $\epsilon_{\rm B}$ of the thermal energy density ($\epsilon_{\rm B}$ being highly uncertain), this then leads to the following constraint on the CSM density
\begin{equation}
\begin{split}
    \rho_{\rm csm}(r=R) &= {B^2\over 16\pi \epsilon_{\rm B} v^2}\simeq {B^2 t^2\over 16\pi \epsilon_{\rm B} R^2}\\
    &\simeq 1.0\times10^{-18}\mr{\,g\,cm^{-3}}\,  \xi_{\rm eB}^{-6/17} \lrb{\epsilon_{\rm B}\over 10^{-2}}^{-1} \lrb{t\over 20\mr{\,d}}^{2} L_{\nu_{\rm a}, 30}^{-20/17} \nu_{\rm a,11}^{4}.
\end{split}
\end{equation}
If the CSM is produced by a quasi-steady outflow with mass-loss rate $\dot{M}_{\rm csm}$ and velocity $v_{\rm csm}$ long before the explosion, we can infer the following constraint on the mass-loss rate
\begin{equation}
    \dot{M}_{\rm csm} = {4\pi R^2 \rho(R) v_{\rm csm}} = 6.3\times 10^{-5}\msunyr\,  {v_{\rm csm}\over 100\mr{\,km\,s^{-1}}} \xi_{\rm eB}^{-8/17} \lrb{\epsilon_{\rm B}\over 10^{-2}}^{-1} \lrb{t\over 20\mr{\,d}}^{2} L_{\nu_{\rm a}, 30}^{-4/17} \nu_{\rm a,11}^{2}.
\end{equation}
For our fiducial parameters and expansion speed $v_{\rm csm}\sim 100\rm\, km\,s^{-1}$, the inferred mass-loss rate is much higher than that of typical massive stars or even Wolf-Rayet stars (which would have higher $v_{\rm csm}$ and hence higher $\dot{M}_{\rm csm}$). 

For known $t$, $\nu_{\rm a}$, and $L_{\nu_{\rm a}}$, the CSM density inferred with the above model depends strongly on $\epsilon_{\rm B}$ and $R$ as $\rho_{\rm csm}(r=R)\propto \epsilon_{\rm B}^{-1} R^6$. The mass-loss rate scales as $\dot{M}_{\rm csm}\propto \rho R^2 \propto \epsilon_{\rm B}^{-1} R^8$, and the swept-up CSM mass near radius $r\sim R$ scales as $M_{\rm csm}\propto \rho R^3\propto \epsilon_{\rm B}^{-1} R^9$. We see that the radio-inferred CSM properties have very large uncertainties (orders of magnitude) unless the emitting radius $R$ is accurately measured from direct observations. Nevertheless, one may still go ahead and try to infer the density profile of the CSM using radio data collected at multiple epochs, with each providing a measurement $\rho_{\rm csm}(r=R(t))$. The hope is that, if the factors $\epsilon_{\rm B}$ and $\xi_{\rm eB}$ remain constant over different epochs, then the inferred shape of the CSM density profile is still more-or-less reliable despite the highly uncertain overall normalization. When the CSM density is very high such that the emission measure of the ionized gas exceeds $\sim 10^{25}\rm\, cm^{-5}$, free-free absorption may become important at low frequencies, and this may provide another constraint on the physical parameters.

As shown in the right panel of Figure~\ref{fig:radio-modeling}, the radio-inferred CSM density profile is quite spatially confined --- nearly all the CSM mass is concentrated within a radius of $R_{\rm csm}\sim 4\times10^{16}\rm\, cm$ and the density rapidly drops beyond this radius. A striking property of the LFBOT population is the \textit{relative} uniformity of the CSM profiles inferred from the radio observations of different objects \cite{2022ApJ...926..112B, 2024A&A...691A.329C, 2025ApJ...993L...6N}.

Physically, this phenomenon has been attributed to the mass loss episode from the progenitor system shortly before the explosion (e.g., \cite{2025ApJ...993L...6N}). The time delay between the mass loss and the explosion may be inferred by assuming a constant expansion speed of $v_{\rm csm}$,
\begin{equation}
    t_{\rm csm} \sim {R_{\rm csm}\over v_{\rm csm}} \sim 100\mr{\,yr}\, {R_{\rm csm}\over 4\times10^{16}\mr{\,cm}} {100\mr{\,km\,s^{-1}}\over v_{\rm csm}},
\end{equation}
where we have taken a fiducial value of $v_{\rm csm}\sim 100\mr{\,km\,s^{-1}}$ as that corresponds to that of the circum-binary outflow expected from the mass loss from the outer Lagrangian point (\cite{2025ApJ...990..172S, 2025arXiv251024127S}). However, the CSM may be formed in ways other tha from the outer Lagrangian point (e.g., winds along the polar directions), and $v_{\rm csm}$ may be different.

The possibility that CSM is formed in late phases of binary mass transfer $\sim 10^2$ up to $10^4\rm\, yrs$ before supernova explosion has been pointed out by \cite{2022ApJ...940L..27W} (see also \cite{2025A&A...696A.103E}). Another possibility was recently proposed by \cite{2026ApJ..1005....2K} where the CSM is produced by  $\sim \rm kyr$-timescale binary mass transfer preceding the final merger between a stellar-mass BH and a stripped helium star, but that requires a much slower $v_{\rm csm}\sim 10\rm\, km\,s^{-1}$ (the physical reason of which is not clearly explained). A third possibility, as proposed by \cite{2022ApJ...933..203K, 2023MNRAS.524.6358K}, is when a star is partially disrupted by a stellar-mass BH (in a micro-TDE) resulting in a remnant core in a bound orbit, the disk wind from the accretion episode associated with an earlier pericenter passage would serve as the CSM for the new episode --- however, in this case, we expect very diverse CSM profiles from one source to another.

\subsubsection{Multi-zone model for flat radio spectra}\leavevmode\\

\noindent In some cases shown in Figure~\ref{fig:cow-radio}, the radio spectra are not well-described by the simplest one-zone, broken power-law model with a self-absorbed segment of $F_\nu\propto \nu^{5/2}$ at $\nu < \nu_{\rm a}$ and optically thin segment $F_\nu\propto \nu^{-(p-1)/2}$ at $\nu > \nu_{\rm a}$, where $p$ is the power-law index describing the Lorentz factor distribution of electrons with $\gamma\gtrsim \gamma_{\rm a}$. The main discrepancy is at high frequencies: for $p \geq 2$ (as expected from particle acceleration by shocks), the high frequency spectral index is steeper (smaller) than $-0.5$, but the observed spectrum can sometimes be rather flat with spectral index closer to 0.

It is possible to generalize the one-zone model outlined in the previous section to a continuous multi-zone picture to explain such flat spectra. Such a multi-zone model has been developed for active galactic nuclei (AGN) with powerful relativistic jets (the so-called flat spectrum radio quasars) by Blandford \& K\"onigl \cite{1979ApJ...232...34B}, and an important observational confirmation of this picture is the ``core shift'' effect \cite{1984ApJ...276...56M, 1998A&A...330...79L, 2011A&A...532A..38S}. 

Extension from one-zone to multi-zone is straightforward from eq. (\ref{eq:onezone_model_R_scalings}). The key idea is that for each logarithmic radial zone near radius $R$ (considering a radial thickness of the order $\Delta R\sim R$), there is a radius-dependent electron population $\gamma_{\rm a}(R)$ and $N_{\gamma_{\rm a}}(R)$ and magnetic field strength $B(R)$. For simplicity, let us assume that the emitting region at different radii spans a similar solid angle, as is the case for a conical, continuous jet, so we do not have to worry about the radius dependence of the solid angle factor and the emitting volume scales as $V\propto R^3$. Even in the absence of jets, the picture still roughly holds as long as the solid angle spanned by the emitting regions at different radii do not differ by more than a factor of a few.

In the original Blandford \& K\"onigl model, the authors assume $B\propto R^{-1}$, as is the case for a Poynting-dominated conical jet (as the Poynting flux scales as $R^{-2}$ according to the inverse-square law). This can be combined with the magnetic field strength scaling law in our one-zone model (eq. \ref{eq:onezone_model_R_scalings}) to obtain $R\propto L_{\nu_{\rm a}}^{2/5} \nu_{\rm a}^{-1}$. They also assume equipartition such that $\xi_{\rm eB}\sim \mc{O}(1)\sim\,$const, so the radius scales as $R\propto L_{\nu_{\rm a}}^{8/17} \nu_{\rm a}^{-1}$ (cf. \ref{eq:radius_from_equipartition}). These two conditions can then be combined to yield
\begin{equation}
    L_{\nu_{\rm a}}\propto \nu_{\rm a}^0,\ \ \nu_{\rm a}\propto R^{-1}.
\end{equation}
This means that higher-frequency emission is dominated by the plasma at smaller radii and that the overall spectrum is flat $L_\nu\propto \nu^0$. 

More generally, one can turn the above argument around: if we measure a relatively flat spectrum with $L_\nu \propto \nu^\alpha$ with $-0.5 < \alpha < 2$ (inconsistent with our one-zone model), it is possible to use this information to infer the radial scalings of other parameters. It is straightforward to carry out this calculation, and the result is
\begin{equation}
\begin{split}
    \gamma_{\rm a} &\propto R^{-{\alpha\over 17-8\alpha}} \xi_{\rm eB}^{2-\alpha\over 17-8\alpha}, \\
    N_{\gamma_{\rm a}} &\propto R^{-{17-19\alpha\over 17-8\alpha}} \xi_{\rm eB}^{5-3\alpha\over 17-8\alpha}, \\
    B &\propto R^{-{17-2\alpha \over 17- 8\alpha}} \xi_{\rm eB}^{-{5-2\alpha\over 17-8\alpha}}.
\end{split}
\end{equation}
In the above expressions, we have retained the dependence on the unknown $\xi_{\rm eB}$, which could be a function of radius if we relax the equipartition assumption. Unfortunately, it is not possible to break the $\xi_{\rm eB}$ degeneracy without a direct measurement of the emitting radii at different frequencies. This issue may potentially be resolved in the future with multi-frequency VLBI observations of nearby LFBOTs. If we na\"ively ignore the $\xi_{\rm eB}$ dependence (assuming equipartition), the radial dependence of the magnetic field strength is $B\propto R^{-1}$ and $\propto R^{-15/9}$ for $\alpha=0$ and 1, respectively. For now, the physical realization of multi-zone synchrotron emission in LFBOTs is observationally unconstrained. It could be due to a continuous jet or shocks at different radii driven by outflows of different velocities interacting with a non-spherical CSM.



\subsubsection{Persistently blue optical continuum:
a continuous, optically thick outflow?}
\label{sec:blue_continuum}
\leavevmode\\

\noindent
A major difference between LFBOTs and various types of SNe is the \textit{persistently blue} optical spectra --- they are initially nearly featureless in the first week or so and later on show intermediate width H and He recombination lines. Here, we discuss the long-lasting blue optical continuum\footnote{So far, the only high SNR HST UV spectrum is from AT2024wpp \cite{2026MNRAS.549ag678P}, which was entirely featureless at $t\simeq 20\rm\, d$. It is possible that other LFBOTs also have long-lasting featureless UV spectra, but this has not been established.} and leave the recombination lines from H-rich gas to \S \ref{sec:Balmer_lines}. Phenomenologically, as the UV/optical luminosity of the source drops over time, the persistently blue color (or hot color temperature) requires a \textit{shrinking} photospheric radius. This is in stark contrast with the evolution of SNe --- in the early photospheric phase, we expect increasing photospheric radius as the ejecta expands, whereas when the outer layers of the ejecta become increasingly optically thin, there is a relatively brief period of receding photospheric radius \cite{2018ApJ...868L..24L} and then the spectrum transitions to the nebular phase with weak or non-existing photospheric continuum emission. 

An attractive explanation for the persistently blue optical continuum is provided by Piro \& Lu (2020) \cite{2020ApJ...894....2P} and Uno \& Maeda (2020) \cite{2020ApJ...897..156U} in the picture of ``wind-reprocessed transients''. In this picture, the system launches a continuous, non-homologous outflow and at late times (when the initial ejecta and CSM become optically thin) the photosphere forms in the outflow. Recently, Aspegren \& Kasen (2026) \cite{2026arXiv260100947A} showed that the emerging optical spectrum from the wind becomes featureless at sufficiently high luminosities $L\gtrsim10^{44}\rm\, erg\,s^{-1}$ due to strong ionization. 

We note that the lack of cooling in the optical SED is also a defining signature of TDEs, which are believed to be powered by accretion onto supermassive BHs (e.g., \cite{2019ApJ...872..151M}). Although it is not clear if the similarity in color evolution necessarily translates to similarity in engine nature, the continuous wind model in Aspegren \& Kasen \cite{2026arXiv260100947A} can be naturally realized if the central engine is a stellar-mass compact object undergoing super-Eddington accretion.

\subsubsection{Delayed Balmer emission:
H-rich, intermediate-velocity outflow?}
\label{sec:Balmer_lines}
\leavevmode\\

\noindent
Another important feature seen in many LFBOTs is the appearance of intermediate width emission lines after a delay of $t\sim 15$--$30\rm\, d$ (the exact delay for the first detection depends on the SNR of the spectra and varies from source to source). These emission lines have typical full width at half maximum (FWHM)  $3000$--$10^4\rm\, km\,s^{-1}$, which is much slower than the ejecta velocity ($\sim0.1c$) inferred from the expanding optical photosphere. In the following, we discuss the implications of the H$\alpha$ line, which most likely comes from recombination in photo-ionized H-rich gas.

The equivalent widths of the H$\alpha$ line in different LFBOTs (AT2018cow, CSS161010, AT2024wpp) are typically in the range $10\lesssim W\lesssim 10^2\,\angstrom$ \cite{2026ApJ...997L..10L}. Given the specific luminosity of the continuum $L_{\lambda_0} = \lrb{\d L/\d\lambda}_{\lambda_0}$ near the line center wavelength $\lambda_0 = 6563\,\angstrom$, total line luminosity is given by $L_{\rm line} = L_{\lambda_0} W$, and we obtain the ratio between the line luminosity $L_{\rm H\alpha}$ and the UVOIR bolometric luminosity $L_{\rm bol}$,
\begin{equation}
    {L_{\rm H\alpha}\over L_{\rm bol}} = {W\over \lambda_0} {\lambda_0 L_{\lambda_0}\over L_{\rm bol}}.
\end{equation}
Since the UV/optical continuum can be well described by a blackbody at the photospheric temperature $T_{\rm ph}$, we find
\begin{equation}
    {\lambda_0 L_{\lambda_0}\over L_{\rm bol}} \simeq {\nu_0 B_{\nu_0}\over \sigma_{\rm SB}T_{\rm ph}^4/\pi} = {3.56\over \mr{e}^{h\nu_0/\kB T_{\rm ph}} - 1} \lrb{T_{\rm ph}\over 10^{4}\rm K}^{-4},
\end{equation}
where $\nu_0 = c/\lambda_0$ is the line center frequency. As an example, for typical late-time photospheric temperature $T_{\rm ph}\simeq 2\times10^4\rm\, K$, we find ${L_{\rm H\alpha}/L_{\rm bol}} \sim 10^{-4}$ to $10^{-3}$ for $10 \lesssim W\lesssim 100\,\angstrom$, which means that roughly 0.01\% to 0.1\% of the bolometric luminosity is emitted in H$\alpha$ line. 

Since H$\alpha$ comes from recombination of H-rich gas under ionization-recombination equilibrium, the line luminosity is given by
\begin{equation}
    L_{\rm H\alpha} = \dot{N}_{\rm ion, H} h\nu_0 {\alpha_{\rm B}^{\rm H\alpha}\over \alpha_{\rm B}},
\end{equation}
where $\dot{N}_{\rm ion, H}$ is the H ionization rate, $h\nu_0=1.89\rm\,eV$ is the line photon energy, $\alpha_{\rm B}^{\rm H\alpha}/\alpha_{\rm B}\simeq 1/3$ is the H$\alpha$ branching ratio from the recombination cascade (which is weakly dependent on the electron temperature \cite{2006agna.book.....O}). In the above expression, we have assumed that the ionized H-rich gas is optically thin to H$\alpha$ photons.

We consider that a fraction $\Phi_{\rm ion, H}$ of $L_{\rm bol}$ is used to ionize the H-rich gas and that the average energy cost per ionization is $\lara{\epsilon_{\rm ion, H}}\simeq 30\rm\, eV$ (appropriate considering He ionization and cooling effects), and this leads to $\dot{N}_{\rm ion, H} = \Phi_{\rm ion, H} L_{\rm bol}/\lara{\epsilon_{\rm ion,H}}$ and hence
\begin{equation}
    {L_{\rm H\alpha}\over L_{\rm bol}} = \Phi_{\rm ion, H} {h\nu_0\over \lara{\epsilon_{\rm ion, H}}} {\alpha_{\rm B}^{\rm H\alpha}\over \alpha_{\rm B}} = 2\times10^{-3} {\Phi_{\rm ion,H }\over 0.1} \lrb{\lara{\epsilon_{\rm ion, H}}\over 30\mr{\,eV}}^{-1} {\alpha_{\rm B}^{\rm H\alpha}/\alpha_{\rm B}\over 1/3}.
\end{equation}
We find that the observed $10^{-4}\lesssim L_{\rm H\alpha}/L_{\rm bol}\lesssim 10^{-3}$ in LFBOTs requires $10^{-2}\lesssim \Phi_{\rm ion,H}\lesssim 10^{-1}$ -- a very small fraction of the total energy output is spent on the ionization of the H-rich gas. We also note that some SLSNe-I also showed intermediate-width H$\alpha$ emission after much longer delays of the order $100\,\rm d$ but with much higher ${L_{\rm H\alpha}/L_{\rm bol}}\sim \mc{O}(1\%)$ \cite{2015ApJ...814..108Y, 2017ApJ...848....6Y}. 

In the following, we discuss some potential scenarios that may lead to (1) $\Phi_{\rm ion, H}\ll 1$, (2) the intermediate linewidth ($3000$--$10^4\rm\,km\,s^{-1}$), and (3) the delayed onset of $\sim 20\rm\,d$. 

If the H-rich gas is part of the CSM located \textit{beyond} the ejecta at $t\gtrsim 20\rm\, d$ (at radii $r_{\rm line}\gtrsim 5\times10^{15}\rm\, cm$), the small $\Phi_{\rm ion, H}$ is likely the result of either the H-rich gas being optically thin to ionizing photons (with bound-free optical depth $\ll 1$) or geometrically confined within a small solid angle like in an equatorially concentrated outflow. In this picture, the delayed onset of H$\alpha$ emission may be explained by delayed CSM interaction, but it may be difficult to explain the intermediate linewidths. This is because, if the shock-heated H-rich gas expands at $v\sim$a few $\times10^3\rm\, km/s$, the electron temperatures would be of the order $10^8$--$10^9\rm\, K$, and the gas does not produce strong recombination line emission due to inefficient cooling. If we attribute the intermediate linewidths to the \textit{original} expansion speed of the CSM, it is also difficult for stellar/binary evolution to produce a sufficiently dense CSM expanding at $3000$--$10^4\rm\, km\,s^{-1}$ via stellar winds or L2 mass loss. For the above reasons, we favor the scenario where the H-rich gas is embedded within the ejecta.


If the H-rich gas is embedded within the ejecta, it may originate from slower outflows from the central engine. Such a picture has been proposed by Tsuna \& Lu (2025) \cite{2025ApJ...986...84T} and Klencki \& Metzger (2026) \cite{2026ApJ..1005....2K}. For a super-Eddington accretion flow with outer disk radius $r_{\rm d}$, the outflow originating from radius $r<r_{\rm d}$ has characteristic velocity
\begin{equation}
    v_{\rm out}(r) \sim \sqrt{GM\over r} = 4\times10^3\mr{\,km\,s^{-1}}\, \lrb{M\over 10 M_\odot}^{1/2} \lrb{r/r_{\rm d}\over 0.1}^{-1/2} \lrb{r_{\rm d}\over R_\odot}^{-1/2}.
\end{equation}
We see that the disk wind may produce a wide range of outflow velocities depending on the launching radius $r$. The fraction of the bolometric luminosity that is spent on H ionization, $\Phi_{\rm ion, H}$, would depend on the detailed geometry of the system. The slowest outflows ($\sim10^3\rm\, km\,s^{-1}$) from $r\sim r_{\rm d}$ are much behind the expanding ejecta, so they occupy a very small volume that makes the reprocessing efficiency very low. The fastest outflows ($\gtrsim0.1c$) from the inner regions of the accretion disk carry a very small fraction of the total mass-loss rate from the disk, so their density is likely too low to efficiently produce recombination lines (they are likely in the density-bounded regime). It is possible that intermediate-velocity outflows from e.g., $r\sim 0.1r_{\rm d}$, may expand to a sufficiently large volume while remaining dense enough to efficiently produce the observed H$\alpha$ line. However, this so far has not been conclusively demonstrated in a detailed calculation.

In the disk wind picture, one can place a lower limit on the total H mass based on the observed H$\alpha$ luminosity (following \cite{2026arXiv260715464L}). The line formation region is located near where $\tau_{\rm H\alpha}(r_{\rm H\alpha}) \simeq \rho \kappa_{\rm H\alpha, eff} r_{\rm H\alpha} \sim 1$, where $\kappa_{\rm H\alpha,eff}$ is 
an effective opacity controlling the escape of H$\alpha$ photons, including true continuum absorption, resonant trapping, and electron-scattering redistribution. For simplicity, we take fiducial value for $\kappa_{\rm H\alpha,eff}$ comparable to that of electron scattering but note that it depends on continuum absorption as well as line photon trapping due to electron scatterings and resonant scatterings in an expanding medium. The total H$\alpha$ luminosity is then given by recombination at radii $r > r_{\rm H\alpha}$, i.e.,
\begin{equation}
\begin{split}
    L_{\rm H\alpha} &\simeq \int_{r_{\rm H\alpha}}^\infty \alpha_{\rm B} f_{\rm H\alpha} n_{\rm e} n_{\rm p} 4\pi r^2 \d r \simeq {\alpha_{\rm B} f_{\rm H\alpha} X_{\rm H}\over \mu_{\rm e} \mproton^2} {\dot{M}_{\rm out}/v_{\rm out}\over \kappa_{\rm eff}}\\
    &\simeq 2.1\times10^{39}\mr{\,erg\,s^{-1}}\, T_{\rm e,4.3}^{-0.8} \lrb{\kappa_{\rm eff}\over 0.3\mr{\,cm^2\,g^{-1}}}^{-1} \lrb{v_{\rm out}\over 3000\mr{\,km\,s^{-1}}}^{-1} {\dot{M}_{\rm out}\over 0.1\,\msunyr},
\end{split}
\end{equation}
where we have taken the gas density $\rho = \dot{M}_{\rm out}/(4\pi r^2 v_{\rm out})$ for an outflow rate $\dot{M}_{\rm out}$, electron number density $n_{\rm e}\simeq \rho/(\mu_{\rm e} \mproton)$ (with fiducial $\mu_{\rm e}\simeq 1.2$ for fully ionized H-rich gas), proton number density $n_{\rm p} = X_{\rm H} \rho/\mproton$ (with fiducial H mass fraction $X_{\rm H}=0.7$ for solar composition), Case-B recombination rate coefficient $\alpha_{\rm B}=1.4\times10^{-13}\mr{\, cm^3\,s^{-1}} T_{\rm e,4.3}^{-0.8}$, and H$\alpha$ branching ratio $f_{\rm H\alpha}\sim 1/3$ \cite{2011piim.book.....D}. We find that, in this picture, the observed H$\alpha$ line luminosities of the order $10^{39}\rm\, erg\,s^{-1}$ require a mass outflow rate of the order $0.1\,\msunyr$. For $L_{\rm H\alpha}\sim 10^{39}\rm\, erg\,s^{-1}$ lasting for a duration of $20\rm\, d$, we obtain a lower limit for the H-rich gas to be $M_{\rm H,min}\sim 3\times10^{-3}M_\odot$. This lower limit is conservative in that it does not account for potentially much higher outflow rates at earlier time ($t\ll 20\rm\, d$) before the emergence of the H$\alpha$.

Finally, for an embedded H$\alpha$ emitter, the line photons will initially be strongly absorbed by the fast ejecta or Compton-scattered to much broader widths, making them not easily observable. Thus, it is expected that intermediate width H$\alpha$ emission would be delayed until the ejecta has become optically thin to electron scattering \cite{2025ApJ...986...84T}. For a homologously expanding ejecta, the time at which the scattering optical depth drops to $\tau \simeq 1$ is roughly given by
\begin{equation}
    t_{\rm H\alpha} \simeq {1\over 4} \lrb{\kappa M_{\rm ej}^2 \over E_{\rm ej}}^{1/2} \simeq 30\mr{\,d}\, \lrb{\kappa\over 0.1\mr{\,cm^2\,g^{-1}}}^{1/2} {M_{\rm ej}\over 0.5M_\odot} \lrb{E_{\rm ej}\over 10^{51}\mr{\,erg}}^{-1/2},
\end{equation}
where the prefactor of $1/4$ weakly depends on the density profile of the SN ejecta \cite{2019ApJ...885L..23M} and $\kappa$ here is the scattering opacity. Here, $M_{\rm ej}$ is the UVOIR diffusion-inferred ejecta mass introduced in Section~\ref{sec:low_ejecta_mass}. This estimate assumes
that the material overlying the line-forming region is approximately
homologous and can be represented by the same $M_{\rm ej}$ and
$E_{\rm ej}$. We see that the delay time of $t_{\rm H\alpha} \sim20\rm\, d$ in LFBOTs is consistent with a relatively low ejecta mass $M_{\rm ej}\sim 0.3M_\odot$ (required by the short diffusion timescale, see \S \ref{sec:low_ejecta_mass}), whereas in SLSNe-I with H$\alpha$ lines, the delay time of $\sim 100\rm\, d$ is consistent with a larger ejecta mass $M_{\rm ej}\sim 5M_\odot$ but also larger ejecta energy $E_{\rm ej}\sim 10^{52}\rm\, erg$.

\subsubsection{NIR excess from dust echo?}\leavevmode\\

\noindent 
A NIR excess above the Rayleigh-Jeans extrapolation of the optical continuum has been detected in AT2018cow, AT2024wpp, and AT2024puz. The excess must emerge outside the optical photosphere $R_{\rm ph}\sim 10^{15}\rm\, cm$, as the emission from smaller radii will be self-absorbed.

Two broad mechanisms have been proposed. In the dust-echo scenario \cite{2023ApJ...944...74M, 2025ApJ...989...27T, 2025ApJ...992...25L}, UVOIR radiation is absorbed by pre-existing dusty CSM at radii $\sim10^{16}$--$10^{17}\rm\,cm$ and reradiated in IR bands. Alternatively, the excess may be free-free emission \cite{2025ApJ...991..180C, 2025ApJ...995..228S, 2026arXiv260118887G} from ionized gas in the ejecta, a continuous engine-driven outflow, or external CSM. We review the dust echo scenario and discuss possible observational tests in this subsection. The free-free emission scenario will be discussed in the next subsection.




Perley et al. (2019) \cite{2019MNRAS.484.1031P} initially argued against the dust echo interpretation, because the implied best-fit blackbody temperature exceeds the dust sublimation temperature $T_{\rm sub}\sim 2000\rm\, K$. However, later Metzger \& Perley (2023) pointed out that the dust emission spectrum is better described by a ``modified blackbody'' with the emission extending more to the blue, because dust grains are poor emitters at wavelengths longer than their own sizes. In the dust echo model proposed by Metzger \& Perley \cite{2023ApJ...944...74M}, the gas density profile of the CSM is taken to be a power-law form
\begin{equation}
    \rho = \rho_0 (r/r_0)^{-k},
\end{equation}
where $r_0=10^{16}\rm\, cm$ is a reference radius (of no physical importance) and $\rho_0$ is the gas density normalization at $r_0$. We restrict our discussion to the case of $k>1$, although it is possible (but not straightforward) to generalize the model to the case of $k<1$ if it is proven to be relevant. The dust opacity in the UV band can be taken as a constant $\kappa_{\rm d}\sim 10^2$--$10^3\rm\, cm^2\,g^{-1}$, which depends on the highly uncertain dust-to-gas ratio and grain-size distribution. The dust echo model has degeneracies such that the density normalization $\rho_0$ and dust opacity $\kappa_{\rm d}$ can be combined into a single parameter --- the characteristic UV optical depth near the reference radius
\begin{equation}
    \tau_0 = \rho_0 r_0 \kappa_{\rm d}.
\end{equation}
For a power-law density profile, the UV optical depth near radius $r$ is roughly given by (ignoring a factor of order unity involving $k$)
\begin{equation}
    \tau(r) \simeq \tau_0 (r/r_0)^{1-k}.
\end{equation}
The radius corresponding to UV optical depth of unity is given by
\begin{equation}\label{eq:r_thin}
    r_{\rm thin} \simeq r_0 \tau_0^{1\over k-1} = 10^{16}\tau_0^{1\over k-1}\mr{\,cm} .
\end{equation}
For an unattenuated source with UV luminosity $L=10^{43}L_{43}\rm\,erg\,s^{-1}$, the corresponding dust sublimation radius for sublimation temperature $T_{\rm sub}\simeq 2000\rm\, K$ and grain size $a\simeq 1\rm\, \mu m$ is given by
\begin{equation}
    r_{\rm sub}(L) \simeq 2\times10^{16} L_{43}^{1/2}\mr{\,cm}.
\end{equation}
Near the peak of the LFBOT lightcure ($L_{\rm pk}\gtrsim \mr{few}\times10^{44}\rm\, erg\,s^{-1}$), the sublimation radius reaches $10^{17}\rm\,cm$ (corresponding to a light-crossing time of $\sim $month), so nearly all the dust grains in the vicinity of the source would be destroyed by peak light. 

However, Metzger \& Perley \cite{2023ApJ...944...74M} pointed out that, during the rise of the light curve, dust grains at smaller radii can temporarily survive and radiate in the NIR. There is a critical luminosity $L_{\rm thin}$ at which the corresponding sublimation radius $r_{\rm sub}(L_{\rm thin}) = r_{\rm thin}$ (eq. \ref{eq:r_thin}), and we obtain
\begin{equation}\label{eq:L_thin}
    L_{\rm thin} \simeq \lrb{{1\over2} \tau_0^{1\over k-1}}^2 \times10^{43}\mr{\, erg\,s^{-1}} = 2.5\tau_0^{2\over k-1}\times10^{42}\mr{\,erg\,s^{-1}}.
\end{equation}

The early rise of the UV light curves of most LFBOTs has so far not been well constrained by observations. As an illustrative  example, Metzger \& Perley \cite{2023ApJ...944...74M} considered a quadratic power-law rise
\begin{equation}\label{eq:Lt_Lpk_tpk}
    L(t) = (t/t_{\rm pk})^2 L_{\rm pk},
\end{equation}
which is expected from a uniformly expanding spherical photosphere with a constant color temperature (although this may not be realistic). For this light curve model, the time the source spends near a given luminosity $L$ is roughly given by
\begin{equation}
    \Delta t_{L} \simeq {L\over \d L/\d t} = {t_{\rm pk}\over 2} \lrb{L/L_{\rm pk}}^{1/2}.
\end{equation}
The amount of energy reprocessed into the IR by surviving dust grains near radius $r\leq r_{\rm thin}$ comes from the early part of the rising light curve with $L\leq L_{\rm thin}$ such that $r\simeq r_{\rm sub}(L)$ and is given by
\begin{equation}
    E_{\rm IR}(L\leq L_{\rm thin}) \simeq L \Delta t_{\rm L} \simeq {t_{\rm pk}\over 2 L_{\rm pk}^{1/2}} L^{3/2} \propto L^{3/2},
\end{equation}
and the duration of the corresponding IR echo is given by (in the limit $t_{\rm IR}(L)\gg \Delta t_{\rm L}$)
\begin{equation}
    t_{\rm IR}(L\leq L_{\rm thin})\simeq {2r_{\rm sub}(L)\over c} \propto L^{1/2}.
\end{equation}
The corresponding IR luminosity is
\begin{equation}
    L_{\rm IR}(L\leq L_{\rm thin}) \simeq {E_{\rm IR}\over t_{\rm IR}} = {t_{\rm pk} c\over 4 L_{\rm pk}^{1/2} r_{\rm sub}(L)} L^{3/2} \simeq 3\times10^{41}\mr{\,erg\,s^{-1}}\, {t_{\rm pk}\over 3\rm\,d} L_{\rm pk,44}^{-1/2} L_{43}.
\end{equation}
The fact that $L_{\rm IR}\propto L$ means that the contribution to the IR luminosity from the much earlier parts of the light curve with $L \ll L_{\rm thin}$ is subdominant. Thus, a characteristic IR luminosity can be obtained by taking $L=L_{\rm thin}$ (eq. \ref{eq:L_thin}),
\begin{equation}
    L_{\rm IR}(L_{\rm thin}) \simeq 8\times10^{40}\mr{\,erg\,s^{-1}}\, {t_{\rm pk}\over 3\mr{\,day}} L_{\rm pk,44}^{-1/2} \tau_0^{2\over k-1},
\end{equation}
and this IR emission comes from the dust grains near radius $r_{\rm thin}$, and the corresponding duration of the dust echo is
\begin{equation}\label{eq:tIR_dust_echo}
    t_{\rm IR}(L_{\rm thin}) \simeq 8\tau_0^{1\over k-1}\mr{\,d}.
\end{equation}
For a spherical dusty CSM, the IR light curve from the emission near radius $r_{\rm thin}$ is nearly flat as it is dictated by the well-known flat response function of a spherical shell.

On the other hand, the amount of reprocessed energy from dust grains near radii $r\gtrsim r_{\rm thin}$ comes from the brighter portion of the rising light curve with $L\gtrsim L_{\rm thin}$. For $L \lesssim L_{\rm peak}$ and considering that the dust layer near radius $r$ is optically thin to UV photons with $\tau(r)<1$, we find $$E_{\rm IR}(L_{\rm thin} < L < L_{\rm pk})\simeq L \Delta t_{\rm L} \tau(r) \propto L^{3/2} (r_{\rm sub}(L))^{1-k}\propto L^{2-k/2}, $$ and $$t_{\rm IR}(L_{\rm thin} < L < L_{\rm pk}) \simeq {2r_{\rm sub}(L)/c} \propto L^{1/2}.$$ Thus, the contribution to the IR luminosity is given by $$L_{\rm IR}(L_{\rm thin} < L < L_{\rm pk}) \simeq E_{\rm IR}/t_{\rm IR}\propto L^{(3-k)/2}.$$ Therefore, the contribution to the IR luminosity from the dust grains near $r\sim r_{\rm sub}(L_{\rm pk})$ is given by
\begin{equation}
    L_{\rm IR}(L_{\rm pk}) \simeq L_{\rm IR}(L_{\rm thin})\, (L_{\rm pk}/L_{\rm thin})^{{(3-k)/2}}.
\end{equation}
For $1<k<3$, the factor of $(L_{\rm pk}/L_{\rm thin})^{{(3-k)/2}}$ makes the IR echo brighter than that given by $L_{\rm IR}(L_{\rm thin})$. The corresponding IR light curve will be longer lived with a duration
\begin{equation}
    t_{\rm IR}(L_{\rm pk}) \simeq {2r_{\rm sub}(L_{\rm pk})\over c} \simeq 50\mr{\,d}\, L_{\rm pk,44}^{1/2}.
\end{equation}
For a relatively shallow CSM density profile with $1 < k < 3$ (e.g., the case of a wind with constant mass-loss rate and velocity corresponds to $k=2$), we find that most of the IR luminosity comes from radius $\sim\! r_{\rm sub}(L_{\rm pk})$, so we expect to see a nearly flat IR light curve with luminosity $L_{\rm IR}(L_{\rm pk})$ lasting for a duration of $t_{\rm IR}(L_{\rm pk})$. However, this simplified spherically symmetric model is not immediately consistent with observations of AT2018cow, which shows an NIR excess with $L_{\rm IR}$ declining from $\sim\!10^{42}\rm\, erg\,s^{-1}$ in the first $\sim\!3\,$d to $\simeq 5\times 10^{40}\rm\, erg\,s^{-1}$ at $t\sim 50\,$d \cite{2023ApJ...944...74M} --- roughly following a power-law of $L_{\rm IR, obs}\propto t^{-1}$. It should be noted that the observationally inferred $L_{\rm IR}$ in the first few days may have large uncertainties, because it is non-trivial to separate the energetically subdominant IR component ($<1\%$ of the bolometric luminosity) from the ``photospheric emission'', which may not take the form of a blackbody spectrum if the photospheric radius is a function of frequency (as is the case for an ionized wind that is optically thick to free-free emission \cite{2025ApJ...991..180C}).

It is possible to modify the base model by considering a non-spherical CSM density profile (e.g., an equatorial disk geometry \cite{2023ApJ...944...74M, 2025ApJ...989...27T}) and the corresponding NIR light curve may be in better agreement with observations. In fact, the CSM density profile must be non-spherical so as to avoid early-time reddening by the dust along the LOS --- meaning that $t_{\rm IR}(L_{\rm thin})$ (eq. \ref{eq:tIR_dust_echo}) along the LOS is much shorter than that along the directions where most of the IR emission is produced. 

If the CSM density profile is steeper ($k>3$), then the brightest portion of the IR light curve comes from radius $\sim\!r_{\rm thin}$ and it produces a plateau lasting for $t_{\rm IR}(L_{\rm thin})$. At later time $t_{\rm IR}(L_{\rm thin})<t < t_{\rm IR}(L_{\rm pk})$, the IR light curve declines as a power-law $L_{\rm IR}\propto L^{(3-k)/2} \propto t_{\rm IR}^{3-k}$ (using $t_{\rm IR}(L)\propto L^{1/2}$). This scenario may be consistent with the IR behavior of AT2018cow if $\tau_0\lesssim 1$ so the IR plateau only lasts for $t_{\rm IR}(L_{\rm thin})\sim\,$a few days, but the corresponding plateau luminosity would be of the order $L_{\rm IR}(L_{\rm thin})\sim 10^{41}\rm\, erg\,s^{-1}$ --- perhaps dust echo only contributes to a fraction of the early-time IR emission. At $t > t_{\rm IR}(L_{\rm thin})$, this model can then reproduce the $L_{\rm IR, obs}\propto t^{-1}$ behavior if $k\simeq 4$. 

There are three complementary ways of further testing the dust echo model for the observed NIR excess.

The first one is to carefully study the NIR variability, as motivated by the fact that the light-crossing time corresponding to $r_{\rm sub}(L)$ for a given luminosity is always much greater than the intrinsic variability timescale of the source $\Delta t_{\rm L}$. In our simplified bolometric light curve model of $L(t)\propto t^2$ (eq. \ref{eq:Lt_Lpk_tpk}), the ratio between the two is given by
\begin{equation}
    {t_{\rm IR}(L)\over \Delta t_{\rm L}} \simeq 30\, \lrb{t_{\rm pk}\over 3\mr{\,days}}^{-1} L_{\rm pk,44}^{1/2},
\end{equation}
and we expect ${t_{\rm IR}(L)/\Delta t_{\rm L}}\gg 1$ even in other possible bolometric light curve models. This means that the intrinsic source variability is largely removed by the light-travel effect, and hence the variability timescale for the NIR light curve should be comparable to the time since the explosion, i.e. $t_{\rm var}\sim t$. This prediction seems to be mildly inconsistent with the observed variability timescale of 2--3 d in AT2018cow even at later time $t\sim 20\rm\, d$ \cite{2019MNRAS.484.1031P}, but a detailed statistical analysis has not yet been carried out. For instance, one can quantify the variability timescale of the \textit{NIR excess} by subtracting the (potentially model-dependent) ``photospheric'' contribution from the total NIR flux. Another possibility is to cross-correlate the NIR light curve with the X-ray or optical light curve to identify or rule out correlations.

The second way of testing the dust echo model is to carry out mid-IR observations for future LFBOTs with e.g., \textit{James Webb Space Telescope}. Depending on the dust composition, the dust echo spectrum may show spectral features (e.g., the silicate feature near 10$\rm\,\mu m$) that clearly distinguish the dust echo from gas emission and can be detected either through photometry or spectroscopy (see \cite{2025ApJ...988L..48M} for the mid-IR spectra of dust echoes from TDEs).

The third way is to carefully look for disappearing dust reddening in the early, rising part of the UV/optical lightcurve. If some LFBOTs are seen at viewing angles with high dust columns (e.g., for an equatorially concentrated CSM density profile), the early UV/optical lightcurve will show significant dust reddening when $L(t)\lesssim L_{\rm thin}$ much before the peak time, see the numerical simulations by Tuna et al. \cite{2025ApJ...989...27T}.

\subsubsection{NIR excess from free-free emission?}\leavevmode\\

\noindent
An alternative scenario for the NIR excess in LFBOTs is free-free emission from ionized gas either in the CSM or the ejecta/wind launched from near the center of explosion. In detail, the system may either be optically thin or thick to free-free absorption in the NIR. In this section, we discuss these two possibilities.

For a fully ionized gas whose composition is dominated by He (the results are similar for a H-dominated composition), the free-free opacity is given by (taking Gaunt factor $\simeq 1$ for simplicity)
\begin{equation}
    \kappa_{\rm ff,\nu} \simeq 1.06\times10^{-5}\mr{\,cm^2\,g^{-1}} \nu_{14.5}^{-2} n_{\rm e,11}T_6^{-3/2},
\end{equation}
where $n_{\rm e} = \rho /(\mu_{\rm e}\mproton) = 10^{11} n_{\rm e,11}\rm\,cm^{-3}$ is the free electron density, $T = 10^6T_6\rm\, K$ is the electron temperature, and we take $\mu_{\rm e} = 2$ for He-dominated composition. Our fiducial electron temperature is appropriate for X-ray ionized gas, whereas if the ionization rate is dominated by photons in the extreme-UV band, the electron temperature will likely be lower $T\sim 10^5\rm\, K$. Our fiducial frequency $\nu = 10^{14.5}\rm\, Hz$ corresponds to a wavelength of $\approx 1\rm\, \mu m$ in the NIR. For our fiducial parameters, the free-free opacity is much smaller than the Thomson scattering opacity $\kappa_{\rm s} \approx 0.2\, \rm cm^2\,g^{-1}$, so the gas is scattering dominated.

For the system to be effectively optically thick at frequency $\nu$ near radius $r$, we require the effective optical depth to be of order unity, i.e.,
\begin{equation}
    \tau_{\rm eff}(\nu)\simeq \sqrt{\tau_{\rm ff,\nu}\tau_{\rm s}} \sim \sqrt{\kappa_{\rm ff,\nu}\kappa_{\rm s}} {M_r\over 4\pi r^2} \sim 0.16 \lrb{M_r\over M_\odot}^{3/2} \nu_{14.5}^{-1} r_{15}^{-7/2} T_6^{-3/4} \sim 1,
\end{equation}
where $M_r\sim 4\pi \rho r^3$ is the mass within a logarithmic radial interval near radius $r$ (up to a factor of order unity). We express the free electron number density in terms of $M_r$ and then obtain an estimate of the ionized gas mass for $\tau_{\rm eff} = 1$,
\begin{equation}
    M_r(\tau_{\rm eff}=1) \sim 3.5\,M_\odot\, r_{15}^{7/3} \nu_{14.5}^{2/3} T_6^{1/2}.
\end{equation}
Comparing $M_r(\tau_{\rm eff}=1)$ with the maximum mass of the X-ray ionized gas (eq. \ref{eq:Mionmax}), we find that it may be difficult for the X-ray photons to ionize their way through such a large mass unless $L_{\rm ion}\gtrsim 10^{46}\rm\, erg\,s^{-1}$.

Another issue for the optically thick case is that it substantially overpredicts the NIR luminosity. For an emitting gas that is (at least marginally) effectively optically thin with $\tau_{\rm eff}\leq 1$, the free-free spectral luminosity tracks the mass of the ionized gas $M_r$ near radius $r$, and is given by
\begin{equation}\label{eq:NIRfreefree_nuLnu_general}
    \nu L_\nu (\tau_{\rm eff}\leq 1) \sim 4\pi \nu j_\nu 4\pi r^3
    \sim 1.2\times10^{42}\mr{\,erg\,s^{-1}}\, \lrb{M_r\over M_\odot}^{2} r_{15}^{-3} \nu_{14.5} T_6^{-1/2},
\end{equation}
where we have taken an emitting volume of $4\pi r^3$ and used Kirchhoff's law for the free-free emissivity $j_\nu = \rho \kappa_{\rm ff,\nu} B_\nu(T)$ and RJ limit ($h\nu\ll \kB T$) for the Planck function $B_\nu(T) \approx 2\nu^2 \kB T/c^2$. In the optically thick case, we plug in $M_{r}(\tau_{\rm eff}=1)$, and then obtain the corresponding spectral luminosity
\begin{equation}\label{eq:NIRfreefree_nuLnu_taueff1}
    \nu L_\nu(\tau_{\rm eff}=1) \sim 1.4\times10^{43}\mr{\,erg\,s^{-1}}\, r_{15}^{5/3} \nu_{14.5}^{7/3} T_6^{1/2}.
\end{equation}
We find that the predicted luminosity in the effectively optically thick case is much higher than observed $L_{\rm NIR}\lesssim 10^{42}\rm\, erg\,s^{-1}$ in AT2018cow, unless the emitting radius is much smaller than the optical photospheric radius of $R_{\rm ph}\sim 10^{15}\rm\, cm$ near peak luminosity (which requires a more complicated geometry). Additionally, the declining NIR luminosity $L_{\rm IR}\propto t^{-1}$ in AT2018cow \cite{2023ApJ...944...74M} would require effective emitting radius (where $\tau_{\rm eff}=1$) to shrink over time, roughly as $r\propto t^{-3/5}$.

Chen \& Shen (2025) \cite{2025ApJ...991..180C} considered the optically thick case in the picture of a continuous dense wind with time-dependent mass-loss rate $\dot{M}_{\rm w}$. Based on detailed SED fitting for AT2018cow, Chen \& Shen (2025) inferred a total outflow mass of $M_{\rm tot} = \int_0^\infty \dot{M}_{\rm w} \d t\simeq 5.7\pm 0.4M_\odot$ --- this is physically plausible for massive star progenitors but the associated radiative diffusion timescale will likely be much longer than the peak duration of LFBOTs (see eq. \ref{eq:diffusion_time_one_snapshot}). Moreover, their model also requires a total wind kinetic energy of the order $10^{53}\rm\, erg$, which may be difficult to achieve from disk accretion onto a stellar-mass compact object (see eq. \ref{eq:total_disk_mass_from_energy}).

Additionally, the optically thick case also faces another serious difficulty in fitting the late-time NIR emission from AT2018cow at $t\gtrsim 20\rm\, d$, which shows a nearly flat SED $F_\nu\propto \nu^{\approx 0}$. In general, such a flat SED is not expected from a quasi-steady wind, which has a density profile of $\rho\propto r^{-2}$ (as long as the dynamical timescale $t_{\rm dy}\sim r/v$ is much shorter than the time since explosion). For a density profile of $\rho\propto r^{-2}$, the predicted NIR SED is $F_\nu\propto \nu^\alpha$ and $0.4\lesssim \alpha\lesssim 0.6$ depending on whether electron scattering dominates opacity in the NIR (see \cite{2020MNRAS.492..686L}).

The above discrepancies motivate us to consider effectively optically thin ($\tau_{\rm eff} < 1$) free-free emission from the ionized regions of either the expanding ejecta or the CSM beyond the ejecta. In this case, free-free emission arises only from the ionized portion of the gas.






In terms of ionization, there are two regimes depending on whether the entire gas column is ionized or has a significant neutral layer: density-bounded regime (fully ionized) and ionization-bounded regime (with a neutral column).

In the density-bounded regime, the free-free luminosity strongly depends on the density profile and is given by eq. (\ref{eq:NIRfreefree_nuLnu_general}) --- the observed NIR luminosity $L_{\rm IR}\sim 10^{41}$ to $10^{42}\rm\, erg\,s^{-1}$ requires an ionized gas mass of the order $M_r\sim0.3$--$1M_\odot$ near $r\sim 10^{15}\rm\, cm$.

This ionized mass should be compared with the maximum mass ionized by the embedded X-ray source (eq. \ref{eq:Mionmax})
\begin{equation}
    M_{\rm ion,max, He} \simeq 1.0 M_\odot\,
    \lrsb{A_{\rm He}/4\over X_{\rm He} \lrb{\lara{\epsilon_{\rm ion, He}}/100\mr{eV}}}^{1/2}
    \lrb{Z_{\rm He}/2}^{-1.3} L_{\rm ion,44}^{1/2} r_{15}^{3/2} T_6^{0.4},
\end{equation}
where we have considered He ionization which has a lower average ionization cost than oxygen, $\lara{\epsilon_{\rm ion, He}} \sim 100\rm\, eV$ (slightly larger than the ionization threshold of 54.4 eV). We see that this required ionized mass to explain the NIR emission is comparable to the maximum ionized mass $M_{\rm ion,max, He}$ near $r\sim 10^{15}\rm\, cm$. This suggests that the system may be near the boundary between density- and ionization-bounded regimes.


If the system is in the ionization-bounded regime, we expect the free-free luminosity to track the luminosity of the ionizing photons, which is insensitive to the gas density profile. This is because free-free emission comes from 2-body encounters between free electrons and ions and the corresponding luminosity is proportional to the product of the ionized gas mass $M_{\rm ion, He}$ and the electron density $n_{\rm e}$. The same scaling applies to the recombination rate, which is proportional to the ionizing luminosity in the ionization-bounded regime. Following eq. (\ref{eq:Mion}), we can estimate the ionized mass
\begin{equation}
    M_{\rm ion, He} \simeq 0.5 M_\odot\,
    {A_{\rm He}/4\over X_{\rm He} \lrb{\lara{\epsilon_{\rm ion, He}}/100\mr{eV}}}
    \lrb{Z_{\rm He}/2}^{-2.6} n_{\rm e,11}^{-1} L_{\rm ion,44}T_6^{0.8},
\end{equation}
Using the mass emissivity $j_\nu/\rho = \kappa_{\rm ff,\nu} B_\nu$ (and we take the Rayleigh-Jeans limit for the Planck function $B_\nu$), we can then estimate the free-free specific luminosity in the \textit{ionization-bounded regime}
\begin{equation}
    \nu L_\nu (\tau_{\rm eff}< 1) = 4\pi \nu B_\nu \kappa_{\rm ff,\nu} M_{\rm ion, He} \simeq 1.4\times10^{42}\mr{\,erg\,s^{-1}}\, \lrb{\lara{\epsilon_{\rm ion, He}}/100\mr{\,eV}}^{-1} L_{\rm ion,44} \nu_{14.5} T_6^{0.3},
\end{equation}
where we have taken a He mass fraction of $X_{\rm He}\simeq 1$, a fiducial average cost of $\lara{\epsilon_{\rm ion, He}}\sim 100\rm\, eV$ per ionization. The electron temperature in the He-ionization zone may be $T\sim 10^5\rm\, K$ (instead of $10^6\rm\, K$ in the oxygen-ionization zone), but our result only depends weakly on the electron temperature.
We find that the free-free spectral luminosity $\nu L_\nu$ at $\nu=10^{14.5}\rm\, Hz$ (or $\lambda \approx 1\rm\, \mu m$) is of the order $1\%$ of the ionizing luminosity $L_{\rm ion}$, provided that the system is in the ionization-bounded regime.

If the system is indeed in the ionization-bounded regime, an important prediction is that the NIR emission may fluctuate rapidly together with the rapidly variable ionization luminosity $L_{\rm ion}$ (X-rays) --- this may be used to differentiate from the dust echo model (which predicts a variability timescale of the order $10\rm\, d$, see eq. \ref{eq:tIR_dust_echo}). However, if the system is in the density-bounded regime, the NIR variability will be suppressed, as the gas is nearly entirely ionized. 






A testable prediction in the free-free emission model is that, in the effectively optically thin case, the spectrum should be nearly flat $L_\nu\propto \nu^{\approx 0}$ in the optical and NIR bands, which seems to be in good agreement with the NIR SED at $t\gtrsim 20\rm\, d$; whereas at longer wavelengths the spectrum should turn downward as the gas becomes effectively optically thick. For a one-zone source with a fixed projected area, the asymptotic spectrum is $L_\nu\propto \nu^2$; a stratified wind generally produces a shallower slope because its effective emitting radius depends on frequency. At earlier time $t\lesssim 20\rm\, d$, the free-free spectrum depends on the transition between the density- and ionization-bounded regimes as well as the effective optical depth at different wavelengths. A more careful analysis is required.

We conclude that optically thin free-free emission is a plausible model for the excess emission in the NIR bands.




\subsubsection{Late-time UV plateau in AT2018cow: a long-lived accretion disk}\label{sec:theory-late-accretion}\leavevmode\\

\noindent
At very late times ($t\gtrsim 1\rm\, yr$), the UV/optical light curves of AT2018cow showed a plateau lasting for at least $5\rm\, yrs$ with slow evolution (fading). The SED is blackbody-like with a hot temperature $T_{\rm BB}\gtrsim 2\times10^4\mr{\,K}$ and the optical bands are near the Rayleigh-Jeans (RJ) tail with $\nu L_\nu\simeq 4\times 10^{38}\mr{\,erg\,s^{-1}}$ near $\nu\simeq 5.6 \times 10^{14}\rm\, Hz$ (HST F555W filter). This indicates an optically thick emitter with a characteristic emitting radius
\begin{equation}\label{eq:RBB_UV_plateau}
    R_{\rm BB} = \lrb{L_\nu c^2\over 8\pi^2\nu^2 \kB T_{\rm BB}}^{1/2} = 44 R_\odot \lrb{\nu L_\nu\over 4\times 10^{38}\mr{\,erg\,s^{-1}}}^{1/2} \lrb{\nu\over 5.6\times 10^{14}\mr{\,Hz}}^{-3/2} \lrb{T_{\rm BB}\over 2\times10^4\rm\, K}^{-1/2},
\end{equation}
where we have assumed an isotropic blackbody emitter with surface area $4\pi R_{\rm BB}^2$ and surface flux density of $\pi B_\nu \approx 2\pi \nu^2\kB T_{\rm BB}/c^2$ in the RJ limit. Detailed geometrical considerations will make the inferred size of the emitter differ from $R_{\rm BB}$ by a factor of order unity.

The above estimate indicates that the UV/optical plateau is produced by a relatively compact, optically thick gas distribution that rules out any astrophysically plausible scenario of CSM interaction (which should occur at much larger radii). Recent interpretations \cite{2023ApJ...955...43C, 2024ApJ...963L..24M, 2024A&A...691A.329C, 2025MNRAS.544L.108I} of the UV/optical plateau attribute the emission to a geometrically thin accretion disk with a finite outer radius $r_{\rm d}$ which is of the order $R_{\rm BB}$ --- the detailed geometrical factor depends on the temperature profile of the disk surface and the viewing angle.

In this picture, the peak UV/optical emission in the first days to months would then correspond to the early-time super-Eddington phase of the disk evolution, and hence LFBOTs are energetically accretion powered \cite{2024ApJ...963L..24M}. It is interesting to note that the late-time UV/optical plateau is phenomenologically similar to that seen in TDEs by supermassive BHs where the favored interpretation is also the emission from the outer regions of a geometrically thin disk \cite{1990ApJ...351...38C, 2024MNRAS.527.2452M}. 

Time-dependent disk models have been constructed by many authors to fit the late-time UV/optical plateau data \cite{2023RNAAS...7..126M, 2025MNRAS.544L.108I, 2026OJAp....965434W, 2026A&A...706A.327C}. In the following, we review the analytical arguments by Lu \& Piro (2026) \cite{2026arXiv260715464L} with conclusions similar to those of the more detailed disk evolution modeling by Winter-Granic \& Quataert (2026) \cite{2026OJAp....965434W}.

A geometrically thin but optically thick accretion disk emits a multicolor blackbody powered by accretion. For a constant (radius-independent) mass accretion rate $\dot{M}$ at $r\lesssim r_{\rm d}$ ($r_{\rm d}$ being the outer disk radius), the effective temperature profile is given by
\begin{equation}
    \sigma_{\rm SB} T_{\rm eff}^4(r) = {3GM\dot{M}\over 8\pi r^3}.
\end{equation}
The flux density from the disk surface is given by $F_\nu(r) = \pi B_\nu(T_{\rm eff})$ (ignoring limb darkening), so the total emission spectrum is given by
\begin{equation}
    L_\nu = 2\int_{r_{\rm in}}^{r_{\rm d}}\pi B_\nu(T_{\rm eff}) 2\pi r \d r,
\end{equation}
where $r_{\rm in}$ is the inner edge of the geometrically thin part of the disk (unimportant for the UV/optical plateau). The RJ-like SED of the UV/optical plateau in AT2018cow motivates us to consider the RJ limit $B_\nu\approx 2\nu^2\kB T_{\rm eff}/c^2$ for all disk radii and obtain
\begin{equation}\label{eq:outer_disk_emission_accretion_only_RJ_limit}
\begin{split}
    \nu L_\nu &\simeq \nu^3 \lrb{3GM\dot{M}\over 8\pi \sigma_{\rm SB}}^{1/4} {32\pi^2\kB \over 5c^2} r_{\rm d}^{5/4}\\
    &= 4.0\times10^{38}\mr{\,erg\,s^{-1}}\, \lrb{\nu\over 5.6\times 10^{14}\mr{\,Hz}}^3\lrb{M\over 10M_\odot}^{1/4}\lrb{\dot{M}\over 0.1M_\odot\,\mr{yr}^{-1}}^{1/4} \lrb{r_{\rm d}\over 40R_\odot}^{5/4},
\end{split}
\end{equation}
where we have taken the limit of $r_{\rm in}\ll r_{\rm d}$ (the exact inner radius only makes a negligible difference). We find that the emission at a given frequency in the RJ limit depends strongly on the outer disk radius but weakly on the compact object mass and accretion rate.

Another constraint on the outer disk radius comes from the requirement that the effective temperature near the outer disk is sufficiently high such that the optical bands are in the RJ limit, and we require
\begin{equation}
    T_{\rm eff}(r_{\rm d}) > {h\nu \over \kB} \ \ \Rightarrow \ \ {3GM\dot{M}\over 8\pi \sigma_{\rm SB} r_{\rm d}^3} > \lrb{h\nu\over \kB}^4,
\end{equation}
and this leads to
\begin{equation}\label{eq:RJ_condition}
    r_{\rm d} < 17R_\odot \lrb{\nu\over 5.6\times 10^{14}\mr{\,Hz}}^{-4/3} \lrb{M\over 10M_\odot}^{1/3} \lrb{\dot{M}\over 0.1\,\msunyr}^{1/3}.
\end{equation}

A final constraint comes from the slow decay of the UV/optical plateau, which implies that the outer disk is evolving on a viscous timescale of the order $t_{\rm vis}\sim 10\rm\, yr$. We then express the disk mass in terms of the accretion rate $\dot{M}$ and the viscous timescale of the outer disk
\begin{equation}
    M_{\rm d}\simeq \dot{M} t_{\rm vis}.
\end{equation}
The viscous timescale is physically connected to the disk pressure scale-height $H$ by
\begin{equation}
    t_{\rm vis}\simeq {1\over \alpha} \lrb{H\over r_{\rm d}}^{-2} \sqrt{r_{\rm d}^3\over GM} \ \ \Rightarrow \ \
    {H\over r_{\rm d}} \simeq {\lrb{r_{\rm d}^3/GM}^{1/4}\over (\alpha t_{\rm vis})^{1/2}}.
\end{equation}
The pressure scale-height is related to the (isothermal) sound speed near the disk mid-plane (based on hydrostatic equilibrium in the vertical direction)
\begin{equation}
    c_{\rm s} \equiv \sqrt{P\over \rho} = \sqrt{GM\over r_{\rm d}} {H\over r_{\rm d}} \simeq {\lrb{GMr_{\rm d}}^{1/4}\over (\alpha t_{\rm vis})^{1/2}}.
\end{equation}
This should be compared with the (isothermal) radiation-pressure sound speed near the disk midplane
\begin{equation}
    c_{\rm sr} \equiv \sqrt{P_{\rm rad}\over \rho} \simeq \sqrt{aT^4/3\over \Sigma(r_{\rm d})/(2H)},
\end{equation}
and the gas-pressure sound speed
\begin{equation}
    c_{\rm sg} \equiv \sqrt{P_{\rm gas}\over \rho} = \sqrt{\kB T\over \mu \mproton},
\end{equation}
where $\mu$ is the mean molecular weight, $T\simeq (3/4) \tau^{1/4}T_{\rm eff}(r_{\rm d})$ is the midplane temperature, $\tau$ is the optical depth of the outer disk
\begin{equation}
    \tau = {\kappa \Sigma(r_{\rm d})\over 2} = {3\kappa M_{\rm d}\over 8\pi r_{\rm d}^2} = 6.1\times10^6 {\kappa\over 0.2\mr{\,cm^2\,g^{-1}}} {M_{\rm d}\over M_\odot} \lrb{r_{\rm d}\over 40R_\odot}^{-2}.
\end{equation}
For outer disk densities and temperatures relevant for the UV/optical plateau, the Rosseland-mean opacity is dominated by Thomson scattering, so we take $\kappa\approx \kappa_{\rm s}=0.2\rm\,cm^2\,g^{-1}$ for a fully ionized He-dominated composition as a fiducial value. For the disk model to be self-consistent, we require that neither the radiation pressure nor gas pressure in the disk midplane should exceed the total pressure
\begin{equation}
    c_{\rm sr} < c_{\rm s}, \ \ c_{\rm sg} < c_{\rm s}.
\end{equation}
These two conditions lead to very stringent constraints on the disk radius
\begin{equation}\label{eq:rd_from_csr_cs}
    r_{\rm d} > 200\,R_\odot \left(\frac{\alpha}{0.1}\right)^{2/7} \left(\frac{\kappa}{0.2\, {\rm cm^2\,g^{-1}}}\right)^{4/7}\left(\frac{M_{\rm d}}{M_\odot}\right)^{4/7}\left(\frac{M}{10M_\odot}\right)^{1/7}\left(\frac{t_{\rm vis}}{10\,\rm yr}\right)^{-2/7},
\end{equation}
\begin{equation}\label{eq:rd_from_csg_cs}
    r_{\rm d} > 350R_\odot \left(\frac{\alpha}{0.1}\right)^{4/7} \left(\frac{\kappa}{0.2\, {\rm cm^2\,g^{-1}}}\right)^{1/7} \lrb{\mu\over 4/3}^{-4/7}\left(\frac{M_{\rm d}}{M_\odot}\right)^{2/7}\left(\frac{M}{10M_\odot}\right)^{-1/7}\left(\frac{t_{\rm vis}}{10\,\rm yr}\right)^{3/7},
\end{equation}
where we have taken $\mu=4/3$ as a fiducial value for fully ionized He-dominated composition. For solar composition ($\kappa\approx 0.34\rm\, cm^2\,g^{-1}$ and $\mu\approx 0.61$), the above constraints are slightly more stringent.
Combining both radiation- and gas-pressure contributions, the pressure self-consistency requirement is
\begin{equation}\label{eq:csr_csg_cs}
    c_{\rm sr}^2 + c_{\rm sg}^2 \leq c_{\rm s}^2,
\end{equation}
where the ``$<$'' sign allows for additional pressure contributions such as magnetic fields and turbulence \cite{2007MNRAS.375.1070B, 2023MNRAS.524.1269K}. Under our fiducial parameters, the lower limits for the outer disk radius from pressure self-consistency constraints show a strong tension with the compact disk size of $r_{\rm d}\sim 40R_\odot$ inferred earlier. Combining the constraints based on (\ref{eq:outer_disk_emission_accretion_only_RJ_limit}), (\ref{eq:RJ_condition}), (\ref{eq:csr_csg_cs}), one can see that the only possible way to remove the tension is to increase the compact object mass to $M\gg 10M_\odot$ while decreasing the viscosity parameter to $\alpha\ll 0.1$.




From the inequalities (\ref{eq:rd_from_csr_cs}) and (\ref{eq:rd_from_csg_cs}), it does not seem that the constraints depend strongly on the compact object mass $M$. However, since the disk outer radius of $r_{\rm d}\sim 40R_\odot$ is firmly constrained by observations (eq. \ref{eq:RBB_UV_plateau}), a higher compact object mass requires a lower disk mass, roughly as $M_{\rm d}\simeq M_\odot (M/10M_\odot)^{-1}$ (for a fixed viscous timescale $t_{\rm vis}\simeq 10\rm\, yr$), and hence the RHS of the inequality (\ref{eq:rd_from_csr_cs}) scales as $M^{-3/7}\alpha^{2/7}$. 

\begin{figure}[b]
    \centering
    \includegraphics[width=0.7\linewidth]{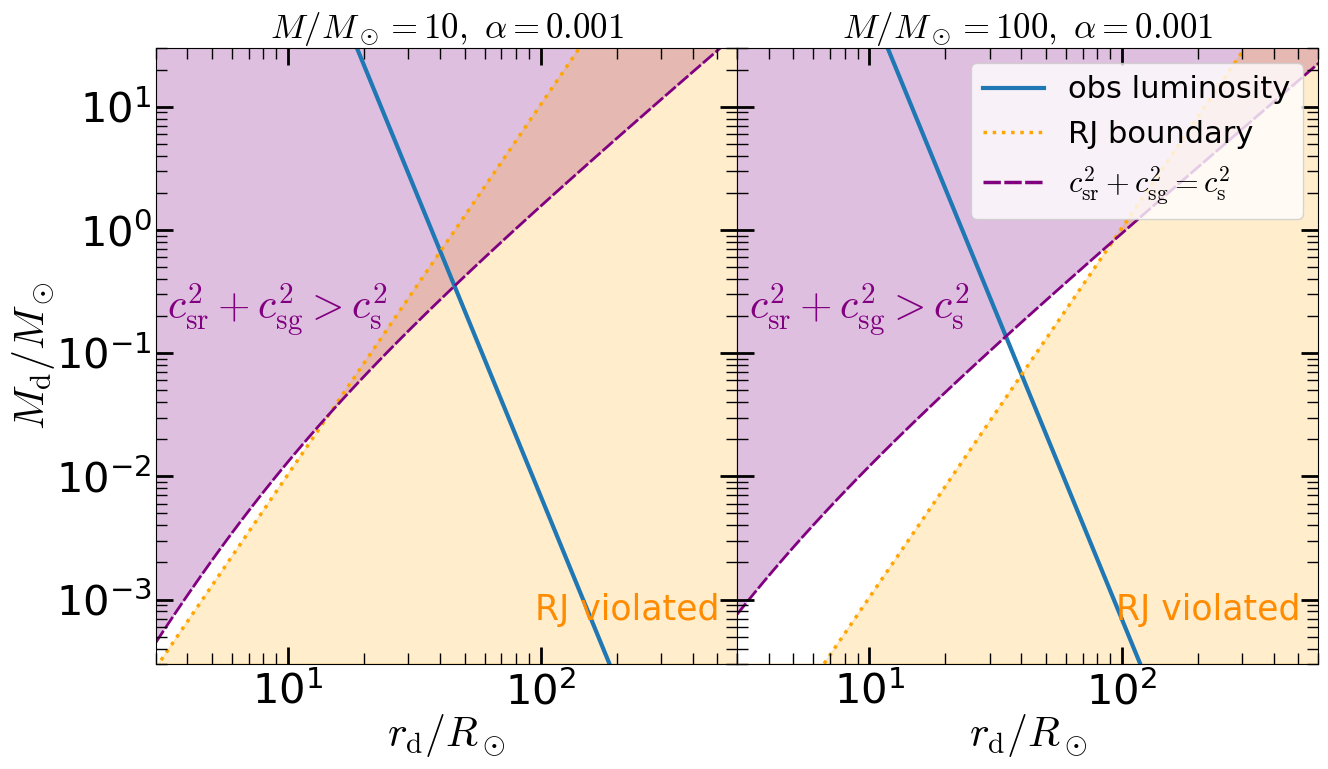}
        \caption{Constraints on disk accretion model for the UV/optical plateau in AT2018cow, reproduced from Lu \& Piro (2026; in submission) \cite{2026arXiv260715464L} with permission. The left panel shows the constraints for accretor mass $M=10M_\odot$ and the right panel is for $M=100M_\odot$. Each panel shows three different constraints based on observed luminosity of AT2018cow near $\nu=5.6\times10^{14}\rm\, Hz$ at $t=1453\rm\, d$ \cite{2023ApJ...955...43C, 2025MNRAS.544L.108I} (HST F555W, blue solid line), self-consistency of disk pressure ($c_{\rm sr}^2 + c_{\rm sg}^2 \leq c_{\rm s}^2$, purple-shaded region),  and the observing frequency $\nu$ being on the RJ limit (orange-shaded region). For both panels, the viscous timescale and viscosity parameter are taken as $t_{\rm vis} = 10\rm\, yr$ and $\alpha=0.001$. The physical solutions lie where the blue solid line passes through the unshaded (white) region.
        }
    \label{fig:Md_rd_acc}
\end{figure}

The joint constraints based on (\ref{eq:outer_disk_emission_accretion_only_RJ_limit}), (\ref{eq:RJ_condition}), and (\ref{eq:csr_csg_cs}) are shown in Figure~\ref{fig:Md_rd_acc}, where the right panel shows that physical solutions are allowed for $M=100M_\odot$ and $\alpha=0.001$. In fact, the above joint constraints require a minimum compact object mass
\begin{equation}
    M > M_{\rm min} \simeq 200 M_\odot \lrb{\alpha\over 0.01}^{2/3} \simeq 40 M_\odot \lrb{\alpha\over 0.001}^{2/3}.
\end{equation}
For physically motivated viscosity parameters under the magneto-rotational instability $0.01\lesssim \alpha\lesssim 0.1$ (e.g., \cite{2010ApJ...713...52D, 2011ApJ...738...84H, 2020MNRAS.494.3656L}), the standard geometrically thin, optically thick, viscously powered accretion disk model requires a compact object mass of the order $10^2M_\odot$ or higher. Note that this lower limit on the BH mass does not depend on the disk formation process (e.g., TDE or stellar core-collapse), as the constraints we are using are only based on the \textit{current} state of the disk. This mass constraint stays unchanged even if magnetic pressure dominates over gas/radiation pressure (e.g., \cite{2007MNRAS.375.1070B}), as our requirement of $c_{\rm sr}^2 + c_{\rm sg}^2 \leq c_{\rm s}^2$ is conservative.

The requirement of a relatively high BH mass ($\gtrsim 10^2M_\odot$ for $\alpha\gtrsim 0.01$) potentially implies a TDE by an intermediate-mass black hole (IMBH) in AT2018cow. 

Migliori et al. (2024) \cite{2024ApJ...963L..24M} presented an X-ray detection of AT2018cow around $t\simeq 1350\rm\, d$ after the explosion, and their X-ray flux measurement of $L_{\rm X}\simeq 5\times10^{38}\rm\, erg\,s^{-1}$ can be considered as an upper limit for the disk emission (considering potential contributions from other sources within the XMM-Newton angular resolution).
Winter-Granic \& Quataert (2026) \cite{2026OJAp....965434W} showed that the emission from the super-Eddington wind launched from the inner regions of the disk may violate the X-ray constraints for $M\gtrsim 10^2M_\odot$. For this reason, they prefer a more ordinary stellar-mass BH of $10\lesssim M\lesssim 10^2\,M_\odot$ but with an unconventional viscosity parameter $\alpha \sim 10^{-3}$. 

A potential solution proposed by Lu \& Piro (2026) \cite{2026arXiv260715464L} is that the UV/optical plateau may be produced by the disk wind emission or the outer disk irradiated by the wind emission. In this scenario, the luminosity from the disk wind is only mildly super-Eddington ($L\sim\,$a few--10$L_{\rm Edd}$) even when the accretion rate is highly super-Eddington. The UV/optical emission is not directly powered by viscous accretion in the outer regions of the disk, which may have a sub-Eddington power and this relaxes the constraints on the compact object mass based on $c_{\rm sr}^2 + c_{\rm sg}^2 \leq c_{\rm s}^2$ (as the disk midplane temperature may be much lower). In fact, \cite{2026arXiv260715464L} found that the disk wind from even an NS accretor (with $M\simeq 1.4M_\odot$) can potentially be consistent with the UV/optical plateau observations. A prediction from the disk wind picture is that free-free emission from the ionized wind may produce an excess in the near-IR and mid-IR as compared to the RJ extrapolation of the UV/optical SED. Such a prediction may be tested by future JWST observations.

Finally, an alternative way to power late-time emission is to decouple the late-time engine from the compact object formed at the center of the original explosion. Lazzati et al. (2024) \cite{2024ApJ...972L..17L} proposed that delayed emission in LFBOTs may arise from accretion onto a BH companion in a binary system. In this picture, the main LFBOT is produced by an explosion in a binary containing a pre-existing BH companion; after a delay set by the binary separation and ejecta velocity, a fraction of the ejecta is gravitationally captured by the companion and powers late-time emission. If the ejecta is clumpy, the accretion can be highly episodic, potentially explaining the rapid late-time optical flares of AT2022tsd \cite{2023Natur.623..927H}, while smoother or more extended accretion may instead produce a persistent late UV/X-ray source relevant to AT2018cow. This model naturally produces a delay between the main transient and the late activity. 
However, it does not by itself explain LFBOT observations near the peak luminosity, and it requires a BH companion with favorable separation, ejecta velocity, and ejecta clumpiness.



\subsubsection{Summary of physical requirements}\label{sec:summary_requirements}
\leavevmode\\

\noindent
Table~\ref{tab:physical_requirements} summarizes the principal
observationally inferred requirements that any successful global model
of LFBOTs should satisfy.

\begin{landscape}
\begin{table*}
\centering
\caption{
Summary of the physical requirements inferred from LFBOT observations.
``Confidence'' refers to the robustness of the inference, not to the quality
of the available observations. A successful global model does not need to
produce every component in exactly the same physical region, but it should
account for their coexistence without violating the mass, energy,
angular-momentum, and timescale constraints.
}
\label{tab:physical_requirements}
\small

\begin{tabularx}{\linewidth}{
    >{\raggedright\arraybackslash}p{0.15\linewidth}
    >{\raggedright\arraybackslash}p{0.19\linewidth}
    >{\raggedright\arraybackslash}X
    >{\raggedright\arraybackslash}p{0.23\linewidth}
}
\toprule
\textbf{Observable} &
\textbf{Characteristic scale} &
\textbf{Physical requirements} &
\textbf{Confidence; caveat} \\
\midrule

Peak UVOIR luminosity and duration &
$L_{\rm pk}\gtrsim 10^{44}\ {\rm erg\,s^{-1}}$,
$t_{1/2}\sim{\rm few\ d}$,
$E_{\rm rad}\sim10^{50}$--$10^{51}\ {\rm erg}$ &
Supply $E_{\rm eng}\gtrsim10^{50}\ {\rm erg}$ on a timescale of days
and efficiently convert the available energy into escaping radiation &
High \\

\midrule

Low radiating mass &
$M_{\rm ej}\lesssim0.5$--$1\,M_\odot$ for standard optical opacities &
Short diffusion time despite high radiated energy.
 ``Dark'' mass must not participate in observed diffusion
and thermalization &
Moderate--high; opacity, density profile, geometry \\

\midrule

Fast outflow &
$v\sim0.1c$ from the optical photosphere and radio expansion &
Generate a fast, energetic outflow through a compact-object engine &
High for existence of fast outflow; its mass and kinetic energy less secure \\

\midrule

Compact and variable X-ray source &
$t_{\rm var}\sim{\rm few\ d}$,
$R_{\rm X}\lesssim10^{15}\ {\rm cm}$; soft continuum plus an early
hard X-ray hump &
Produce X-rays near or beneath the optical photosphere and allow them
to escape through a scattering and ionized medium without erasing
the variability &
Moderate; $R_{\rm X}$ follows if variability traces the local dynamical
scale; emission mechanism remains debated \\

\midrule

X-ray reprocessing and ionization breakout &
Ionized layer with isotropic- equivalent mass
$\lesssim0.3\,M_\odot$ along escaping directions &
Provide escape channels for soft X-rays while retaining denser regions
that absorb and reprocess part of the engine or shock power into the
UVOIR bands &
Moderate; strongly geometry- and ionization-dependent \\

\midrule

Radio-emitting shock and outer CSM &
$R\sim10^{16}$--$10^{17}\ {\rm cm}$,
shock velocity $\sim0.1c$; CSM confined within a few
$\times10^{16}\ {\rm cm}$ &
Launch fast outflow that interacts with a dense, radially confined CSM
produced shortly ($\lesssim{\rm kyr}$) before the transient &
High for source radius; low to moderate for CSM density,
mass, velocity due to microphysics and geometric uncertainties \\

\midrule

Persistently blue optical continuum &
$T\sim{\rm few}\times10^4\ {\rm K}$ and a shrinking effective
photospheric radius after peak &
Maintain an optically thick, reprocessing photosphere, plausibly in a
continuous non-homologous outflow (ordinary SN ejecta would normally
cool and approach the nebular phase) &
High observationally; continuous-wind interpretation is model-dependent \\

\midrule

Delayed H and He line emission &
Onset at $t\sim15$--$30\ {\rm d}$,
$v_{\rm line}\sim3000$--$10^4\ {\rm km\,s^{-1}}$;
$M_{\rm H}\gtrsim{\rm few}\times10^{-3}\,M_\odot$ in the disk-wind model &
Supply intermediate-velocity H-rich gas and explain both the delayed
visibility and the asymmetric/shifted profiles &
High for line properties; emitting mass and location depend on
radiative transfer and geometry \\

\midrule

NIR excess &
$10^{41}$--$10^{42}\ {\rm erg\,s^{-1}}$ in events with adequate coverage &
Produce either a dust echo at large radii or free--free emission from
ionized CSM, ejecta, or disk wind, including the observed spectral
shape and variability &
Secure detection in several events; physical origin remains uncertain \\

\midrule

Late-time UV plateau &
For AT2018cow, $t\gtrsim1\ {\rm yr}$, Rayleigh--Jeans-like SED, and
$R_{\rm BB}\sim40\,R_\odot$ &
Leave a compact, optically thick, long-lived source, plausibly an
accretion disk, irradiated outer disk, or disk-wind photosphere &
High for AT2018cow only; connection to early engine and generality
across LFBOTs remain uncertain \\

\midrule

Environment and volumetric rate &
Young star-forming environments, sub-solar metallicity;
$\lesssim10^{-4}$ of core-collapse SN rate &
Identify an evolutionary channel consistent with young stellar
populations while explaining the rarity of LFBOTs &
Moderate; limited by small sample and observational selection effects \\

\bottomrule
\end{tabularx}

\end{table*}
\end{landscape}

\subsection{Proposed models}\label{sec:proposed_models}

In \S \ref{sec:basic_constraints}, we have discussed the constraints on the progenitor system based on individual components without considering the global picture. In this section, we review the proposed global models for the progenitor systems of LFBOTs. A summary of these global models is provided in Table \ref{tab:model_comparison}.

Before going into individual models, we briefly re-iterate the environmental constraints: observational studies of the host galaxy environments (e.g., offset distribution, star-formation history) suggest that they prefer star-forming galaxies with slightly sub-solar metallicities and high specific star-formation rates. In terms of metallicities, specific star-formation rates, and galactocentric offsets, the current population of LFBOTs lie between those of LGRBs/SLSNe-I and normal CCSNe \cite{2026ApJ..1006...75N}. We conclude that LFBOTs are most likely related to young massive stars. Their stellar evolution channel(s) likely involve binary interactions --- at lower metallicities, stars are more compact (enabling closer binary separations) and the core-envelope structures of the donor stars are such that they are more likely to survive the common-envelope evolution (e.g., \cite{2021A&A...645A..54K}).

However, not all proposed progenitor models for LFBOTs are based on massive stars, and in the non-massive star models (e.g., TDEs by intermediate-mass BHs) the environmental properties remain to be explained.

\subsubsection{Compact object-star mergers or micro-TDEs}\leavevmode\\

\begin{figure}[b]
    \centering
    \includegraphics[width=0.5\linewidth]{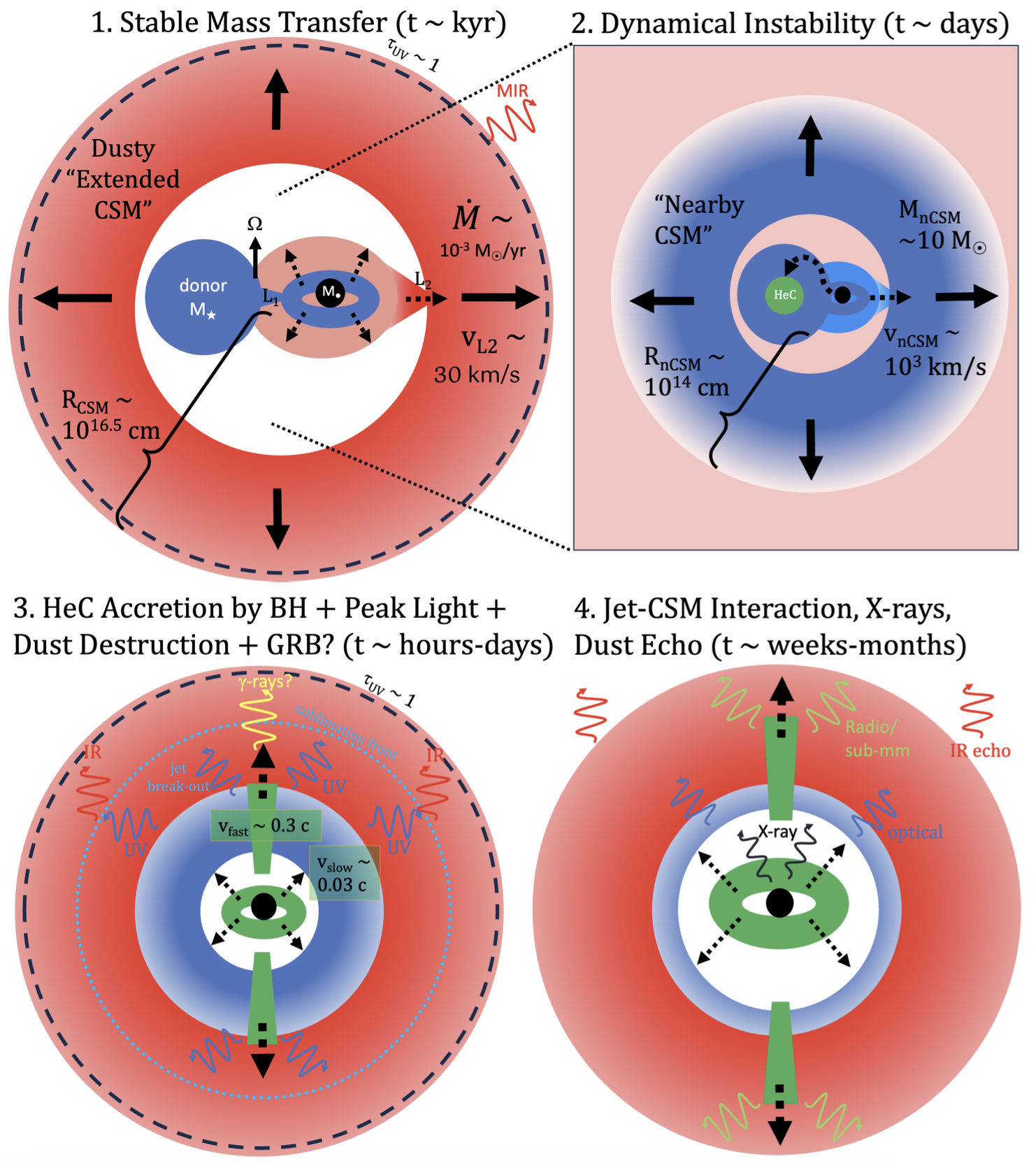}
        \caption{Model proposed by Klencki \& Metzger (2026) \cite{2026ApJ..1005....2K}. A binary system initially undergoes stable mass transfer from a donor star to a BH accretor. At some point, the mass transfer becomes unstable, which causes the BH to spiral into the donor star's remaining envelope and merge with its He core. The resulting accretion disk then produces powerful wind and likely a jet, which interact with the CSM launched during earlier stages of mass transfer and power X-ray and UVOIR emission. An IR echo is produced when the UV photons heat up the dust grains in the dense CSM. Figure reproduced from Klencki \& Metzger (2026) \cite{2026ApJ..1005....2K} with permission.
        }
    \label{fig:KM25_model}
\end{figure}

\noindent
Metzger (2022) \cite{2022ApJ...932...84M} proposed that LFBOTs may be produced by a merger between a He star and a stellar-mass BH. A conceptually similar model was proposed by Soker et al. (2019) \cite{2019MNRAS.484.4972S} that involves a merger between a NS and a massive giant star. In the more recent model by Klencki \& Metzger (2026) \cite{2026ApJ..1005....2K}, this scenario is made more specific by considering the cases where the Case-B mass transfer from a redgiant donor to a BH can transition from a dynamically stable phase to an unstable phase near the end. Their schematic picture is shown in Figure~\ref{fig:KM25_model}.

The long-lived ($t\sim \rm\, kyr$) dynamically stable mass-transfer phase produces an extended, aspherical, dusty, and H-rich CSM up to radii $r\sim 10^{16}$--$10^{17}\rm\, cm$. In this phase, the mass-transfer rate is sufficiently high such that a large fraction of the transferred mass is likely lost in an equatorially concentrated outflow through the L2 point of the binary \cite{2023MNRAS.519.1409L, 2025ApJ...990..172S, 2025arXiv251024127S}. On the other hand, the short-lived ($t\sim\rm\, days$) dynamically unstable mass-transfer or common-envelope (CE) phase produces ``nearby CSM'' up to radii of the order $10^{14}\rm\, cm$, and the ejected (still H-rich) stellar envelope is also likely to be aspherical as it is shaped by the binary's potential. Afterwards, if the BH fails to eject the entire envelope (in the so-called ``failed CE'' case), we expect a merger between the helium core (HeC) and the BH. The accretion disk launches powerful jets and outflows, which then interact with the nearby CSM ($r\lesssim 10^{14}\rm\, cm$) to produce the X-rays and UV/optical emission near the peak luminosity. Afterwards, the interactions between the jets and the extended CSM produce the bright radio/mm afterglow emission. The IR excess is produced by the dust echo. Finally, at late times $t\gtrsim \rm months$, the disk settles down to a geometrically thin one and the UV/optical plateau is produced by viscous accretion on timescales of years to decades.

\begin{figure}[b]
    \centering
    \includegraphics[width=0.65\linewidth]{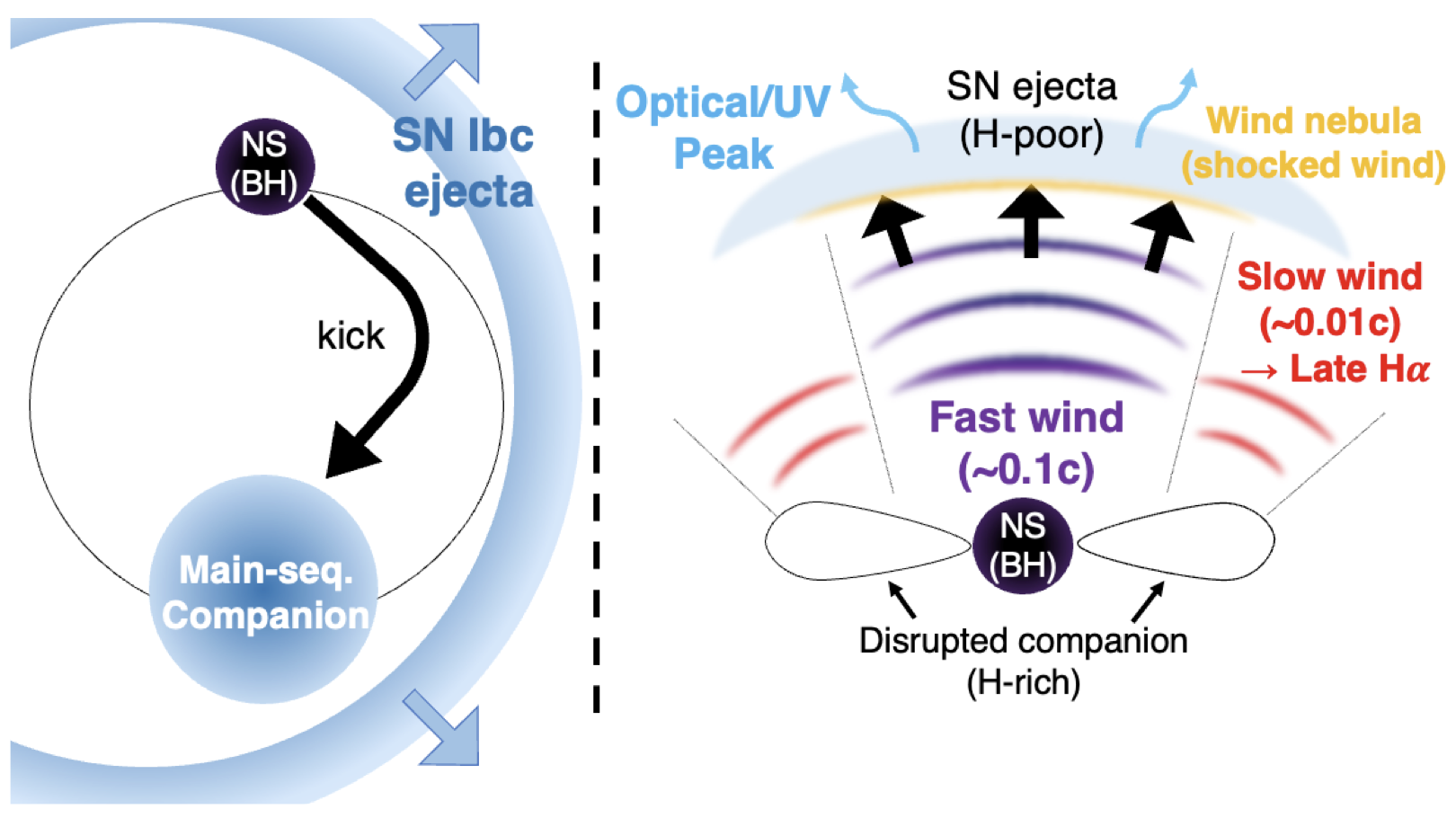}
        \caption{Model proposed by Tsuna \& Lu (2025) \cite{2025ApJ...986...84T}. Following a stripped-envelope supernova, the newly born compact object (NS or BH) receives a fortunate kick that brings it into a close encounter with a main-sequence companion star. Tidal disruption of the companion star leads to the formation of an accretion disk. The powerful disk wind then collides with the supernova ejecta launched earlier. The shock interactions then power X-ray and UVOIR emission. Figure reproduced from Tsuna \& Lu (2025) \cite{2025ApJ...986...84T} with permission.
        }
    \label{fig:TL25_model}
\end{figure}

Tsuna \& Lu (2025) \cite{2025ApJ...986...84T} proposed that LFBOTs may be produced by a micro-TDE following a fortunate SN natal-kick that sends the compact object to the vicinity of a main-sequence companion star --- this follows the earlier hydrodynamic study by Hirai \& Podsiadlowski (2022) \cite{2022MNRAS.517.4544H}. The schematic picture for this model is shown in Figure~\ref{fig:TL25_model}. A micro-TDE differs from the conventional TDE by supermassive BHs or IMBHs in their modest mass ratio $M/M_*\lesssim 10$, which means that the stellar debris after tidal disruption have relatively low-eccentricity orbits, so we expect rapid disk formation shortly after the TDE ($t\lesssim\rm 10\, hrs$ for the disruption of main-sequence stars) \cite{2016ApJ...823..113P, 2022ApJ...933..203K, 2024A&A...685A..45V, 2024ApJ...961..149X}. Subsequently, viscous accretion launches powerful outflows\footnote{It has been proposed that the emission from the escaping radiation carried by the disk outflows may be sufficiently bright to explain the peak luminosity of LFBOTs, provided that a large fraction of the disk gas reaches the BH --- for $\dot{M}\propto r^p$ with $p\simeq 0.2$ \cite{2021ApJ...911..104K, 2023MNRAS.524.6358K}. However, recent numerical simulations of radiative inefficient accretion flows favor $p\simeq 0.5$, which corresponds to a much lower accretion efficiency and hence fainter peak luminosity of $L_{\rm pk}\lesssim 10^{43}\rm\, erg\,s^{-1}$ than in the case of $p\simeq 0.2$. } that interact with the earlier SN ejecta, and the shocks then power the multi-wavelength emission near the peak luminosity. Part of the fast outflows that may break out from the ejecta's confinement will interact with the CSM produced by pre-SN binary evolution (e.g., \cite{2022ApJ...940L..27W}) and produce bright radio/mm emission. The IR excess is not directly addressed by the original model of Tsuna \& Lu (2025) \cite{2025ApJ...986...84T}, but it may be produced by the dust echo as proposed by Metzger \& Perley (2023) \cite{2023ApJ...944...74M}. At late time $t\gtrsim \rm months$, the disk becomes geometrically thin and may produce the UV/optical plateau by long-term accretion.

The above models based on the BH-star merger and micro-TDE scenarios fit well with the environmental constraints (as they are based on massive stars) and can in principle explain many features of LFBOTs (but not all clearly demonstrated yet), including the high peak luminosity, short peak duration, hot featureless early spectra as due to high photospheric temperatures, dense CSM and radio/mm emission, IR excess, Balmer lines, and late-time UV/optical plateau. Although these two models are both based on super-Eddington accretion, there are some important differences that may be used to differentiate between them.

First, in the picture of Klencki \& Metzger (2025), as their delayed BH+He-core merger scenario requires a BH-to-star mass ratio $M/M_* \sim 0.3$, for a $M=10M_\odot$ BH, the typical mass for the donor star is $M_*\sim 30M_\odot$. Thus, their model predicts that LFBOTs should be found in regions with the youngest stellar population ($\lesssim 4\rm\, Myr$), unless the binary system received a substantial kick associated with the BH formation --- a systemic kick of 30--$100\rm\, km/s$ may displace the system by a few $10^2\rm\, pc$ from the birth location. In the micro-TDE model by Tsuna \& Lu (2025) \cite{2025ApJ...986...84T}, the SN progenitor is a He star in a close orbit with a main-sequence companion. A low-mass ($\lesssim 5M_\odot$) He star is preferred, as its late-phase evolution may involve mass transfer at high rates ($\gtrsim 10^{-4}\,\msunyr$) and potentially a dense CSM via L2 mass loss. The longer lifetimes of the binaries in the micro-TDE scenario mean that LFBOTs should be found in regions with somewhat older stellar population, with ages of $10$--$30\rm\,Myr$ and with larger offsets from their birth location.

Second, the model of Klencki \& Metzger (2025) predicts a kyr-long phase of stable mass transfer during which the system manifests itself as an ultra-luminous X-ray (ULX) source. For a limiting flux sensitivity of $2\times10^{-14}\rm\, erg\,cm^{-2}\,s^{-1}$ (anticipated eRASS4 sensitivity), the eROSITA survey can detect bright ULXes with $L_{\rm x}\sim 10^{40}\rm\, erg\,s^{-1}$ up to a distance of 60 Mpc. Thus, eROSITA may detect or place useful constraints on the ULX precursor for future LFBOTs as close as AT2018cow ($D\simeq 60\rm\, Mpc$).

Finally, in the model of Tsuna \& Lu (2025), the micro-TDE occurs after some delay due to the time it takes for the NS/BH to encounter the companion star. For plausible pre-SN orbital separations of $10R_\odot$ up to 1 AU, the delay time is of the order 1--$30\rm\, d$, with the shortest delay times preferred as the probability of having a micro-TDE increases for the closest orbital separations. Their model predicts that, during this delay phase, stripped-envelope SN emission should be observed, with luminosities in the range $10^{40}$--$10^{42}\rm\, erg\,s^{-1}$, depending on the ejecta mass, delay time, and potential CSM interaction. On the other hand, the model by Klencki \& Metzger (2025) predicts the onset of a luminous red nova (LRN) caused by rapid envelope ejection during unstable mass transfer days before the final BH+He-core merger. These two different precursors may be tested by future observations with the Vera Rubin Observatory. 





\subsubsection{Magnetar-powered SNe}\leavevmode\\

\noindent
The possibility of a millisecond magnetar central engine, as originally proposed by Kasen \& Bildsten (2010) \cite{2010ApJ...717..245K} and Woosley (2010) \cite{2010ApJ...719L.204W}, has been considered by many authors \cite{2018ApJ...865L...3P, 2019ApJ...872...18M, 2019ApJ...871...73H, 2019ApJ...878...34F, 2019arXiv190409604W, 2020ApJ...888L..24M, 2024ApJ...963L..13L}. This is a natural extension from the magnetar-powered SLSNe-I to LFBOTs with unusually low-mass ejecta of $M_{\rm ej}\sim 0.3$ to $0.5M_\odot$. The magnetar model can reproduce the bolometric light curve in the first month or so. However, as has been pointed out \cite{2023ApJ...955...43C}, it is difficult to explain the late-time plateau in the same framework. The very flat light curve in the plateau phase is inconsistent with the expected rapid decay of the magnetar powered UV/optical light curve --- as we expect $L\propto t^{-4}$ due to the combination of the decaying spin-down power $L_{\rm sd}\propto t^{-2}$ and the optical depth ($\sim$ reprocessing efficiency) of the ejecta $\tau\propto t^{-2}$ \cite{2015ApJ...799..107W, 2015MNRAS.452.1567C}. Another difficulty is the RJ-like SED of the late-time plateau, which requires a compact ($\sim40R_\odot$), optically thick source --- this is more consistent with an accretion disk. It is possible that the central engine is powered by disk accretion onto a magnetar, and in this case, the engine power depends on the interactions between the disk and the NS magnetosphere (see e.g., \cite{2025arXiv250714284L}).


\subsubsection{Failed SNe, fallback accretion, and very massive stars}\leavevmode\\

\begin{figure}[tb]
    \centering
    \includegraphics[width=0.7\linewidth]{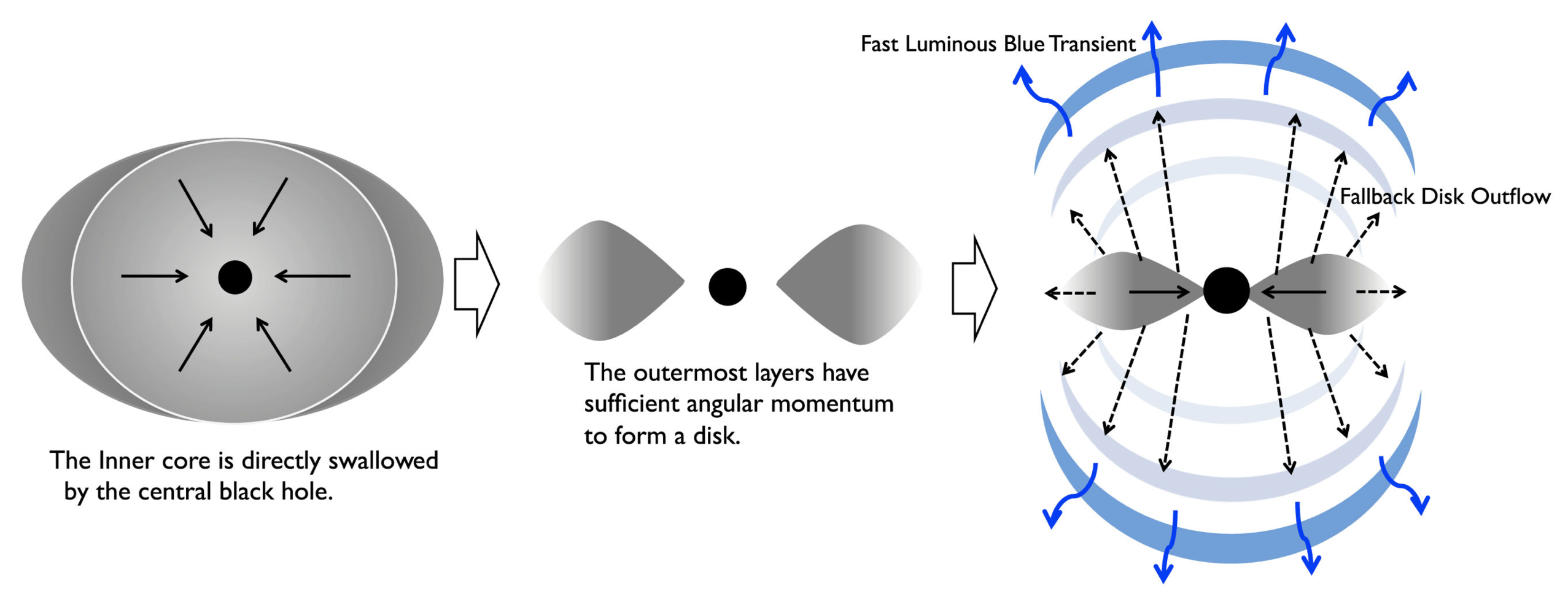}
        \caption{Model proposed by Kashiyama \& Quataert (2015) \cite{2015MNRAS.451.2656K}. Following the core-collapse of a rapidly rotating star, the inner regions directly fall into a black hole whereas the outer regions have sufficient angular momentum to form an accretion disk. The emission from the accretion disk wind then powers a fast blue optical transient. Figure reproduced from Kashiyama \& Quataert (2015) \cite{2015MNRAS.451.2656K} with permission.
        }
    \label{fig:KQ15_model}
\end{figure}

\noindent 
LFBOTs may be the signature of massive-star collapse where an ordinary core-collapse SN fails, leaving a newly born BH whose fallback accretion, disk winds, or jets power the fast blue emission and late activity. Such a model, schematically shown in Figure~\ref{fig:KQ15_model}, has been considered by Kashiyama \& Quataert (2015) \cite{2015MNRAS.451.2656K} before LFBOTs emerged as a new class of transients. After the discovery of the LFBOT prototype AT2018cow, many other follow-up studies have considered the Kashiyama-Quataert picture \cite{2019MNRAS.484.1031P, 2019ApJ...872...18M, 2024A&A...691A.329C, 2025arXiv250821116T, 2026A&A...706A.327C}. For instance, Chrimes et al. (2026) \cite{2026A&A...706A.327C} explored whether LFBOTs could arise from the core collapse of rapidly rotating very massive stars, and concluded that such a channel is plausible for at least some LFBOTs but faces challenges, including host metallicities, dense environments, and the need for large fallback masses/angular momentum to power long-lived emission.

The Kashiyama \& Quataert model applies if the outer layers of the collapsing star have sufficient angular momentum (due to stellar rotation) to circularize beyond the ISCO radius of the newly formed BH. Subsequently, super-Eddington accretion onto the BH leads to fast disk winds and, as the wind expands to larger radii, the initially trapped radiation can diffuse out near the so-called photon trapping radius where the diffusion timescale becomes comparable to the dynamical expansion timescale. The escaping radiation produces a thermal, fast blue optical transient. As the disk viscously spreads over time, the outer regions will eventually collapse to a geometrically thin structure which could produce a long-lived UV/optical plateau.

Confronted with LFBOT data, the original Kashiyama \& Quataert model has two major difficulties. First, the radiative efficiency is generally very low, because the disk wind needs to expand from near the circularization radius (near the ISCO) to the photon trapping radius (much larger), and the radiation energy is largely lost due to adiabatic expansion --- nearly all the energy output from the disk wind is in the form of kinetic energy. For instance, in the fiducial parameters of Kashiyama \& Quataert (2015) \cite{2015MNRAS.451.2656K}, the circularization radius is $r_{\rm cir}\sim 10r_{\rm g}\sim 10^7\mr{\,cm}$ (where $r_{\rm g}=GM/c^2$) and the trapping radius is $r_{\rm tr}\sim 10^{14}\rm\, cm$, and this leads to an adiabatic loss factor of $(r_{\rm tr}/r_{\rm cir})^{-2/3}\sim 10^{-5}$, where the power of $-2/3$ comes from the expansion of the wind (see e.g., \cite{2020ApJ...894....2P}). For an initial disk mass of $M_{\rm d,0}\sim 1M_\odot$, the resulting peak bolometric luminosity is only $L_{\rm pk}\sim 10^{43}\rm\, erg\,s^{-1}$ and the total radiated energy (over the peak duration of $\sim 1\rm\, d$) is only $\sim 10^{48}\rm\, erg$, even though the total accretion energy from the system is $(r_{\rm g}/r_{\rm cir})M_{\rm d,0}c^2\sim 10^{53}\rm\, erg$. 

The second issue of the failed SN model, at least in the cases where the circularization radius is close to the ISCO, is that the disk likely does not carry sufficient angular momentum to explain the late-time UV/optical plateau. For an initial disk mass of $M_{\rm d,0}\sim 1M_\odot$ and radius of $r_{\rm cir}\sim 10^7\rm\, cm$, we find the initial angular momentum to be $J_{\rm d,0}\sim \sqrt{GMr_{\rm cir}} M_{\rm d,0} \simeq 2\times 10^{50} \mr{\, erg\,s}, (M/10M_\odot)^{1/2} (r_{\rm cir}/10^7\mr{\,cm})^{1/2} (M_{\rm d,0}/M_\odot)$. This should be compared with the disk angular momentum in the plateau phase $J_{\rm d}\sim 10^{51}\mr{\, erg\,s}\, (M/10M_\odot)^{1/2} (r_{\rm d}/40R_\odot)^{1/2} (M_{\rm d}/10^{-2}M_\odot)$, where we have taken a minimum \textit{current} disk mass of $M_{\rm d}\sim 10^{-2}M_\odot$ required by the disk accretion model. We see that the \textit{initial} disk angular momentum in the original Kashiyama \& Quataert is insufficient. To make matters worse, as the disk viscously expands to larger radii, the disk wind carries away angular momentum such that the disk angular momentum years after the core collapse will have decreased substantially from the original one. It is possible that the outer layers of the progenitor star have more angular momentum (as pointed out by Chrimes et al. 2026 \cite{2026A&A...706A.327C}), and hence the circularization radius is further out and the angular momentum issue becomes less severe. However, in such cases, the accretion efficiency of the material in the \textit{outer} disk would be much lower, and the resulting peak luminosity would be lower.

The same issue of low radiative efficiency applies to the class of models based on \textit{only} the disk accretion in micro-TDEs \cite{2021ApJ...911..104K}. This issue can be addressed if the disk winds can strongly interact with and deposit its kinetic energy into some dense matter at larger radii, e.g., the CSM due to pre-core collapse mass loss or the ejecta in the picture of failed SNe or fallback disk. For instance, Tsuna et al. (2025) \cite{2025arXiv250821116T} proposed that the core collapse of rapidly rotating blue supergiants lead to long-lived ($10^4$--$10^5\rm\, s$) accretion onto spinning BHs and that the interactions between the powerful disk wind and a low-mass ejecta may power transients similar to SN2011kl associated with an ultra-long GRB \cite{2015Natur.523..189G}. Many of their models have fairly large disk circularization radii $r_{\rm cir}\sim 10^9\rm\, cm$, which corresponds to $J_{\rm d,0}\sim\,$a few$\times10^{51}\rm\, erg\,s$, which is closer to the minimum angular momentum required to explain the UV/optical plateau in AT2018cow. However, this model does not easily explain the formation of a large-scale ($\gtrsim 10^{16}\rm\,cm$) CSM required to explained the radio emission.

\subsubsection{CSM interactions}\label{sec:CSM_interactions}
\leavevmode\\

\noindent
CSM interactions provide another possible way to generate fast, blue, luminous emission: the conversion of kinetic energy into radiation is efficient if a fast, low-mass ejecta/disk wind component collides with dense material at radii $10^{14}$--$10^{15}\rm\,cm$. Models include pulsational pair-instability SNe \cite{2020A&A...640A..56R, 2020ApJ...903...66L}, extreme tail of Type Ibn/Icn SNe \cite{2019MNRAS.488.3772F, 2021ApJ...910...42X, 2022ApJ...926..125P}, and mergers between a BH and a He star/core \cite{2022ApJ...932...84M, 2026ApJ..1005....2K}. Such models are attractive because several LFBOTs show direct evidence for fast shocks interacting with dense CSM at larger radii $\gtrsim 10^{16}\rm\,cm$ through their luminous radio/mm emission --- it is possible that even denser material exists at smaller radii in an aspherical distribution. However, a pure CSM-interaction model must also account for the hot (initially featureless) spectra, rapid light curve evolution, and luminous variable X-rays.

\begin{figure}[b]
    \centering
    \includegraphics[width=0.4\linewidth]{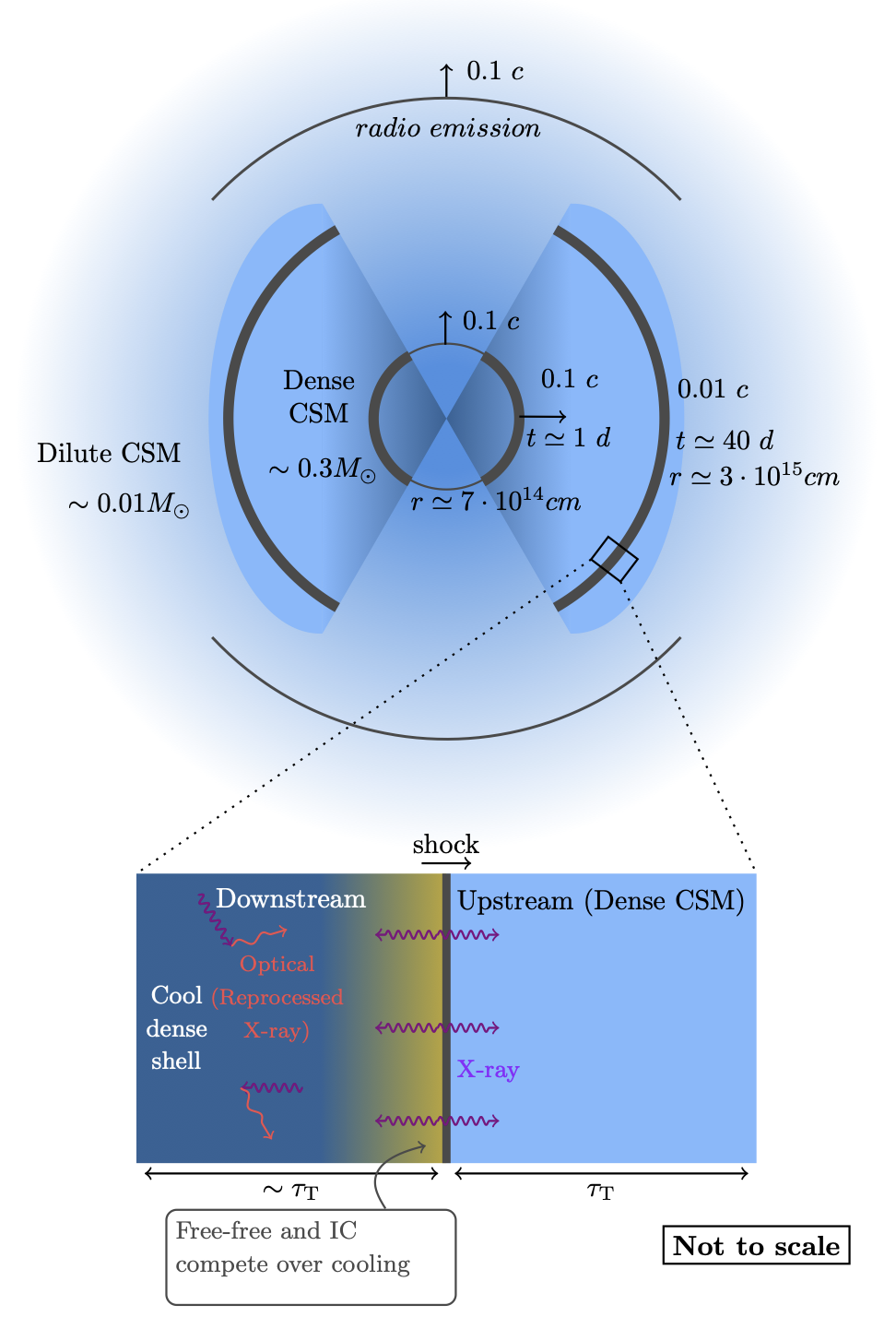}
        \caption{Model proposed by Govreen-Segal et al. (2026) \cite{2026arXiv260118887G}. An explosion or engine-driven outflow drives a forward shock into an aspherical, equatorially concentrated CSM. Near the denser equatorial regions, the shock- heated electrons produces X-ray emission and a cool dense shell in the far downstream reprocesses the X-ray photons into the UV/optical bands. Near the dilute polar regions, the non-thermal electrons accelerated by the forward shock produce radio emission. Figure reproduced from Govreen-Segal et al. (2026, in submission) \cite{2026arXiv260118887G} with permission.
        }
    \label{fig:GS26_model}
\end{figure}

CSM interactions have recently been developed into a much more comprehensive model for AT2018cow by Govreen-Segal et al. (2026) \cite{2026arXiv260118887G}. They proposed that the UV/optical and X-ray emission can be powered by a shock propagating through an aspherical CSM   --- as schematically shown in Figure~\ref{fig:GS26_model}. In their picture, hot post-shock electrons produce X-rays via free-free emission, while soft X-ray photons are reprocessed by the cool dense shell \textit{downstream} of the shock into the UV/optical emission. They attempt to explain the puzzling features of AT2018cow with a fast powerful ejecta of energy $E_{\rm ej}\sim 1$--$5\times 10^{50}\rm\, erg$ and velocity $v_{\rm ej}\sim 0.1c$ (corresponding to ejecta mass of $M_{\rm ej}\sim 0.01$--$0.05M_\odot$) interacting with a dense asymmetric CSM with mass of the order $M_{\rm CSM}\sim 0.3M_\odot$ extending to a few $\times10^{15}\rm\,cm$. Their model explains the broadband emission (radio to X-rays) after the shock breakout.

Several aspects of the model are worth emphasizing.  First, the soft X-ray spectrum, approximately $F_\nu\propto\nu^{-\beta}$ with $\beta\simeq0.6$, is attributed to thermal free-free emission from a distribution of electron temperatures expected in the fast-cooling post-shock gas (see \S \ref{sec:X-ray_emission}).  Second, the ratio of UV/optical to soft X-ray luminosity after shock breakout is controlled by radiative transfer through the scattering-dominated CSM. When the scattering optical depth $\tau_{\rm s}\gg1$, only a fraction $\sim\tau_{\rm s}^{-1}$ of the X-ray power escapes directly, while the remaining power is reprocessed by the cool dense shell into the UV/optical bands.  At later times, $t\gtrsim20\rm\,d$, as the CSM becomes optically thin, the UV/optical luminosity is expected to track the soft X-rays. 
Third, the hard X-ray hump above $\sim10\rm\,keV$ is attributed to the opacity transition between bound-free and electron scattering, instead of Comptonization (eq. \ref{eq:Compton_temperature}). X-ray photons at lower energies $\lesssim 10\rm\, keV$ can be efficiently reprocessed by the cool dense shell due to bound-free absorption; whereas the propagation of higher energy photons $\gtrsim 10\rm\, keV$ is mainly controlled by electron scattering, so they can escape the CSM after undergoing many scatterings.
Finally, X-ray variability on timescales of a few days is explained by a global radiative shock instability, while intermediate-width emission lines (e.g. H$\alpha$) may arise once the forward shock decelerates and the shocked CSM enters the line-cooling-dominated regime.

The model of Govreen-Segal et al. (2026) focuses on the detailed shock structure and radiative processes in CSM interaction, but it does not specify the physical origin of the fast ejecta. They mention possible channels such as white-dwarf accretion-induced collapse (AIC; see \S\ref{sec:AIC}) and tidal disruption of a He star by a BH. 
In such scenarios, an accretion disk is naturally formed, and powerful winds from the inner disk may provide the fast ejecta required by the model. 
In a binary-evolution context, the dense asymmetric CSM may be produced by high-rate mass transfer and associated L2 mass loss \cite{2023MNRAS.519.1409L, 2025ApJ...990..172S}. We also note that Govreen-Segal et al. (2026) primarily model the evolution after shock breakout, when the CSM optical depth has dropped to $\tau\lesssim c/v\sim10$. The earlier evolution before and around peak luminosity is therefore not explicitly modeled, and further work is needed to connect this post-breakout CSM-interaction picture to the full rise and peak of the transient. Finally, multi-dimensional hydrodynamic instabilities in the cool dense shell formed behind the radiative shock may strongly modify the X-ray radiative efficiency as well as the X-ray spectrum \cite{2018MNRAS.479..687S}.

We conclude that this model provides an attractive shock-powered alternative to 
central-engine interpretations, but it shifts the main burden to the progenitor: the system must produce both fast low-mass ejecta and a massive, compact, asymmetric CSM shortly before explosion.


\subsubsection{GRB jet cocoon emission}\leavevmode\\

\noindent
Another possible origin of the early fast blue emission is shock-cooling radiation from a GRB jet cocoon. As a relativistic jet propagates through a massive-star envelope, it deposits energy into a hot cocoon; after breakout, this cocoon expands quasi-spherically and can produce a short-lived UV/optical transient \cite{2017ApJ...834...28N, 2022MNRAS.513.3810G, 2022ApJ...931L..16D, 2025ApJ...985...21Z}. This mechanism is physically attractive because it naturally provides fast, hot, mildly relativistic material, and it connects LFBOTs to the broader family of engine-driven stripped-envelope explosions. However, applying this picture to AT2018cow-like events is challenging. 

For a compact Wolf-Rayet progenitor, strong adiabatic losses after breakout suppress the optical luminosity, so the observed luminosity requires either an unusually energetic cocoon or a much larger progenitor/effective breakout radius, as emphasized in the low-luminosity GRB cocoon/shock-breakout context \cite{2015ApJ...807..172N}. Moreover, classical long-GRB/cocoon scenarios are usually associated with broad-lined Type Ic SNe, whereas AT2018cow and the best-studied radio-loud LFBOTs show early blue, largely featureless spectra rather than clear Ic-BL SN features \cite{2019MNRAS.484.1031P,2019ApJ...872...18M}. 
Thus, while cocoon emission may contribute to the early UV/optical light in some engine-driven LFBOTs, a standard GRB jet cocoon from a compact stripped star is unlikely to explain the full AT2018cow phenomenology without additional ingredients such as an extended envelope/CSM, unusually large cocoon energy, or suppression/obscuration of the accompanying Ic-BL supernova.


\subsubsection{IMBH TDE}\label{sec:IMBH_TDE}\leavevmode\\

\noindent
A TDE by an intermediate-mass black hole (IMBH) has also been proposed as an explanation for LFBOTs (e.g., \cite{2019MNRAS.484.1031P,2019MNRAS.487.2505K, 2020ApJ...895L..23C, 2024A&A...691A.228C, 2024ApJ...977..162G}), motivated by the short debris fallback timescales expected for lower-mass black holes and more recently by the long-lived accretion disk model for the UV plateau. Two flavors of TDEs have been discussed based on the disruption of either a main-sequence star or a white dwarf. The main-sequence TDE scenario has the advantage that the H-rich stellar debris and disk may reproduce the Balmer lines (e.g., H$\alpha$) seen in many LFBOTs; whereas even a DA-type white dwarf has very little H, typically $\lesssim 10^{-4}M_\odot$ \cite{2010A&ARv..18..471A}, because CNO shell burning consumes more massive hydrogen envelopes during the hot pre-white-dwarf evolution. Moreover, in the aftermath of a white-dwarf TDE, most H will be lost in the early disk evolution on a viscous timescale $\lesssim10^3\rm\, s$, so we do not expect the delayed Balmer emission as discussed in \S \ref{sec:Balmer_lines}. In the following, we focus on the main-sequence TDE scenario, which is strongly constrained by the fallback and disk formation timescales.

For a main-sequence star disrupted by a black hole of mass $M$, the fallback time of the most bound debris scales approximately as
\begin{equation}
    t_{\rm fb} \simeq 30\,{\rm d}\,
    \left({M\over 10^6M_\odot}\right)^{1/2},
\end{equation}
up to order-unity factors depending on the stellar structure and penetration factor \cite{2022MNRAS.517L..26C,2026ApJ...998...81B}. 
Very massive black holes $M\gg 10^3M_\odot$ are immediately disfavored for LFBOTs with few-day rise times. 

Moving to lower-mass IMBHs shortens $t_{\rm fb}$, but does not by itself guarantee prompt accretion: hydrodynamic simulations and analytic studies of TDE debris have shown that circularization can be inefficient, with the debris remaining on eccentric, extended orbits and the disk-formation/circularization time exceeding $t_{\rm fb}$ by factors of several to tens (e.g., \cite{2009ApJ...697L..77R, 2014ApJ...783...23G, 2015ApJ...804...85S, 2025arXiv251210564M, 2026ApJ..1001...71A}). Recent radiation hydrodynamic simulations \cite{2025arXiv251210564M} suggest that, due to photon trapping by the optically thick wind, the early emission from IMBH TDEs is likely capped near the Eddington luminosity $L_{\rm Edd}\sim 10^{42}\mr{\,erg\,s^{-1}}\, (M/10^4M_\odot)$, which is much below the observed peak luminosity of LFBOTs. This weakens the link between the fallback rate and the observed bolometric light curve, especially for the rapidly evolving optical emission. 

Finally, the environments of AT2018cow-like LFBOTs provide an additional challenge: AT2018cow occurred in a star-forming region whose local stellar population is typical of core-collapse supernova environments \cite{2023MNRAS.519.3785S}, rather than in a dense-cluster-like environment naturally associated with many IMBH TDE scenarios. Inkenhaag et al. (2023) \cite{2023MNRAS.525.4042I} showed that HST photometry in the UV/optical plateau phase allows a star cluster mass up to a few times $10^4M_\odot$ for a cluster age $\lesssim 30\rm\, Myr$ (required by the young stellar environment). Even for a cluster mass near the upper limit, this is not the most natural environment to produce an elevated main-sequence TDE rate. Future observations can test the variability of the plateau emission and hence constrain the fraction of the optical emission contributed by a potential star cluster. 

Therefore, we disfavor the IMBH TDE scenario, because white dwarf TDEs are inconsistent with the detection of Balmer lines in many LFBOTs, and main sequence TDEs face the bottleneck of inefficient circularization of the stellar debris. In either case, a plausible model also needs to place the IMBH in the observed young stellar environment \cite{2020MNRAS.495..992L, 2023MNRAS.519.3785S, 2025MNRAS.544L.108I}. Nevertheless, periodicity/QPO searches in the X-ray lightcurves of LFBOTs provide interesting tests for the IMBH TDE scenario.




\subsubsection{Accretion-induced collapse of white dwarfs}\label{sec:AIC}
\leavevmode\\

Accretion-induced collapse (AIC) of a white dwarf to a NS provides another possible route to a fast, engine-powered transient. 
Canonical AIC models naturally produce a compact remnant, low ejecta mass, and short diffusion time \cite{2006ApJ...644.1063D,2009MNRAS.396.1659M,2010MNRAS.409..846D}. 
However, ordinary radioactive AIC transients are expected to be relatively faint compared with LFBOTs, unless additional energy injection is present. 
Thus, if AIC is relevant to LFBOTs, 
the collapse must either produce a central engine with $E_{\rm eng}\sim 10^{50}$--$10^{51}\rm\,erg$ whose energy can be deposited in, or diffuse out through, the ejecta on a timescale of days, or the AIC ejecta must interact with a dense CSM as in Govreen-Segal et al. (2026) \cite{2026arXiv260118887G}.

An engine-powered version of AIC was proposed by Lyutikov \& Toonen (2019) and Lyutikov (2022) \cite{2019MNRAS.487.5618L,2022MNRAS.515.2293L}. 
In this model, LFBOTs arise from the electron-capture collapse of the remnant of a double white dwarf (WD) merger, in which a massive ONeMg WD merges with another WD. 
The merger disrupts the less massive WD and forms a shell-burning, non-degenerate envelope around the ONeMg core. 
During the subsequent $\sim10^2$--$10^4\rm\,yr$ shell-burning phase, much of the envelope is lost in a fast wind. 
If the ONeMg core reaches the Chandrasekhar mass before the envelope is completely lost, it collapses to a NS with only a small remaining envelope. 
The resulting ejecta can have a low mass, $M_{\rm ej}\sim10^{-2}$--$10^{-1}M_\odot$, mildly relativistic velocities, and kinetic energy of order a few $\times10^{50}\rm\,erg$, giving a diffusion time compatible with a fast optical transient.

The key difference from canonical AIC is that the luminosity is powered primarily by the newborn NS rather than by radioactive heating. 
The newly formed NS is assumed to be rapidly rotating and strongly magnetized, launching a long-lived relativistic wind. 
This wind drives a radiation-dominated shock through the low-mass ejecta, powering the peak optical emission, while the wind termination shock can produce high-energy emission like in a pulsar-wind nebula. 

This scenario has several attractive features: it naturally provides low ejecta mass, fast outflows, a compact central engine, and dense pre-explosion wind material. 
Moreover, the predicted delay time distribution for CO-ONeMg WD mergers peaks at $\sim50$--$100\rm\,Myr$, making their host galaxies more similar to those of core-collapse SNe than to old Type Ia progenitors.

Nevertheless, important tensions remain. The model requires a rather specific WD-merger outcome: the system must avoid thermonuclear disruption, lose most but not all of its envelope, and collapse only when the remaining ejecta mass is small enough to produce an LFBOT-like diffusion time. More importantly, it is unlikely that the system can reproduce the observed Balmer line emission. Even if the disrupted secondary was a DA white dwarf, the H mass ($\lesssim 10^{-4}M_\odot$) is much too small compared to what is needed to power the observed Balmer line luminosities of the order $L_{\rm H\alpha}\sim 10^{39}\rm\, erg\,s^{-1}$ lasting for tens of days. It is also difficult for the accretion disk to carry sufficient angular momentum at late time to explain the long-lived UV plateau seen in AT2018cow. Finally, even a $\sim50$--$100\rm\,Myr$ delay may be too long if the local environments of LFBOTs robustly indicate young massive-star progenitors (as is the case for AT2018cow \cite{2023MNRAS.519.3785S}). Thus, magnetar-boosted AIC is an interesting route to a low-mass, engine-powered transient with long-lived X-rays, but it is inconsistent with AT2018cow-like events in massive-star environments and with the Balmer emission.

\subsubsection{Summary of progenitor models}\label{sec:summary_progenitor_models}
\leavevmode\\

Table~\ref{tab:model_comparison} summarizes the strengths and limitations of the main proposed LFBOT models. No model yet explains the full phenomenology from progenitor evolution and engine formation to the early broadband emission and late-time source. Multiple origins may contribute. 

\begin{landscape}
\begin{table*}
\centering
\caption{
Qualitative assessment of LFBOT models against the requirements summarized
in Table~\ref{tab:physical_requirements}. A check mark (\good) denotes a
feature that is naturally accommodated or has been at least
semi-quantitatively demonstrated; a triangle (\partialok) denotes a plausible
but incomplete explanation; a cross (\bad) denotes a substantial tension; and
a question mark (\unknown) denotes a feature that has not yet been
meaningfully addressed. The symbols summarize the present state of the
models rather than constitute a formal model ranking.
$^\dagger$CSM interaction is primarily an emission mechanism rather than a
complete progenitor channel; the assessment refers mainly to the compact,
asymmetric interaction model discussed in \S~\ref{sec:CSM_interactions}.
}
\label{tab:model_comparison}

\setlength{\tabcolsep}{3pt}
\renewcommand{\arraystretch}{0.95}

\begin{tabularx}{\linewidth}{
    >{\raggedright\arraybackslash}p{0.14\linewidth}
    *{8}{>{\centering\arraybackslash}p{0.047\linewidth}}
    >{\raggedright\arraybackslash}X
}
\toprule
\textbf{Model} &
\textbf{$L_{\rm pk}, t_{\rm pk}$} &
\textbf{X-rays} &
\textbf{Radio/ CSM} &
\textbf{H/He} &
\textbf{NIR} &
\textbf{Late UV} &
\textbf{Young env.} &
\textbf{Rate} &
\textbf{Principal gap or distinctive test} \\
\midrule

Merger between compact object \& He core &
\partialok & \partialok & \good & \partialok & \partialok & \good & \good & \partialok &
Predicts a long-lived ULX phase and an LRN-like precursor; should prefer very
young stellar populations. The full disk/wind/CSM radiation problem remains
uncalculated. \\

\midrule

SN+micro-TDE &
\good & \partialok & \partialok & \good & \unknown & \good & \good & \partialok &
Predicts a faint stripped-envelope SN before the main event and slightly
older local populations than the He-core merger channel. Disk formation and
wind-ejecta interaction require (radiation-)hydrodynamic calculations. \\

\midrule

Magnetar-powered SN &
\good & \partialok & \unknown & \unknown & \unknown & \bad & \good & \unknown &
Can reproduce the early bolometric light curve for low ejecta mass, but the
compact, slowly evolving late plateau is difficult to explain with ordinary
magnetar spin-down. \\

\midrule

Failed SN w/ fallback disk &
\unknown & \partialok & \unknown & \unknown & \unknown & \bad & \partialok & \unknown &
Severe adiabatic losses suppress the radiative peak when the disk forms close
to the ISCO, while the initial disk may lack the angular momentum required
for the late disk and UV plateau. \\

\midrule

Compact CSM interaction$^\dagger$ &
\unknown & \good & \good & \unknown & \partialok & \bad & \partialok & \unknown &
Explains much of the post-breakout broadband emission, but the pre-breakout
rise and peak remain to be modeled. It also requires a progenitor that
produces both fast low-mass ejecta and compact, massive, asymmetric CSM. \\

\midrule

GRB jet cocoon &
\partialok & \unknown & \partialok & \bad & \unknown & \bad & \good & \unknown &
A standard cocoon from a compact Wolf--Rayet star suffers strong adiabatic
losses and would normally accompany an Ic-BL SN or GRB. An extended envelope
or CSM is required. \\

\midrule

Main-sequence IMBH TDE &
\bad & \partialok & \partialok & \partialok & \partialok & \partialok & \unknown & \unknown &
Must circularize and form a disk within days and place an IMBH in the
observed young stellar environment. An underlying star cluster is a key test. \\

\midrule

Accretion-induced collapse &
\partialok & \partialok & \partialok & \bad & \unknown & \unknown & \bad & \unknown &
Naturally gives low ejecta mass and a compact remnant, but lack of sufficient
H-rich material and the older expected delay-time distribution are serious
difficulties. \\

\bottomrule
\end{tabularx}

\end{table*}
\end{landscape}

\section{Future Directions}
\label{sec:future}

\subsection{Suggested observations}\label{sec:future_obs}

To make progress in determining the progenitors of LFBOTs, we recommend several avenues of observational work.

First, while the number of LFBOTs is growing (Table~\ref{tab:summary}), the number with certain types of observations---particularly bolometric UV to NIR light curves and X-ray light curves---remains small, and could be substantially increased with existing facilities. 

Second, certain observations are only possible for the nearest LFBOTs, and should be prioritized at the next opportunity. These include intensive X-ray monitoring to search for variability and/or periodicity, intensive mm/sub-mm band monitoring to search for variability in flux and spectral index observed in AT2018cow, and cm/mm wave VLBI for a direct source size measurement\footnote{At 22d, AT2018cow had a radius of $R\approx7\times10^{15}\,$cm, and was 90 (50) mJy at 100 (230) GHz. The corresponding source size (angular diameter) is $\approx14\,\mu$as. This would be within reach for ground-based VLBI.} and proper motion search. Searches for precursor emission in Rubin Observatory (as the most sensitive wide-field optical survey) data should be carried out. For instance, a micro-TDE may follow a supernova, while something like a luminous red nova might be expected before the merger of a He core and a compact object. Searches in data from deep X-ray surveys could reveal a pre-existing ultraluminous X-ray source, another expectation in the compact object-star merger model. JWST observations during the transient itself could reveal the origin of the NIR excess (dust echo vs. free-free), while observations after the transient fades could help reveal an underlying globular or stellar cluster or free-free emission from the disk wind, which would put significant constraints on the progenitor model. HST should be used to search for late-time plateaus in other $z<0.1$ events. Spectroscopy or narrow-band imaging could help constrain late-time ($t\gtrsim 1\rm\, yr$) H$\alpha$ emission from a potential disk wind.

Third, we recommend certain analyses of the existing data. It would be useful to revisit all the X-ray observations for AT2018cow and come to an understanding of how to reconcile the different variability/periodicity claims.  We suggest quantifying the variability timescale of the NIR excess by subtracting the (potentially model-dependent) ``photospheric'' contribution from the total NIR flux. We also suggest re-modeling the radio evolution of all LFBOTs in a consistent fashion (preferably forward modeling), as different papers make different assumptions and take different approaches. 

We suggest searching for ``LFBOTs'' in other bands of the electromagnetic spectrum, such as the radio and X-rays, in order to understand to what extent the current optical approach imposes a selection bias. Possibilities could include the Deep Synoptic Array (although LFBOTs are much less luminous at 2\,GHz than at higher frequencies), the Simons Observatory (for 100\,GHz), and the Einstein Probe for X-rays. It would be particularly useful to understand if there exist lower luminosity versions in the optical band, which might not be found efficiently by optical surveys. For all objects, we recommend population-level studies of the local host-galaxy environments using HST, to constrain the ages of the stellar populations. 

The discovery rate of LFBOTs has the potential to increase by orders of magnitude in the coming years, with the optical-NIR sensitivity of the Vera Rubin Observatory and the ultraviolet sensitivity of ULTRASAT and the Ultraviolet Explorer (UVEX). With very large samples of objects, it will be possible to measure rates and host-galaxy environments with significantly more precision than what is possible now. Very high cadence surveys (Argus Array, ULTRASAT) can detect LFBOTs much earlier in their evolution, measure their bolometric properties  during the rise phase of the light curve, and study the prevalence of short-duration flaring. 

\subsection{Suggested modeling/theory efforts}
\label{sec:future_theory}

Future theoretical work should increasingly aim to predict the full multiwavelength phenomenology of LFBOTs, rather than treating individual emission components in isolation.

First, end-to-end, multidimensional radiation-hydrodynamic calculations are needed to connect energy injection (e.g., a fast disk wind) by a central engine to the observed UVOIR, X-ray, and radio emission. Such calculations should follow shock propagation through the ejecta and circumstellar medium (CSM), photon diffusion around peak luminosity, and the subsequent emergence of X-rays as the optical depth decreases. Multi-group radiation transport and time-dependent non-LTE ionization calculations will be particularly important for predicting X-ray reprocessing and ionization breakout, the evolution of the UVOIR color temperature, and the delayed appearance of H and He emission lines. These calculations should include asphericity and viewing-angle effects explicitly, because asymmetric ejecta and CSM may simultaneously affect the X-ray escape fraction, optical polarization, and asymmetric line profiles. The same physical models, with additional microphysical parameters on non-thermal particles and magnetic field amplification, should ultimately be used to predict the radio emission, replacing one-zone equipartition estimates with forward calculations of the evolving shock interactions with the CSM. Well-observed events such as AT2018cow and AT2024wpp provide natural test cases for these calculations.

Second, the physics and long-term evolution of the central engine require further investigation. In accretion-powered models, (radiation-)hydrodynamic or (radiation-)magnetohydrodynamic simulations of accretion disk evolution should determine what fraction of the supplied mass reaches the compact object and how the released accretion energy is divided among radiation, winds, and collimated jets. The power and angular distribution of super-Eddington disk winds are needed as inputs to models of wind-ejecta and wind-CSM interactions. Calculations should also follow the disk over a much longer dynamic range in time, from the initial viscous timescale of hours to days, through the highly super-Eddington phase during the first weeks, to a possible geometrically thin phase at $t\gtrsim 1\rm\, yr$. In particular, it remains unclear when and where the (outer) disk transitions from geometrically thick to thin, how this transition affects the accretion rate and wind power, and what late-time spectral energy distribution is produced when the inner disk remains geometrically thick (super-Eddington) while the outer disk has become thin (sub-Eddington). The origin of recombination-line emission in disk winds should also be modeled using radiative transfer models, including the effects of wind stratification and irradiation.

Neutron-star accretors should be subjected to a quantitative consistency test. In particular, can accretion through a magnetosphere and onto a hard stellar surface generate powers of $\sim 10^{45}$--$10^{46}\ {\rm erg\ s^{-1}}$ for several days while launching the required fast and slow outflows? Addressing this question will require understanding the interaction between a highly super-Eddington disk and the neutron-star magnetosphere, including possible propeller phases, magnetic-field burial, and the partition of energy between radiation and mechanical outflows.

Third, progenitor models must be connected to (binary) stellar evolution, CSM formation, and event rates. For compact-object--star mergers and micro-TDE scenarios, hydrodynamic simulations should establish whether and how rapidly a disk forms, the mass and angular momentum retained by the disk, and the properties of the unbound matter during disk formation. Of particular interest are micro-TDEs produced when a newly formed neutron star or black hole interacts with a companion in a tight orbit. Simulations should determine the conditions for tidal disruption, the probability distributions of disk masses and angular momenta from this scenario, and whether the resulting disk is capable of supplying the required engine energy for LFBOTs on a timescale of days.

These calculations should be embedded in models of the preceding binary evolution, including mass transfer, mass loss through the outer Lagrange points, and potentially common-envelope evolution. A principal goal should be to predict the radial extent, density profile, composition, and asymmetry of the CSM, rather than treating these quantities as freely adjustable parameters. The same evolutionary calculations should predict possible precursor signatures. Examples include a luminous-red-nova-like precursor before a compact-object-He core merger, a pre-existing ultraluminous X-ray source, or a supernova preceding a natal-kick-induced micro-TDE. Similar end-to-end consistency is needed for failed-supernova, accretion-induced-collapse, magnetar, CSM-interaction, and jet-cocoon scenarios.

Binary population synthesis can then determine the rates, delay-time distributions, and host-metallicity dependence expected for each channel. This is particularly important given the very low inferred volumetric rate of LFBOTs. A viable progenitor model must explain not only the properties of individual events, but also why the required evolutionary outcome is so rare ($\lesssim 10^{-4}$ of the core-collapse rate).

Finally, theoretical effort is needed to determine how LFBOTs fit into the broader landscape of engine-powered transients. Present samples may select only systems in which the central engine is sufficiently powerful for a fast engine-driven wind or ejecta to escape, for luminous radio emission to be produced, and for X-rays to emerge through ionization breakout. Models should map the intrinsic engine power/duration, ejecta mass, CSM properties, and viewing angle onto the probability that an event would be identified as an optical, radio, and X-ray LFBOT. This would predict whether weaker engines with $E_{\rm eng}\sim10^{48}$--$10^{49}\ {\rm erg}$ produce lower-luminosity counterparts, events visible only in certain wavebands, or transients that are currently classified as more ordinary supernovae.

Engine-driven shocks propagating through dense ejecta or CSM may also accelerate cosmic rays and produce high-energy neutrinos through $pp$ and $p\gamma$ interactions. Quantitative calculations of the particle acceleration efficiency, meson-production optical depth, and neutrino spectrum are needed to determine whether meaningful multi-messenger tests are possible.

\section{Conclusions}

LFBOTs appear to represent a genuinely new phenomenon, with unprecedented characteristics across the electromagnetic spectrum. Although LFBOTs are rare in terms of their volumetric rate, the observations collected are among the most detailed and multiwavelength of any extragalactic transients, in large part because they are among the most luminous transients known from radio to X-ray wavelengths. The fact that they have been so challenging to explain is due to an incomplete understanding of certain physical processes that play a fundamental role in astronomy. 

We conclude that the best-observed LFBOTs (typified by AT2018cow) are engine-powered transients, likely involving accretion onto a black hole or neutron star, placing them alongside GRBs and superluminous supernovae. 
Unlike in GRBs and superluminous supernovae, however, which have a high ejecta mass and therefore an engine shrouded to the observer for years to decades, the central engine in LFBOTs is long-lived and becomes exposed to the observer on timescales of a few days to tens of days. This presents an opportunity to study newly formed accretion disks in systems that underwent super-Eddington accretion ($\gtrsim 10^{5}$ Eddington rates for stellar-mass accretors): a poorly understood regime of accretion that is important for black hole formation and growth at all mass scales, as well as for understanding the progenitors of multiple transient classes. 

The available evidence further suggests that the dominant channel in LFBOTs involves massive stars in binary systems. We expect a broad landscape of transients arising from such systems to exist, from ordinary stripped-envelope supernovae to the most extreme events likely represented by LFBOTs. Determining their occurrence rate, diversity of properties, and connection to other transient classes is therefore important for understanding the progenitors of gravitational-wave sources, and for the physics of binary mass transfer that creates the ambient medium. 

Finally, much of the emission in LFBOTs ultimately arises from shocks. The radio emission probes a poorly understood regime of trans-relativistic shocks, with implications for a wide variety of transient types (e.g., neutron star mergers, tidal disruption events). The X-rays are an interesting probe of radiative processes near shocks as well as ionization breakout from an optically thick medium.

\ack{
We thank Daniel Perley for constructive comments that helped shape the structure of this manuscript. 
We also thank Brian Metzger, Taya Govreen-Segal, Claudia Gutiérrez, Nayana AJ, Anya Nugent, and Daichi Tsuna for helpful comments that improved the scientific content and clarity of this review.
}

\funding{A.Y.Q.H. acknowledges support from a Sloan Research Fellowship (Award Number FG-2024-21320) from the Alfred P. Sloan Foundation, a Packard Fellowship from the David and Lucile Packard Foundation (Grant Number 2024-77386), National Aeronautics and Space Administration (NASA) grant 80NSSC24K0377, an LSST Scialog Early Science grant from the Research Corporation for Science Advancement, and {\it Hubble Space Telescope} ({\it HST}) grant HST-GO-17477.006-A. W. L.'s research is supported by a Sloan Research Fellowship (Award Number FG-2026-79505) from the Alfred P. Sloan Foundation and by an LSST Scialog Early Science grant from the Research Corporation for Science Advancement.
}

\roles{Both authors contributed heavily to the conceptualization, analysis, and writing (both the original draft and reviewing and editing) of the manuscript.}




\bibliography{refs}

\end{document}